\documentclass[12pt, letterpaper]{article}

\usepackage[english]{babel}

\usepackage[letterpaper,paperheight=27.5cm,paperwidth=21.25cm,top=2cm,bottom=2cm,left=2cm,right=2cm,marginparwidth=1.75cm]{geometry}

\usepackage{lastpage}
\usepackage[pagestyles]{titlesec} %

\usepackage{amsmath}
\usepackage{graphicx}
\usepackage{multirow}
\usepackage{array}
\usepackage{subfig}
\newcolumntype{P}[1]{>{\centering\arraybackslash}m{#1}}
\usepackage{authblk}
\usepackage{enumitem}
\usepackage{makecell}
\usepackage{pifont}
\newcommand{\cmark}{\ding{51}} 
\newcommand{\xmark}{\ding{55}} 

\usepackage[colorlinks=true, allcolors=blue]{hyperref}
\usepackage[
sorting=none,
style=numeric,
maxcitenames=2, 
uniquelist=false, 
giveninits=true, 
uniquename=init, 
backend=biber, 
url=false]{biblatex}
\usepackage{csquotes}
\usepackage[normalem]{ulem}
\usepackage[version=4]{mhchem}

\title{Automatic Segmentation of Organs at Risk in Head and Neck Cancer Patients from CT and MR scans}
\author[1,2]{S\'ebastien Quetin}
\author[3]{Andrew Heschl}
\author[3]{Mauricio Murillo}
\author[4]{Rohit Murali}
\author[1]{Piotr Pater}
\author[5] {George Shenouda}
\author[1,2,6,*]{Shirin A. Enger}
\author[3,7,8,*]{Farhad Maleki}
\affil[1]{Medical Physics Unit, Department of Oncology, McGill University, Montreal, QC, Canada}
\affil[2]{Montreal Institute for Learning Algorithms, Mila, Montreal, QC, Canada}
\affil[3]{Department of Computer Science, University of Calgary, Calgary, AB, Canada}
\affil[4]{Cloud Innovation Center, University of British Columbia, BC, Canada}
\affil[5]{Division of Radiation Oncology, Department of Oncology, McGill University, McGill University Health Centre, Montreal, QC, Canada}
\affil[6]{Lady Davis Institute for Medical Research, Jewish General Hospital, Montreal, QC, Canada}
\affil[7]{Department of Diagnostic Radiology, McGill University, Montreal, QC, Canada}
\affil[8]{Department of Radiology, University of Florida, Gainesville, FL, USA}

\affil[*]{Co-supervisors for this research.}
\date{}  

\begin{document}

\maketitle

\newpage 

\section*{Abstract}   

\textbf{Purpose:} To present a high-performing, robust, and flexible deep learning pipeline for automatic segmentation of 30 organs-at-risk (OARs) in head and neck (H\&N) cancer patients, using MRI, CT, or both.\\
\textbf{Method:} We trained a segmentation pipeline on paired CT and MRI-T1 scans from 296 patients. We combined data from the H\&N OARs CT and MR segmentation (HaN-Seg) challenge and the Burdenko and GLIS-RT datasets from the Cancer Imaging Archive (TCIA). MRI was rigidly registered to CT, and both were stacked as input to an nnU-Net pipeline. Left and right OARs were merged into single classes during training and separated at inference time based on anatomical position. Modality Dropout was applied during the training, ensuring the model would learn from both modalities and robustly handle missing modalities during inference. The trained model was evaluated on the HaN-Seg test set and three TCIA datasets. Predictions were also compared with Limbus AI software. Dice Score (DS) and Hausdorff Distance (HD) were used as evaluation metrics.\\
\textbf{Results:} The pipeline achieved state-of-the-art performance on the HaN-Seg challenge with a mean DS of 78.12\% and HD of 3.42 mm. On TCIA datasets, the model maintained strong agreement with Limbus AI software (DS: 77.43\% , HD: 3.27 mm), while also flagging low-quality contours. The pipeline can segment seamlessly from the CT, the MRI scan, or both.\\
\textbf{Conclusion:} The proposed pipeline achieved the best DS and HD scores among all HaN-Seg challenge participants and establishes a new state-of-the-art for fully automated, multi-modal segmentation of H\&N OARs.

\newpage

\section{Introduction} 

Recent advances in computational resources and neural network architectures have facilitated the development of automated segmentation methods leveraging large-scale datasets, leading to more robust and generalizable solutions~\cite{wasserthal_totalsegmentator_2023, totalsegmentator_mri}. While segmentation tools trained on extensive datasets demonstrate strong performance across diverse anatomical regions, comparable efforts for the head and neck (H\&N) region remain notably absent. Although limited training data is a well-recognized challenge in medical AI, segmentation of the H\&N region presents additional hurdles. In particular, H\&N scans raise high privacy concerns due to the presence of identifiable anatomical features, which makes sharing such data substantially more difficult compared to whole-body scans.

H\&N cancer is the seventh most common cancer worldwide and includes a heterogeneous group of malignancies affecting the upper aerodigestive tract~\parencite{argiris_head_2008}. Accurate segmentation of H\&N organs at risk (OARs) is critical for radiotherapy planning~\parencite{caudell_future_2017}, surgical navigation, and patient monitoring~\parencite{Xu_2022}. However, manual segmentation is labor-intensive and prone to inter-observer variability~\parencite{wong_comparing_2020}. 
Although deep learning methods have been applied for automated segmentation in the H\&N region, most are developed for either computed tomography (CT) scans or magnetic resonance imaging (MRI) independently. These approaches are often limited in scope, targeting only a small subset of structures~\parencite{he_multitrans_2023, gao_focusnetv2_2021, tappeiner_multi_organ_2019, luan_accurate_2024, singh_multi_organ_2024, tappeiner_tackling_2022}, and trained on relatively small datasets.



Despite MRI offering superior soft tissue contrast compared to CT~\parencite{khoo_comparison_1999, verhaart_relevance_2014}, not all patients undergo both imaging modalities. This has motivated studies focused on synthesizing MRI from CT in order to leverage MRI-like contrast for segmentation~\parencite{8494797,Li_Yu_Wang_Heng_2020,liu2020head,jiang2020self,lei2020ct,jiang2018tumor,jue2019integrating,yang2020synthetic}. 

Some studies have devised methods capable of segmenting organs interchangeably from both modalities in cardiac~\parencite{8764342} and abdominal images~\parencite{8354170}. Some studies used both modalities for segmenting organs in prostate~\parencite{8933421} and H\&N~\parencite{podobnik2023multimodal} cancer patients. The former directly learns from a weighted sum of the two modalities. The latter learns modality-specific features and merges them to predict OAR contours.\par  


In this retrospective study, we present a robust and flexible deep-learning pipeline for automated segmentation of 30 clinically significant OARs in H\&N cancer patients, using CT, MRI or both modalities in combination. Our approach is designed to leverage the complementary strengths of CT and MRI, namely, the widespread clinical availability of CT and the superior soft tissue contrast of MRI, while maintaining adaptability to scenarios where only a single modality is available. By enabling accurate, multi-modal OAR segmentation, our method addresses key clinical needs in radiation therapy planning and paves the way for more generalizable and modality-agnostic solutions.
%
\section{Materials and Methods}

\begin{figure}
    \centering
    \includegraphics[width=0.95\textwidth]{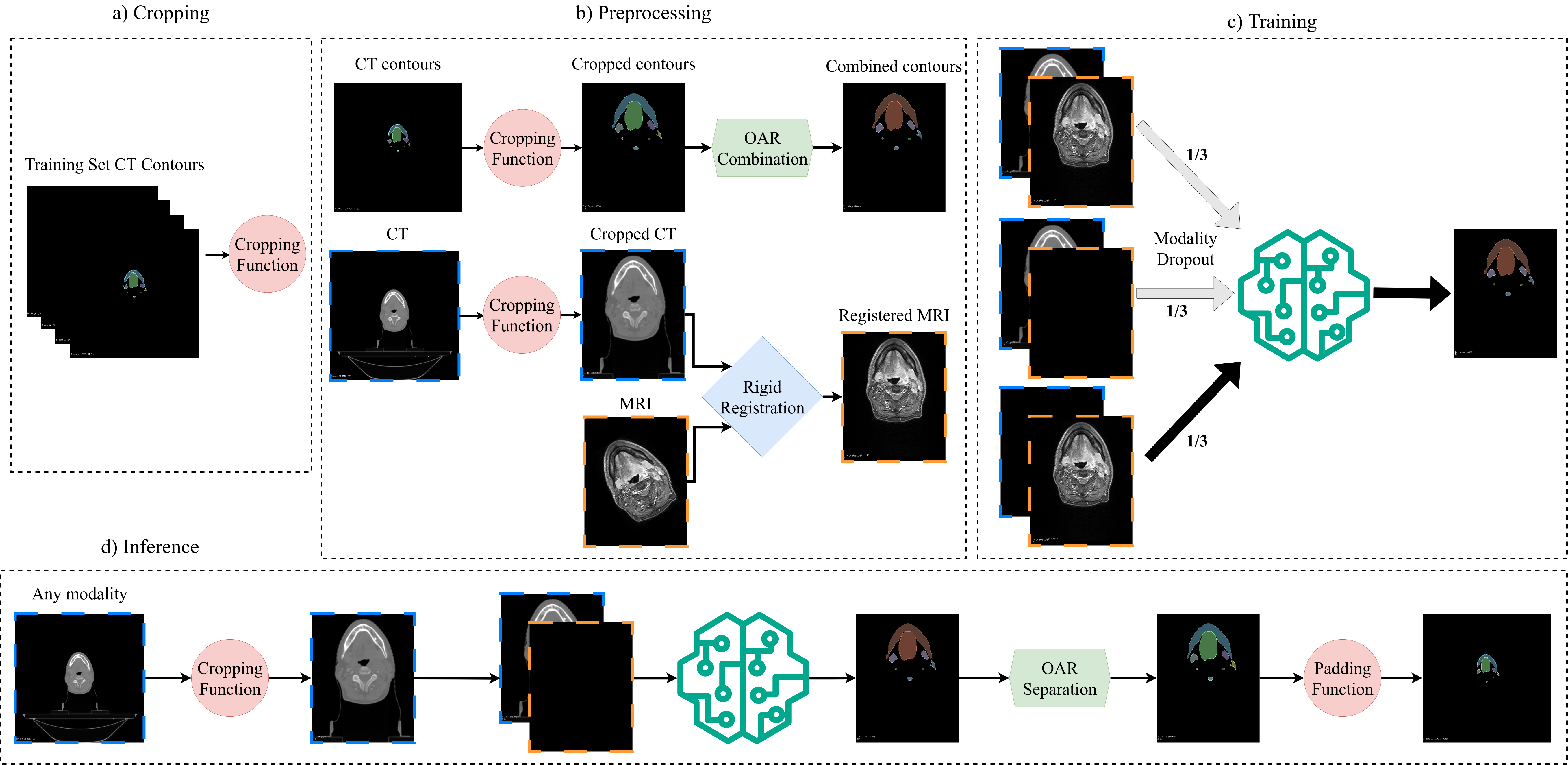}
    \caption{A schematic overview of the proposed approach. When training, modality dropout is employed. When running inference, only one modality is necessary for the pipeline to predict contours. In the case where only one modality is used, the remaining modality is set to zeroes.
    }
    \label{Fig:ProposedApproach}
\end{figure}

\subsection{Datasets} \label{datasetsection}
\subsubsection{Training Datasets}

The training of our segmentation pipeline followed a multistage approach, beginning with supervised training on the publicly available H\&N OARs CT and MR segmentation (HaN-Seg) dataset~\parencite{podobnik_han-seg_2023}. This dataset includes CT and MRI-T1 scans from 42 H\&N cancer patients, along with manual contours for 30 OARs delineated on the CT images, with MRI used to support soft tissue identification and following specific guidelines~\cite{han_contouring_guidelines}. Due to the limited field of view of the MRI volumes, some OARs, specifically the arytenoid cartilage, cricopharyngeal inlet, cervical esophagus, thyroid gland, and supraglottic larynx, were not visible in all MRI scans.

To expand the dataset and improve model generalization, we used the initial model trained on HaN-Seg to generate pseudo-contours on two additional datasets from The Cancer Imaging Archive (TCIA): the Glioma Image Segmentation for Radiotherapy (Glis-RT) dataset~\cite{glisrt_dataset, glisrt_paper} and Burdenko Glioblastoma Progression Dataset (BGPD)~\cite{burdenko_dataset}. Glis-RT contains data from 230 patients with glioblastoma and low-grade glioma treated with surgery and adjuvant radiotherapy at Massachusetts General Hospital (Boston, USA). For each patient, post-surgical MRI T1 and T2 FLAIR scans, used for target definition, CT scans, used for treatment planning, and manually delineated structures for targets and OARs
are available. BGPD contains data from 180 glioblastoma patients treated with radiotherapy at the Burdenko Neurosurgery Center (Moscow, Russia). Each patient has four MRI sequences (T1, T1C, T2, FLAIR), a CT scan, and associated radiotherapy treatment planning files.

Importantly, manual contours from Glis-RT and BGPD were not used during training. Instead, we applied the HaN-Seg-trained model to these datasets to generate pseudo-contours. These pseudo-labeled datasets, combined with the original HaN-Seg dataset, formed the final training set for our segmentation pipeline. The training strategy is detailed in Section~\ref{sec:dataset_expansion}.

\subsubsection{Evaluation Datasets}

Model evaluation was conducted in two stages. First, for official evaluation and ranking within the HaN-Seg challenge, our pipeline was submitted to the challenge platform which evaluated performance on an independent test set of 14 patients, only accessible by the challenge organizers.

Secondly, to further assess generalization, we evaluated our pipeline on manually annotated structures from three TCIA datasets: Glis-RT, BGPD and the Head-Neck-CT-Atlas (HNSCC) collection~\cite{hnscc_dataset, hnscc_paper}. Glis-RT and BGPD manual contours were not used in any training stage, they were retained for evaluation purposes. Glis-RT includes annotations for the brainstem, optic chiasm, optic nerves, eyes, cochleae, and lacrimal glands. BGPD includes annotations for the eyes, lenses, optic nerves, brain, brainstem, and optic chiasm.

HNSCC dataset does not contain MR scans, but contains CT scans and manually delineated contours for 17 OARs from 215 patients with H\&N squamous cell carcinoma treated with radiotherapy at MD Anderson Cancer Center (Houston, USA). Only the 193 patients with H\&N-related OAR contours were included in our evaluation.

\subsection{Cropping}
\label{sec:cropping}
In the HaN-Seg dataset, a substantial portion of both CT and MRI volumes consists of background regions, primarily air, the patient table, or anatomical areas outside the H\&N that do not contribute meaningfully to the segmentation of OARs. To reduce the volume sizes and facilitate model training while preserving the OARs, the CT volumes were cropped using the following mechanism. Outside body volume (air) was first cropped using Otsu thresholding~\cite{otsu_thresholding}. Then, the minimum and maximum distances from the new CT origin to a voxel belonging to any OAR were computed across all patient scans in the HaN-Seg dataset on each axis. Volumes were cropped using these limits, extended with a margin of 20 mm on each side of each axis. When a cropping boundary exceeded the patient's volume, the new cropping boundary became the patient's volume boundary. 
The field of view of scans from Glis-RT and BGPD datasets often did not include the shoulders of the patients, so for these two datasets, only the cropping around the body volume was applied. 

The MRI volumes were not cropped at this stage, as they were implicitly cropped by the registration process (see below).

At inference time, to avoid processing the entire patient scan, such as a whole-body scan in the HNSCC dataset, the cropping strategy described above was applied along the z-axis only.

\subsection{Registration}
A rigid (translation and rotation) registration was performed using the SimpleITK toolkit~\parencite{lowekamp_design_2013} built with SimpleElastix software (\href{https://simpleelastix.github.io}{https://simpleelastix.github.io}), an extension of the Elastix toolkit~\cite{5338015}, to align the cropped CT scans and the MRI. Each MRI was set as the fixed volume, and its corresponding CT was set as the moving volume to learn the transformation that would minimize the negative Mutual Information (MI)~\parencite{mattes_nonrigid_2001, mattes_pet-ct_2003} between the two modalities. This ensured that the points sampled from the fixed image were present in the moving image. After learning the transformation, the inverse transformation was used to register the MRI to the CT. Registration parameters of~\parencite{regparam} were taken from the \href{https://github.com/SuperElastix/ElastixModelZoo/tree/master/models/Par0023}{ElastixModelZoo}. The number of sampling points was updated to match 1\% of the number of voxels in the MRI. 

Registration quality was evaluated on the HaN-Seg dataset by placing markers at six different anatomical positions, as done in~\cite{podobnik_han-seg_2023}, both in the raw CT scan and the raw MRI scan. This was done manually using the \href{https://www.slicer.org/}{3DSlicer software}~\cite{FEDOROV20121323} for all the patients in the HaN-Seg dataset. The MRI marker points were registered using the transformations computed during the registration process. The distance between each CT marker and the corresponding registered MRI marker was computed. The resulting average target registration error was used to evaluate the registration process. 

Non-rigid registration (B-Spline) was not performed as it added computational burden without leading to an improvement in registration quality.

To avoid confounding the performance of the models with the randomness of the registration process, the MR scans of all considered datasets have been registered once, and all models presented thereafter used the same registered MR scans for training and inference. More details on the impact of the registration on the prediction can be found in Supplementary Material Section A.

\subsection{Model training} \label{training section}

OAR categories representing bilateral structures (e.g., right and left optic nerves) were combined as one component during training, resulting in 22 OAR categories, a reduction from the 30 OARs. The cropped CT and registered MRI volumes were stacked in a vector as input to the nnU-Net pipeline~\parencite{isensee_nnu-net_2021,isensee2024nnu} (version 2.5.2). Using the default nnU-Net 3D full-resolution configuration, the automatically adapted architecture for our dataset comprised six stages with 32, 64, 128, 256, 320, and 320 feature maps. All volumes were resampled to the median CT spacing in the training dataset, which is 0.64 mm by 0.64 mm by 2 mm for the HaN-Seg dataset and 0.63 mm by 0.63 mm by 1.25 mm for the expanded dataset presented in Section~\ref{sec:dataset_expansion}. A five-fold cross-validation was used to train five models from scratch for 1000 epochs with a batch size of eight, on different splits (training 80\%/validation 20\%) of the dataset. The Dice-Cross Entropy sum, which is the summation of the Dice loss and Cross Entropy loss, was used as the loss function. All default nnU-Net augmentations, including spatial transformations, were used. Mirroring augmentation was only activated across the sagittal axis.\par
To accommodate learning from both data modalities, the Modality Dropout (MD)~\parencite{7169562, 10.1007/978-3-319-55050-3_8, 10.1007/978-3-031-16443-9_43} augmentation was adapted to the CT-MRI modalities, and incorporated into the nnU-Net pipeline. The developed MD randomly selected one of the following three scenarios with equal probability: (1) only the CT channel was utilized while the MRI channel was set to zero, (2) only the MRI channel was utilized while the CT channel was set to zero, or (3) both channels were used simultaneously. 
\par 
For each fold, the model checkpoint that maximized the exponential moving average of the pseudo-Dice score, as computed by the nnU-Net framework on the internal validation set during training, was selected and used for inference.
Figure~\ref{Fig:ProposedApproach} presents a schematic overview of the proposed approach.

\subsection{Inference and post processing} \label{pred section}

Test-time augmentation was used by flipping volumes across the sagittal axis, resulting in two different volumes. The raw model outputs for these volumes, referred to as logits, were then averaged after resetting the sagittal axes of the logits. 

This operation was performed with the five best models, one from each fold. Mean logits obtained from each of the five models were averaged, and the result was converted to segmentation masks using the argmax operation.

Since bilateral OARs, i.e., OARs with left and right components, were combined into a single component for training and inference, the pipeline's predicted segmentation masks were split to create the left and right components for those OARs. 
To split the bilateral OARs into their left and right components, a center reference position of the bilateral OAR was approximated by fitting a mixture of two Gaussian distributions to the voxel x-axis positions of the predicted bilateral OAR and computing its mean. This center reference cannot be used directly to separate the bilateral OAR into left and right components because parts of some OARs could be on both the left and right of this threshold. To solve the issue, all connected components of a predicted bilateral OAR were studied separately. For each connected component, the mean x-axis position was compared to the center reference position to assign the left or right label. This was repeated for each bilateral OAR.

\subsection{Training dataset expansion} \label{sec:dataset_expansion}

\begin{figure}
    \centering
    \includegraphics[width=0.95\textwidth]{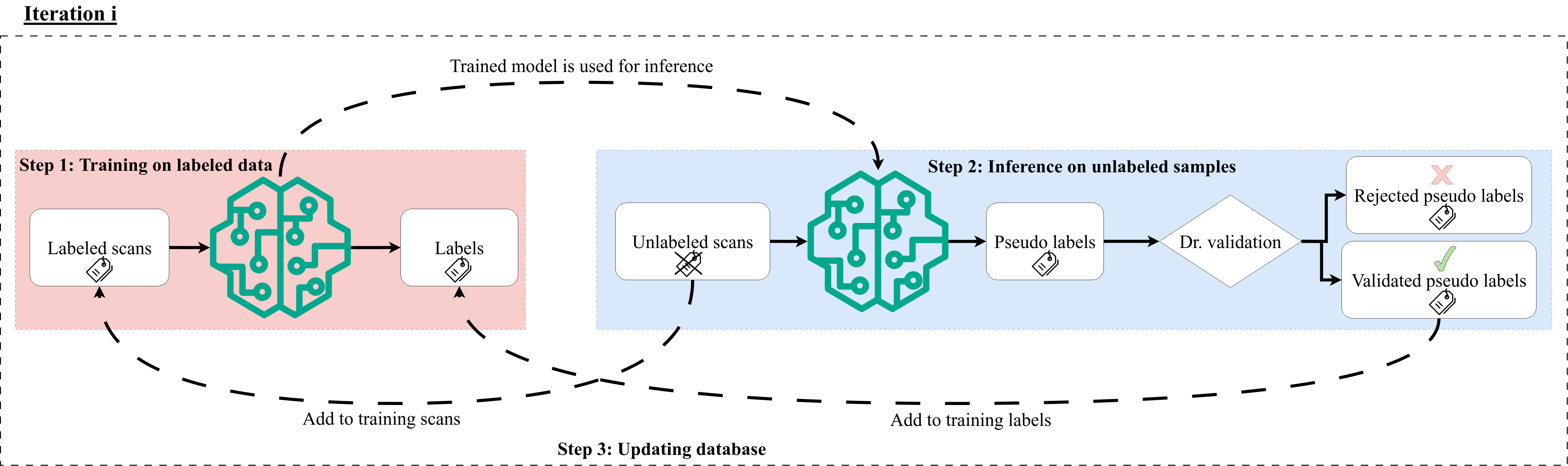}
    \caption{Single iteration from our multi-step training approach for dataset expansion. Pseudo contours are predicted for datasets without manual contours. Approved pseudo contours are used for further model training until all patient contours from unlabeled datasets have been reviewed. The final dataset created will be used for final model training.}
    \label{Fig:mutlistep_training}
\end{figure}

The training strategy presented in Section~\ref{training section} was first used to train five-fold models on the HaN-Seg dataset using the manual contours. Using this first pipeline, contours were predicted for the two other non-labeled datasets: Glis-RT and BGPD. These ``pseudo contours'' were reviewed on the CT scans by an experienced radiation oncologist. If they were deemed good enough to design a treatment plan without manual modifications, they were added to the database of patients with contours. If the pseudo contours were not validated, they were not used in subsequent trainings. New five-fold models were retrained from scratch with the updated dataset before each review session. The pseudo contours generated for the following review were created with a pipeline trained on the most up-to-date dataset of patients with contours. This process was repeated three times, until pseudo contours of all patients from the non-labeled datasets were reviewed. Figure~\ref{Fig:mutlistep_training} illustrates an iteration of the pseudo contours creation process. Exact details on the number of patients used at each iteration are provided in Supplementary Material Section B.

For carotid arteries, the model’s limited generalization across datasets, caused by differences in Hounsfield Unit distributions, was mitigated by temporarily adopting a threshold-based prediction strategy before reverting to the standard argmax approach once training data diversity improved. Details are available in Supplementary Material Section C.

In total, 254 patients contoured with preliminary versions of our segmentation model were added to our dataset. Unless specified otherwise, all results presented correspond to models trained on the expanded dataset of 296 patients, consisting of 42 manually labeled patients and 254 pseudo-labeled patients. 


\subsection{Evaluation}

\subsubsection{Pipeline performance}
An evaluation was performed on the HaN-Seg withheld test set, which was not accessible to the participants. The evaluation was performed by submitting the inference pipeline via the HaN-Seg Challenge \href{https://han-seg2023.grand-challenge.org/}{website}. The developed pipeline was packaged into a Docker container and submitted to the challenge platform to make predictions from the raw medical scans of unseen patients. The pipeline performance was evaluated by the challenge platform, on this external test set, in terms of Dice Score (DS) and Hausdorff Distance at the $95^{th}$ Percentile (HD). The submissions of different pipelines were statistically ranked based on the Wilcoxon signed-rank test with Bonferroni correction~\cite{podobnik_han-seg_2024}.

We conducted further evaluations on the three datasets from TCIA: Glis-RT, BGPD, and HNSCC. For consistency with the reference annotations, our pipeline’s posterior and anterior eyeball predictions were merged into a single eye class during evaluation. Our pipeline was used to predict contours from CT and MR scans for Glis-RT and BGPD, and to predict only from the CT scan for HNSCC. Agreement with available manual contours was evaluated in terms of DS and HD.

Although they are limited to processing either CT or MRI, as they are designed to generate contours from a single imaging modality only, several segmentation tools exist for the head and neck (H\&N) region~\cite{ijerph19159057,chen_validation_2024}. Therefore, our pipeline was ultimately evaluated against the commercially available Limbus AI software~\cite{ijerph19159057} version 1.8.1. This software was chosen due to its current clinical usage at McGill University's teaching hospitals. Limbus did not make the details of their software public \cite{wong_comparing_2020}. This version of the software automatically segments 36 OARs from a CT scan. For a fair comparison, evaluation was performed on the HNSCC dataset, which was not seen by our model during training, and only contains CT scans.
Different parts of the pharynx, larynx, esophagus and spinal cord structures are contoured in our work and Limbus software. Limbus AI contours the full pharynx, full larynx, esophagus and spinal cord up to a certain depth. Our pipeline contours the cricopharyngeus, the larynx supraglottic, the cervical esophagus and a part of the spinal cord up to a lower depth than Limbus AI. Consequently, these structures were not compared. Similarly to the TCIA datasets, our anterior and posterior eyeball segments were merged into a single eye class for comparison with Limbus AI predictions. Limbus software contours the following ten OARs that are not included in our study: Bone\_Hyoid, BrachialPlex\_L, BrachialPlex\_R, Brain, Clavicle\_L, Clavicle\_R, Hippocampus\_L, Hippocampus\_R, Lung\_L and Lung\_R. 
Six contours are included in our study but not contoured by Limbus AI software: A\_Carotid\_L, A\_Carotid\_R, Arytenoid, BuccalMucosa, Glnd\_Lacrimal\_L and Glnd\_Lacrimal\_R. Comparison was performed on the 17 remaining OARs that could be contoured by both Limbus software and our pipeline.

DicomRTTool software~\cite{dicomrttool_paper} version 2.2.0 was used to read DICOM RTSTRUCT files available in TCIA datasets and those created by Limbus AI predictions.

\subsubsection{Ablation study}

As part of our ablation study, internal evaluation was performed locally to validate the design choices of our pipeline. The model from the first fold of the training was chosen for this evaluation. Models from the other four folds were not used. Predictions were made on the corresponding internal testing set for the different experiments. It consisted of 20\% of the patients (n=60) in the dataset, while the remaining 80\% (n=236) was used for model training. 

Experiments were conducted to evaluate the effect of the MD augmentation, the image modality configuration, and finally the combination of the left/right OARs.
First, the model was trained following the training procedure presented in Section~\ref{training section} with different datasets, i.e., CT only and MRI only. When training on MRI-only, and since reference standard contours are in the CT scan coordinate system, registered MR scans were used. Additionally, considering that the MRI field of view was always smaller than the CT scans' in our dataset, the model was trained with input volumes and reference standard contours cropped to the field of view of the MRI scan. 
Then, the model was trained following Section~\ref{training section}, but without the MD augmentation.
Finally, to evaluate the impact of having left and right contours from the same organs, the model was trained following Section~\ref{training section}, with the 30 original OAR reference standard contours without combining left and right organs. For this particular training, and to avoid sending a counterproductive signal to the model considering the left and right position of the OARs, neither mirroring nor test-time augmentation was used for training and inference.

The performance of the models were compared in terms of DS and HD on the internal testing dataset of the first fold. The results were first averaged for each OAR across all cases; then averaged once more across all OARs.

The Wilcoxon signed-rank test with Benjamini-Hochberg correction was used to compare two training results and identify significant differences, for each OAR separately, in terms of both HD and DS.

\section{Results}

\subsection{Registration}


Rigid registration on the full HaN-Seg dataset (n=42) took on average 53 seconds $\pm$ 15 seconds, ranging from 22 to 84 seconds on a machine equipped with a 24-core Intel i9-14900KF processor with 64GB RAM. The mean target registration error averaged 2.45 mm $\pm$ 1.48 mm ranging from 0.16 mm to 9.39 mm.

\subsection{Pipeline performance}

\subsubsection{HaN-Seg challenge}

The strict time requirement of the challenge submissions enforced that the segmentation of the 30 OARs for one patient would not exceed 15 minutes on a machine with 32GB RAM and a GPU with 16GB of VRAM. Table~\ref{table:performance} summarizes the best submissions from the top five teams in the HaN-Seg challenge in May 2025. Although our pipeline was submitted after the final phase, it was evaluated by the challenge platform using the same held-out internal test set and evaluation protocol as those submitted during the final phase. The proposed approach obtained a performance of 78.03\%$\pm$10.85 and 3.44 mm$\pm$1.64 mm as the mean DS and HD $\pm$ their standard deviation, respectively. 
Supplementary Material Section D presents the mean DS and the mean HD obtained by the proposed approach on the test set for each OAR and relates these metrics to the OAR size. The mean DS ranged from 47.05\% for the Optic Chiasm to 94.86\% for the Mandible bone. The mean HD ranged from 1.30 mm for the Mandible bone to 8.32 mm for the Cervical esophagus.
\begin{table}[ht]
    \footnotesize
    \centering
    \begin{tabular}{|P{19ex}|P{10ex}|P{20ex}|P{7em}|P{6em}|P{17ex}|}
    \hline
    \textbf{Algorithm} & \textbf{Inference dataset} & \textbf{Reference standard} &\textbf{Mean Dice Score (\%)} $\uparrow$& \textbf{Mean Hausdorff Distance (mm)} $\downarrow$ & \textbf{Statistical ranking} (phase) \tabularnewline 
    \hline\hline
    hanseg3 & \multirow{ 7}{*}{\makecell{HaN-Seg \\ test set\\ (n=14)}} & \multirow{ 7}{*}{\makecell{HaN-Seg \\ (30 OARs)}} & 75.16 $\pm$ 11.41 [44.42 93.23] & 3.86 $\pm$ 1.80 [1.55 8.05]& $12^{th}$ (final) \\ \cline{1-1} \cline {4-6}
	HaNSeg\_zhack & & & 75.05 $\pm$ 12.26 [41.08 94.24]& 3.71 $\pm$ 1.58 [1.38 7.61]& $7^{th}$ (final) \\ \cline{1-1} \cline {4-6}
    han\_seg & & & 76.77 $\pm$ 11.59 [45.58 95.00]& 3.79 $\pm$ 1.77 [1.20 7.45]&  $5^{th}$ (final) \\ \cline{1-1} \cline {4-6}
    the\_HaN-Seg23\_game & &  & 76.94 $\pm$ 11.54 [44.61 94.33]& 3.52 $\pm$ 1.60 [1.34 7.95] & $1^{st}$ (final) \\ \hline \hline
    Ours: 42 patients & \multirow{ 7}{*}{\makecell{HaN-Seg \\ test set\\ (n=14)}} & \multirow{ 7}{*}{\makecell{HaN-Seg \\ (30 OARs)}} & 78.01 $\pm$ 11.41 [45.49 95.20] &  3.46 $\pm$ 1.70 [1.38 8.57] & $5^{th}$ (post-challenge)  \\ \cline{1-1} \cline {4-6}
    Ours: 42 patients + custom std &  &  & 77.98 $\pm$ 11.35 [45.34 95.31] & 3.43 $\pm$ 1.72 [1.37 8.65] & $3^{rd}$ (post-challenge)  \\ \cline{1-1} \cline {4-6}
    Ours: 296 patients &  &  & 78.03 $\pm$ 10.85 [47.05 94.86]& 3.44 $\pm$ 1.64 [1.30 8.32]& $2^{nd}$ (post-challenge)  \\ \cline{1-1} \cline {4-6}
    Ours: 296 patients + custom std &  &  & \textbf{78.12} $\pm$ 10.74 [46.47 94.88]&  \textbf{3.42} $\pm$ 1.60 [1.32 7.78] & $1^{st}$ (post-challenge)  \\ \hline \hline
    \multirow{ 7}{*}{Ours: 296 patients} & BGPD (n=176)& BGPD (6 OARs) & 64.50 $\pm$ 22.61 [25.81 87.64] & 7.24 $\pm$ 4.68 [3.15 17.00] & \multirow{ 7}{*}{--} \\ \cline{2-5}
    & Glis-RT (n=230)& Glis-RT (10 OARs) & 65.96 $\pm$ 16.4 [37.12 89.95] & 4.14 $\pm$ 2.59 [1.68 9.70]  &  \\ \cline{2-5}
     & HNSCC (n=188) & HNSCC (15 OARs) & 59.32 $\pm$ 22.51 [20.02 87.24] & 8.86 $\pm$ 7.10 [1.99 27.37]  &  \\ \cline{2-5}
     & HNSCC (n=188)& Limbus AI (17 OARs)& 77.41 $\pm$ 15.19 [27.12 92.78]  & 3.32 $\pm$ 1.97 [1.11 7.89]  &  \\ \hline \hline
     Ours: 42 patients & HNSCC (n=188)& Limbus AI (17 OARs)&  76.48 $\pm$ 15.25 [26.61 92.66] & 3.27 $\pm$ 2.03 [1.13 8.37] & -- \\ \hline
    \end{tabular}
    \caption{Segmentation performance on HaN-Seg and TCIA datasets. The reported measures are in the form of mean $\pm$ standard deviation [minimum OAR average value, maximum OAR average value]. The performance measures for each OAR were first averaged across all cases, and then the resulting values were averaged across all organs to calculate the overall model performance. $\uparrow$ indicates that higher values are better and $\downarrow$ indicates lower values are better.}
    \label{table:performance}
\end{table}

\begin{figure}
    \centering
    \includegraphics[width=\textwidth]{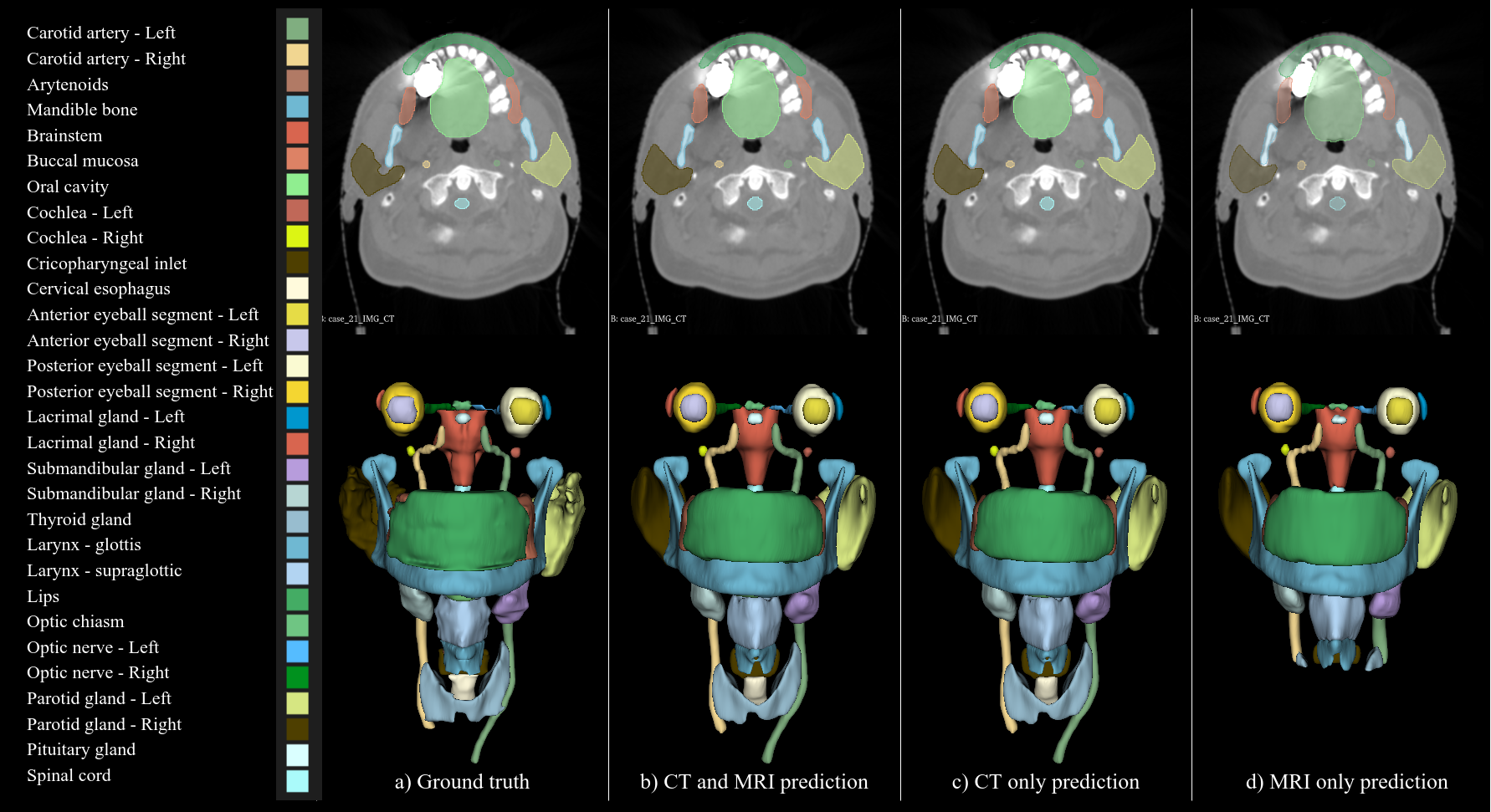}
    \caption{Prediction of our fold 0 model trained with Modality Dropout, on a patient of the Head and Neck Organ-at-Risk CT \& MR Segmentation Challenge in our internal testing set. It can be observed that the model is able to predict contours from different input modalities and that a prediction from MRI only is limited to the field of view of the MR scan.}
    \label{fig:prediction}
\end{figure}

Figure \ref{fig:prediction} presents the pipeline predictions using only the fold 0 model for a patient in the internal testing set of fold 0 belonging to the HaN-Seg training dataset. As shown, if the model only predicts from the MRI modality, it can only predict organs in the field of view of the MRI scan.

A custom standardization tailored to the HaN-Seg dataset could increase the challenge performance further to an average DS of 78.12\% and a HD of 3.42 mm. However, as this preprocessing strategy did not generalize well to larger, more heterogeneous datasets, it is described in the Supplementary Material Section E.


\subsubsection{TCIA datasets}
Table~\ref{table:performance} also presents the average performance of the pipeline on BGPD, Glis-RT and HNSCC datasets. There was an important performance variability between the different structures, which, for instance, for the HNSCC dataset, went from an average DS of 20.02\% for the optic chiasm to 87.24\% for the left eye. The optic chiasm was always the lowest average DS among the datasets with 25.81\% for BGPD, 37.12\% for Glis-RT and 20.02\% for HNSCC. 

\begin{figure}[!tbp] 
  \centering
 \includegraphics[width=0.95\textwidth]{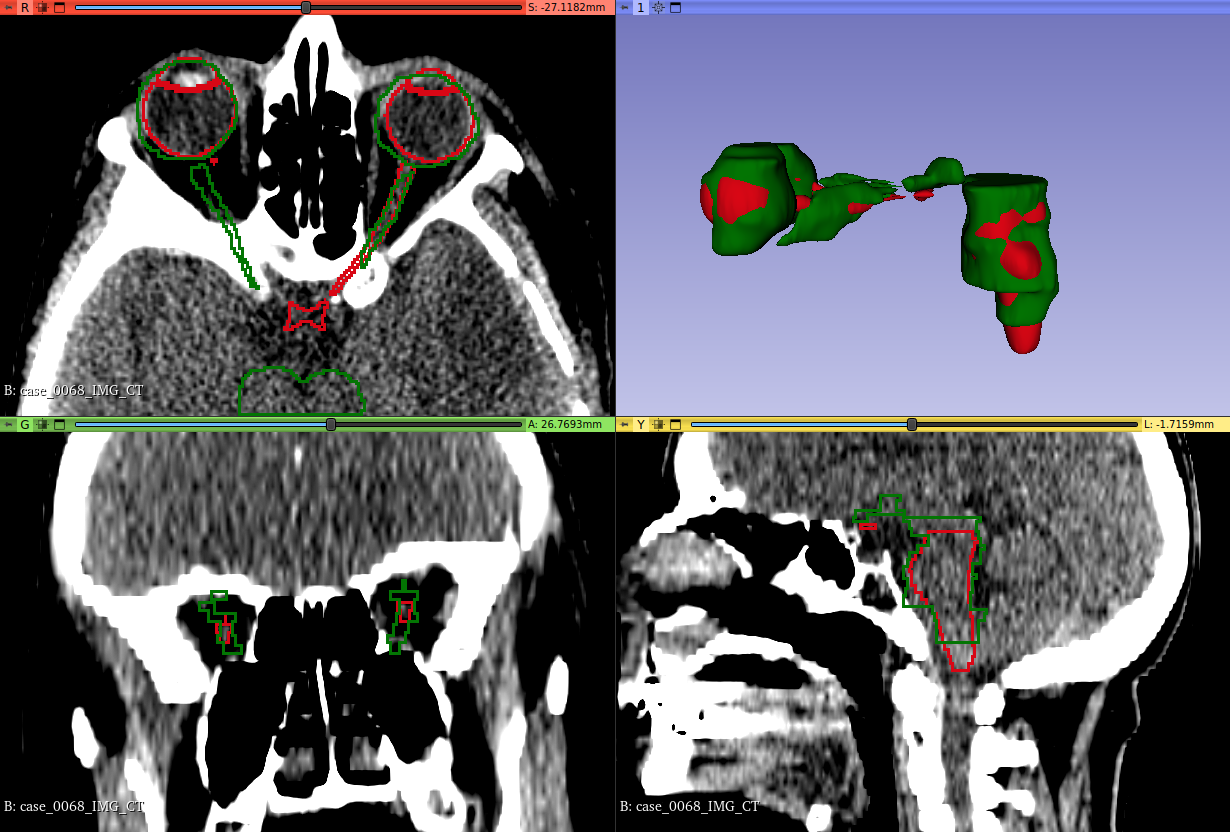}
  \caption{Manual contours along with our pipeline predictions on a patient from BGPD dataset unused in model training. Red represents the prediction of our model while green shows the reference standard contours provided in the BGPD dataset. It can be seen that our model prediction is more faithful to the true anatomy of the patient than contours available in the BGPD dataset.}
  \label{fig:case68_bgpd}
\end{figure}

Optic nerve contours in TCIA datasets also did not align well with the predictions of our pipeline with for instance an average DS of 52.76\% and 53.21\% for left and right nerves in BGPD dataset, respectively. In contrast, the agreement with the Limbus AI software was higher, with average DSs of 69.68\% and 70.13\% for the left and right optic nerves, respectively.
Figure~\ref{fig:case68_bgpd} presents a failure analysis by showing the prediction of our pipeline for a patient from BGPD dataset; this sample was not used in our model training. The DS for the optic chiasm was 0.00\% and for the left and right optic nerves was 29.07\% and 28.66\%. 

A comprehensive performance evaluation on these three datasets can be found in Section F of the Supplementary Material.

\subsubsection{Comparison with a commercially available software}
\begin{figure}
    \centering
    \subfloat{\includegraphics[width=0.327\textwidth]{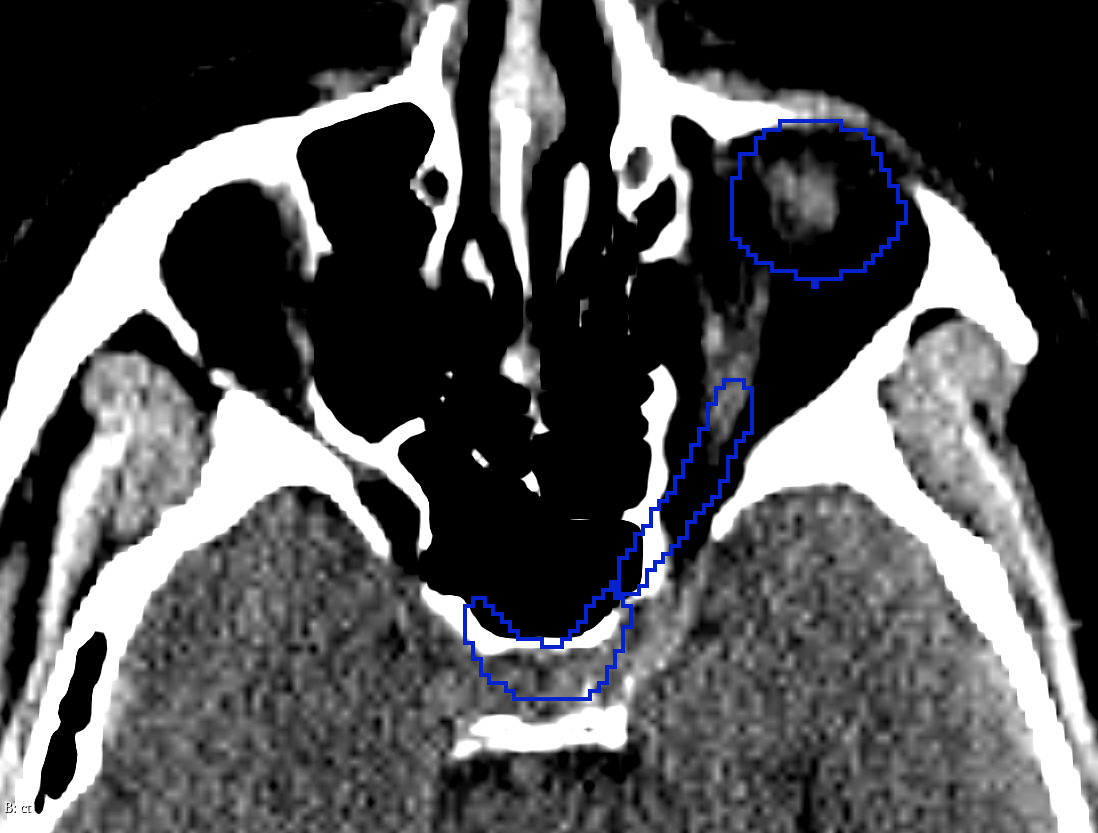}}\label{fig:f1}
    \hfill
   \subfloat{\includegraphics[width=0.327\textwidth]{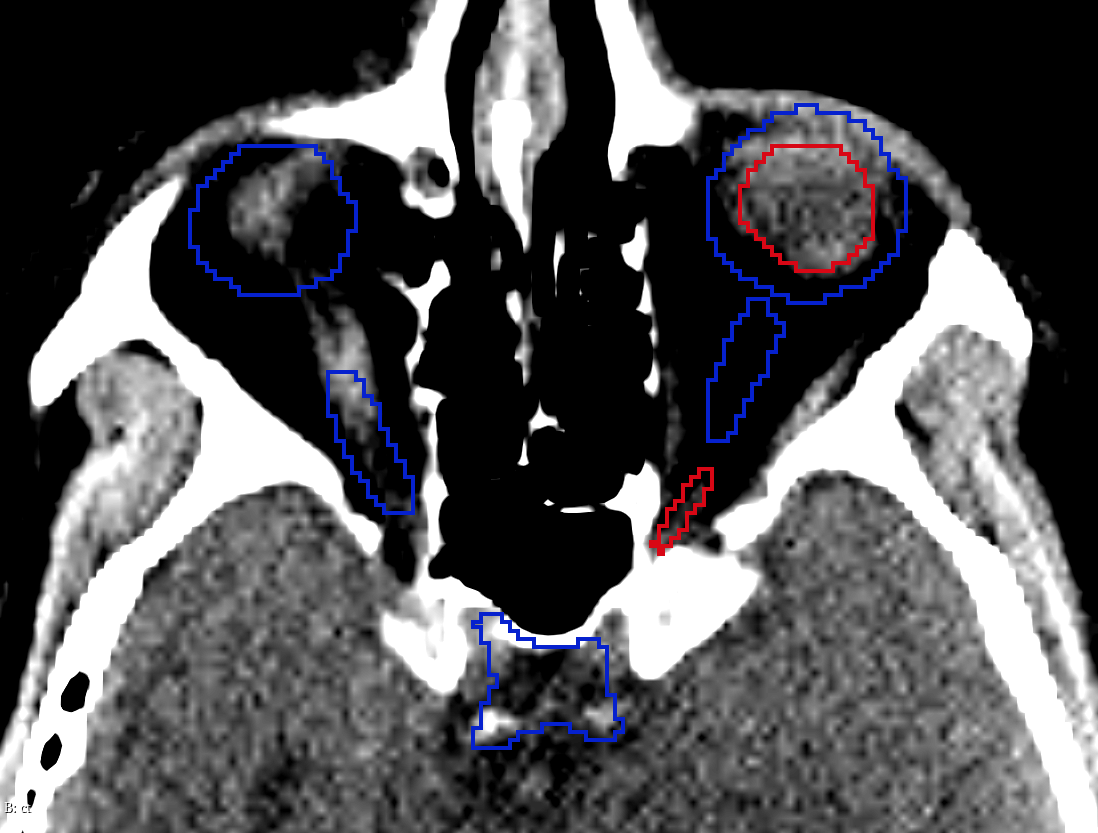}}\label{fig:f1}
    \hfill
     \subfloat{\includegraphics[width=0.327\textwidth]{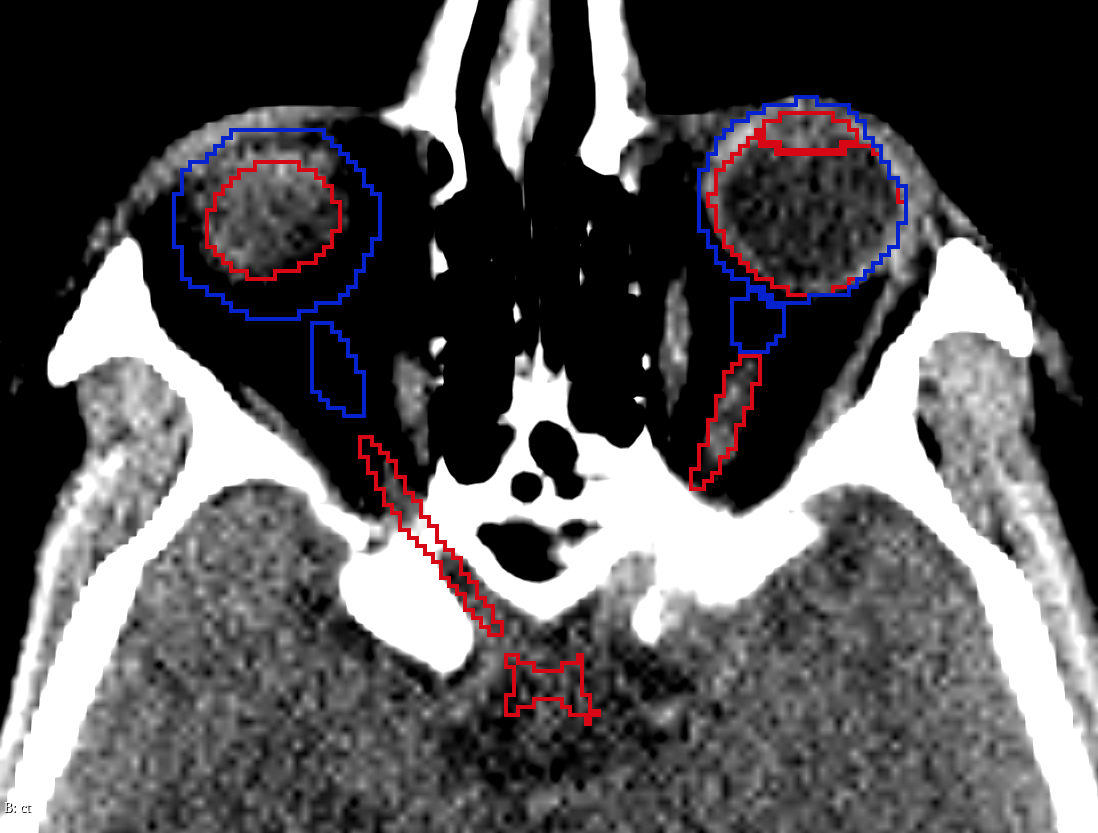}}\label{fig:f1}
    
    \centering 
    \subfloat{\includegraphics[width=0.327\textwidth]{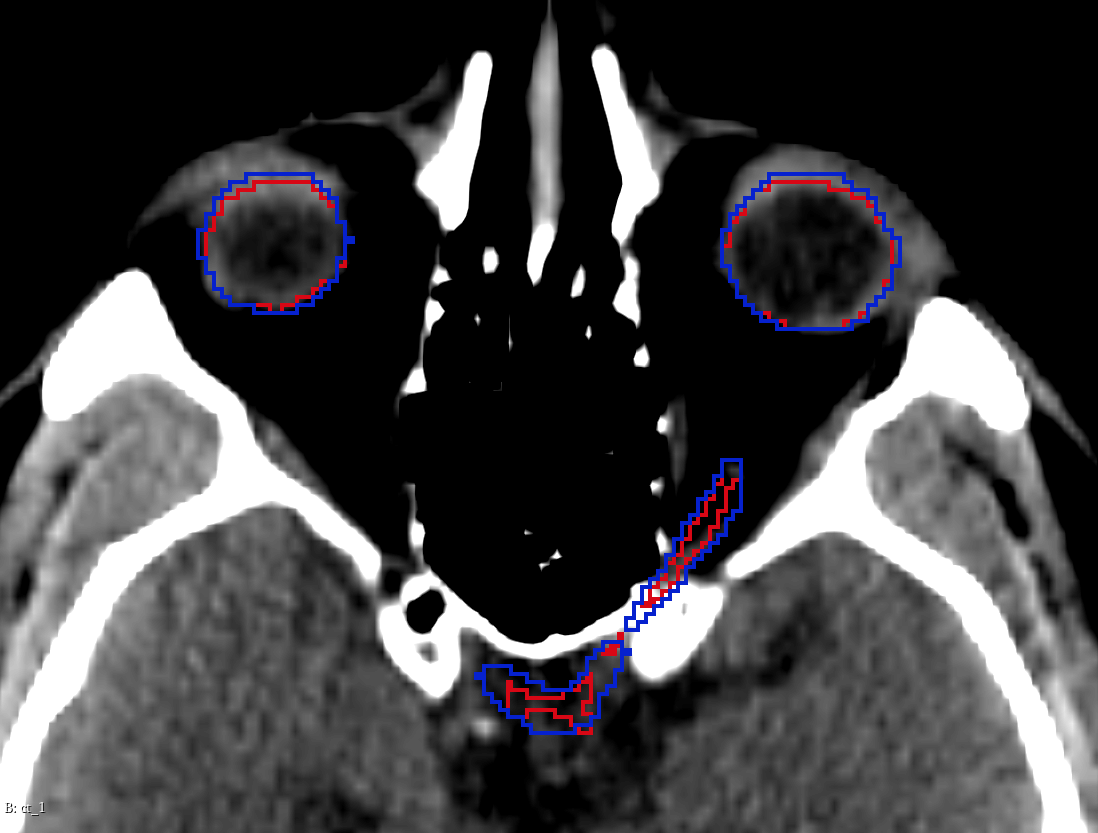}\label{fig:f3}}
    \hspace{0.10em}
    \subfloat{\includegraphics[width=0.327\textwidth]{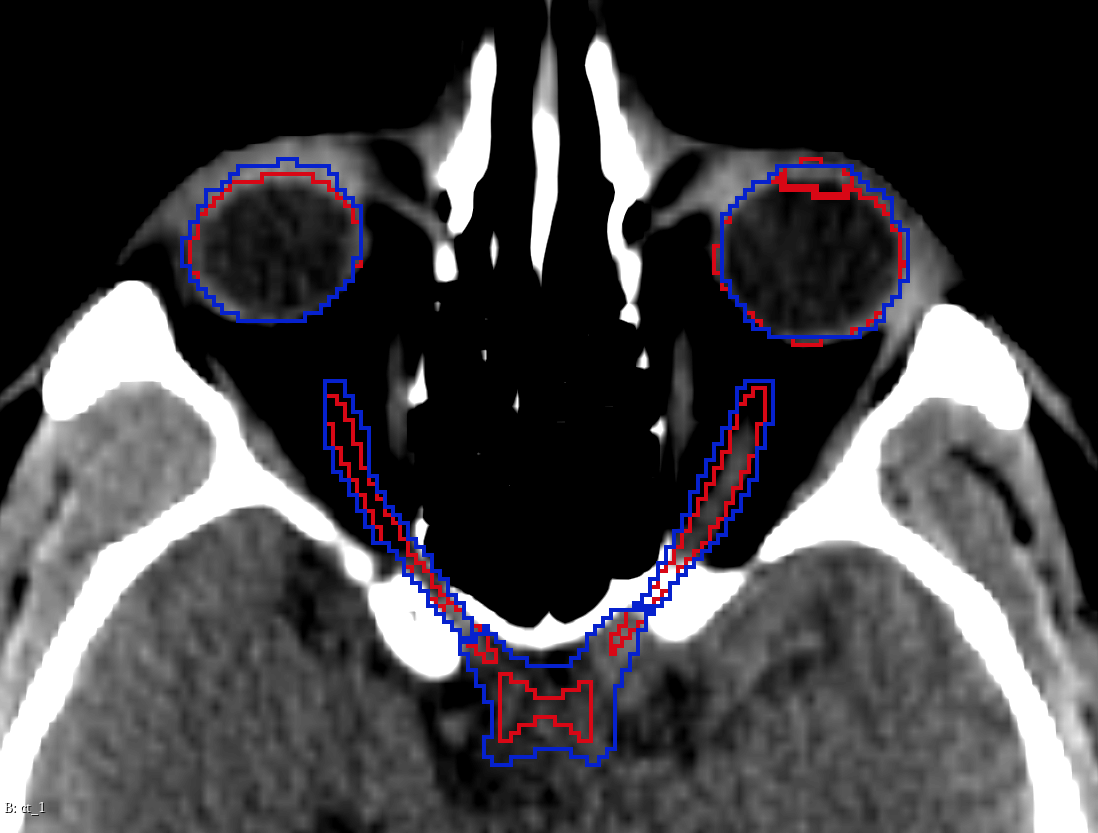}\label{fig:f3}}

  \caption{Comparison of optic structures predicted by Limbus AI (blue) and by our CT-only pipeline (red), overlaid on CT scans from patients in the HNSCC dataset. The top row shows three slices from the patient with the lowest Dice Score for the optic chiasm between Limbus AI prediction and ours, while the bottom row shows two slices from the patient with the highest Dice Score for the optic chiasm. Notably, the low Dice Score is not due to a completely misplaced prediction by our method. In general, Limbus AI contours of the optic chiasm encompass a larger region compared to our predictions.}
  
    \label{fig:optich_chiasm_best_worst}
\end{figure}

Table~\ref{table:performance} shows the agreement between our pipeline predictions and Limbus AI predictions on the HNSCC dataset. A much larger agreement was observed compared to manual contours from TCIA datasets, with an overall mean DS of 77.41\%. The lowest mean DS was observed for the optic chiasm with a value of 27.12\%. Figure~\ref{fig:optich_chiasm_best_worst} presents the predictions of Limbus AI and our pipeline for the optic chiasm, optic nerves, and eyes in patients from the HNSCC dataset with the lowest (0.00\%) and highest (44.79\%) optic chiasm DSs. The figure highlights the different contouring approaches for the optic chiasm that have been learned during training. The highest mean HD of 7.89 mm was observed for the oral cavity. Details for each OAR can be found in the Supplementary Material Section F.

Without the dataset expansion, the average DS decreased to 76.48\%. Ten OARs had a significantly different DS at a 95\% confidence level according to the Wilcoxon signed-rank test with Benjamini-Hochberg correction, and were on average better contoured with the model trained on the expanded dataset. Similarly, ten OARs had significantly different HD, with eight being better contoured by the model trained on the expanded dataset. Details, including p-values, for each OAR whose segmentation performance was significantly different between the base and expanded datasets are provided in the Supplementary Material Section F.

\subsection{Prediction time} \label{sect:predtime}
The time spent on different parts of the pipeline was evaluated when predicting contours with our end-to-end pipeline for the patients of the internal testing set of our first fold (n=60), on a machine equipped with a 24-core Intel i9-14900KF processor with 64GB RAM and a GeForce RTX 4090 Nvidia GPU with 24GB of VRAM. The average time dedicated to the registration process was 29 seconds ± 19 seconds (on CPU), with a duration ranging from a minimum of 3 seconds to a maximum of 96 seconds. Inference with the five-fold models consumed an average of 12 ± 3 seconds (on GPU), spanning from 6 seconds to 19 seconds. The resampling phase of the nnU-Net pipeline that resamples the predicted logits back to the original spacing of the patient CT scan required an average of 16 ± 7 seconds, with observed times ranging between 3 seconds and 52 seconds. Postprocessing tasks that include splitting the OARs into left and right components and padding the contours back to the original CT scan shape, averaged at 2 seconds ± 1 second, with a minimum time of 1 second and a maximum of 6 seconds. Cumulatively, the entire process averaged 60 ± 23 seconds, with a total time range of 22 seconds to 125 seconds, highly correlated with the patient's scan size. Analyzing the percentage distribution of time spent, registration accounted for an average of 45\% of the total pipeline prediction time, while resampling accounted for 29\% on average. 


\subsection{Ablation study} \label{ablationstudy}

\subsubsection{Modality Dropout} \label{ab:study_MD}

Table~\ref{table:ablation} shows the average segmentation performance of the developed pipeline trained with and without the MD augmentation for three inference scenarios: inferring only using CT, inferring only using MRI, and inferring using both CT and MRI jointly. Detailed performance for each OAR can be found in Supplementary Material Section G.

\begin{table}[ht]
    \small
    \centering   
    \begin{tabular}{|P{5em}|P{6em}|P{6em}|P{6em}|P{7em}|P{7em}|}
    \hline
    \textbf{Training inputs} & \textbf{Modality Dropout} & \textbf{Left/Rigth OAR combination} & \textbf{Inference inputs} & \textbf{Dice Score (\%) $\uparrow$} & \textbf{Hausdorff Distance (mm) $\downarrow$} \tabularnewline
    \hline \hline

    \multirow{5}{*}{\textbf{\makecell{CT and \\ MRI}}} & \multirow{5}{*}{\cmark} & \multirow{5}{*}{\cmark} & \textbf{CT only} & 90.06 $\pm$ 5.41 [73.22 97.61] & 1.58 $\pm$ 0.66 [1.04 3.99] \tabularnewline  \cline{4-6}  
    &  &  & \textbf{MRI only} & \textbf{57.00} $\pm$ 26.49 [5.84 92.22] & 9.09 $\pm$ 9.51 [1.47 36.11]*\\  \cline{4-6}  
     &  &  & \textbf{CT and MRI} &  90.16 $\pm$ 5.41 [73.19 97.59]& 1.55 $\pm$ 0.64 [1.03 3.89]\\ \hline 

    \multirow{6}{*}{\textbf{\makecell{CT and \\ MRI}}} & \multirow{6}{*}{\xmark} & \multirow{6}{*}{\cmark} & \textbf{CT only} &  89.85 $\pm$ 5.40 [74.55 97.72] & 1.57 $\pm$ 0.65 [1.02 3.81]\\   \cline{4-6}  
    & &  & \textbf{MRI only} &  \textbf{0.91} $\pm$ 4.88 [0.00 27.19] & 101.89 $\pm$ 64.12 [20.25 206.02]* \\  \cline{4-6} 
    &  & & \textbf{CT and MRI} & 90.24 $\pm$ 5.30 [74.07 97.69] & 1.52 $\pm$ 0.65 [1.02 3.78]\\ \hline 

    \textbf{CT only} & \xmark & \cmark & \textbf{CT only} & 90.06 $\pm$ 5.35 [73.78 97.75] & 1.59 $\pm$ 0.71 [1.02 3.81]  \\  \hline  

    \textbf{CT and MRI} & \cmark & \xmark & \textbf{CT and MRI} & 89.63 $\pm$ 5.52 [72.69 97.49] &  1.74 $\pm$ 0.92 [1.04 4.17] \\  \hline

    \end{tabular}
    \caption{Average segmentation metrics obtained by our fold 0 model on the patients of our fold 0 internal testing set. Results were first averaged for each OAR over all cases and finally over all OARs. Results are presented as: mean across patients and OARs $\pm$ standard deviation among the mean metrics of each OAR [minimum OAR metric averaged over cases and maximum]. $\uparrow$ indicates that higher is best and $\downarrow$ indicates smaller is best. As described in Section~\ref{datasetsection}, segmentation metrics obtained with MRI-only predictions are computed with non-NaN values only, excluding OARs outside of the MRI FOV.}
    \label{table:ablation}
\end{table}

The table shows that a model trained with MD can predict from both modalities used jointly or separately. The model trained without MD resulted in a negligible average DS: 0.91\% when CT information was missing, indicating that the trained model heavily relies on CT images and cannot perform well when missing CT scans. However, using MD during the training makes the model much more robust to missing modalities. The model trained with MD obtained an average DS of 57.00\% when the CT information was missing, indicating that it successfully learned from the MRI scan. A similar behavior was observed when evaluating the model using the HD. Predicting using only the CT scans resulted in similar average HD and slightly better DS, with 90.06\% for the model using MD against 89.85\% when not using MD. Finally, when predicting with both modalities as inputs, the model trained without MD obtained a slightly better average performance with a DS of 90.24\% against 90.16\%, and a HD of 1.52 mm against 1.55 mm. 
As the performance of a model trained with MD could be impacted by the fact that the two modalities do not have the same field of view, results of models trained only on the MRI field of view are provided in Supplementary Material Section H. 

It can also be observed that when only a single modality was used for inference, a model trained without MD exhibited a greater performance drop compared to the one trained with MD. With MD, the DS decreased from 90.16 \% to 90.06 \% and the HD increased from 1.55 mm to 1.58 mm. In contrast, without MD, the DS decreased from 90.24 \% to 89.85 \% and the HD increased from 1.52 mm to 1.57 mm. The performance of the model trained with MD is less impacted by a missing modality. 

When comparing predictions from the model trained with MD, significant differences, at a 95\% confidence level according to the corrected Wilcoxon signed-rank test, were observed in DS for the brainstem and spinal cord, and in HD for the brainstem, depending on whether inference was performed using only the CT scan or both CT and MR scans. These significant differences were also observed for a model trained without MD, with the addition of significant differences in DS for the lacrimal and parotid glands, the larynx supraglottic, the optic chiasm and the eyeballs, and in HD for the lacrimal and parotid glands, the optic chiasm and the spinal cord. Details, including p-values, of the performance for these OARs can be found in Supplementary Material Section H.

When comparing predictions of a model trained with MD and a model trained without MD, predictions using only CT scans were significantly different for 11 OAR DS at a 95\% confidence level according to the corrected Wilcoxon signed-rank test. According to the DS, the model trained with MD outperformed the model trained without MD for the brainstem, the optic chiasm, both anterior and posterior eyes, and the pituitary. For the mandible, the oral cavity, the left submandibular gland and the larynx supraglottic, the model trained without MD was better. According to HD, only the brainstem showed significant differences, with the model trained with MD being better. Details, including p-values, can be found in Supplementary Material Section H.

Comparison of the predictions using MR scans only between models trained with and without MD results in significant differences for all OARs in terms of DS. Statistical tests could not be performed for HD because of all the ``not a number values'' computed for HD with the model trained without MD.

When comparing predictions of a model trained with MD, from CT only and MRI only, all OARs showed significant differences in terms of DS with a level of confidence of 95\%, except the brainstem. The average DS of the brainstem when predicting only with CT was 95.10\% $\pm$ 3.11\% ranging from 84.38\% to 98.59\% and the average HD is 1.36 mm $\pm$ 0.63 mm ranging from 1.0 to 3.32 mm. When predicting with MRI only, the average DS was 92.22\% $\pm$ 12.89\% ranging from 0.0 to 98.15\% and the average HD 3.15 mm $\pm$ 12.69 mm ranging from 1.0 to 100.43 mm. Detailed performance of each OAR can be found in Supplementary Material Section H.

When predicting with CT and MR scans, significant differences between a model trained with and without MD were observed for the mandible bone, the oral cavity, the right lacrimal gland, the left submandibular gland, the optic chiasm, the parotid glands and the spinal cord, with the model trained without MD having the best performance. Details can be found in Supplementary Material Section H.

Figure \ref{fig:reg_impact} shows the prediction of contours using the model trained with MD when CT information is missing. It can be observed that the prediction was faithful to the MRI scan but not to the CT scan and the reference standard. This underscores the limitations of the metrics, which rely on CT-based reference standard, in accurately reflecting the model’s performance when using only the MRI modality, due to imperfect image registration.

\begin{figure}
    \centering
    \includegraphics[width=1\textwidth]{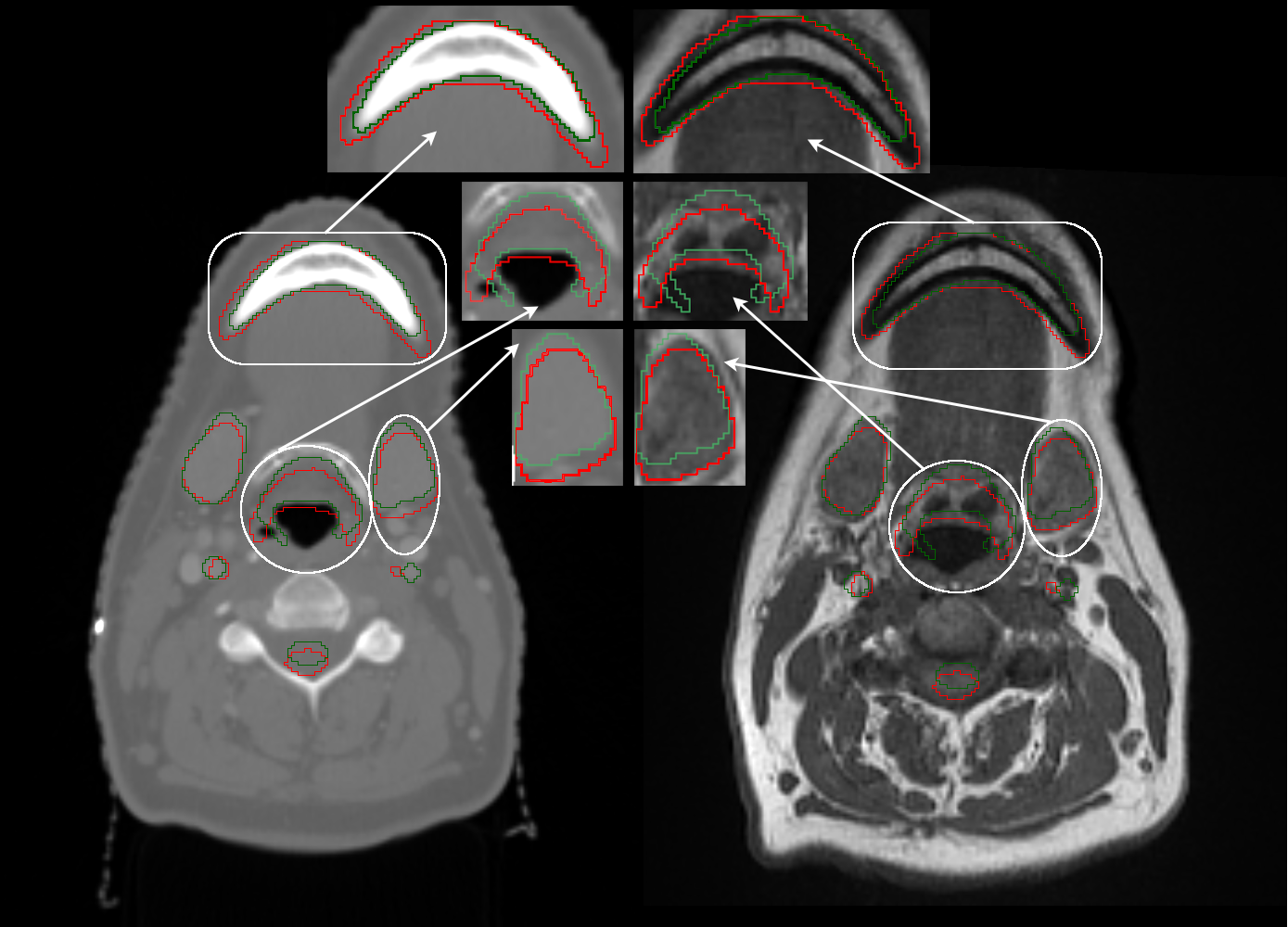}
    \caption{Prediction from MRI only, of a model trained with Modality Dropout. Left: Predicted (Red) and reference standard contours (Green) over the CT. Right: Prediction and reference standard contours overlaying the MRI. As seen on the CT scan, the reference standard matches the organs very well. However, the MRI-only predictions are inaccurate. As observed on the MRI, the reference standard is not properly contouring its target organ, while the MRI-only prediction is more accurate.}
    \label{fig:reg_impact}
\end{figure}

\subsubsection{Dataset composition}

As shown in Table~\ref{table:ablation}, a model trained to predict contours only from CT scan obtained an average DS of 90.06\% $\pm$ 5.35\% ranging from 73.78\% to 97.75\% and an average HD of 1.59 mm $\pm$ 0.71 mm ranging from 1.02mm to 3.81 mm on the internal testing set of our first fold. 
A model trained to predict contours from both CT and MR scans obtained an average DS of 90.16\%$\pm$ 5.41\% ranging from 73.19\% to 97.59\% and an average HD of 1.55 mm $\pm$ 0.64 ranging from 1.03 mm to 3.89 mm on the internal testing set of our first fold.
At a 95\% confidence level, according to the corrected Wilcoxon signed-rank test, DSs were significantly different for the brainstem, where the CT and MRI model obtained a DS of 95.70\% against 95.12\% for the CT only model, the mandible bone, where the CT and MRI model DS was 97.59\% against 97.75\% for the CT only model, and the oral cavity where the CT and MRI model DS was 96.78\% against 96.88\% for the CT only model.

As depicted by Table~\ref{table:ablation}, a model trained to predict raw contours (from both CT and MR scans), without combination of left/right components of the OAR (30 classes) obtained an average DS of 89.63\% $\pm$ 5.52\% ranging from 72.69 to 97.49\% and an average HD of 1.74 mm $\pm$ 0.92 mm ranging from 1.04 mm to 4.17 mm on the internal testing set of our first fold. 
A model trained to predict contours with left/right components of OARs combined (22 classes) from both CT and MR scans performs better with an average DS of 90.16\% and an average HD of 1.55 mm. According to the corrected Wilcoxon signed-rank test, statistical differences were observed in terms of DS for 19 OARs, including 11 left-right OARs. More details, including p-values, can be found in Supplementary Material Section I.

\section{Discussion}


The developed multimodal H\&N OAR segmentation pipeline yielded state-of-the-art results in terms of DS and HD in the \href{https://han-seg2023.grand-challenge.org/}{Head and Neck Organ-At-Risk CT \& MR Segmentation Challenge}.

Our pipeline enhances nnU-Net preprocessing by integrating several advanced processing steps: tailored cropping strategies, rigid registration, and OAR combination. The integration of these steps into the pipeline proved beneficial, as evidenced by our higher statistical ranking compared to other teams in the challenge who also based their work on the nn-UNet pipeline~\cite{podobnik_han-seg_2024}.

Our improved pipeline addresses the absence of annotations in certain modalities through the extrapolation of annotations from an alternative modality. Given the labor-intensive and costly nature of manual annotation across multiple data modalities, particularly for medical images, the proposed pipeline is anticipated to accelerate the deep-learning-based approaches for multimodal segmentation tasks. 

Evaluation of the proposed pipeline using manual contours from TCIA datasets highlights the approximate delineation of certain structures. On the other hand, the strong agreement between our pipeline’s predictions on an external, unseen dataset and those from Limbus AI software supports the validity of our predicted contours.

Although all models were trained for the same number of iterations, training on a larger and more diverse dataset, through our proposed dataset expansion, led to a modest performance gain on the HaN-Seg test set and a more substantial improvement on an external, unseen dataset, highlighting the advantages of large-scale training data.


The nnU-Net pipeline generates sub-patches from the input volumes. In our pipeline, the input volumes are cropped to algorithmically determined regions for OARs. It was empirically observed that incorporating this cropping approach accelerates the convergence and improves model performance. This acceleration results from the increased frequency of sub-patches containing OARs being fed to the model, leading to more frequent and relevant parameter updates.


Our top-performing solution for the HaN-Seg challenge incorporated a HaN-Seg-specific standardization method, which outperformed nnU-Net’s standardization on that specific test dataset (see Table~\ref{table:performance} and Supplementary Material Section E). However, this improvement did not generalize to a larger, more diverse test set that included TCIA patients imaged with different scanners. This underscores the critical impact of standardization on model performance in the target dataset.




The standard nnU-Net pipeline has been designed for unimodal image segmentation tasks and does not address scenarios with multimodal segmentation with missing data modalities. By incorporating MD, the proposed pipeline extends the application of nnU-Net to multimodal image segmentation allowing for missing data modalities, which is a common scenario in the medical imaging field.

In this study, the multimodality problem was approached through early fusion, whereby different imaging modalities were stacked together as the input to the segmentation model. Intermediate and late fusion in the latent space could mitigate the need for registration, which is time-consuming and computationally intensive. Late fusion in the latent space was studied by ~\textcite{podobnik2023multimodal} for the same problem but showed no improvement compared to CT alone. In this study, the effectiveness of early fusion was observed both with and without MD; however, further exploration of intermediate and late fusion for multimodal image segmentation is suggested for future research.

The registration process of scans acquired at different times, whether rigid or non-rigid, is inherently imperfect due to the patient's body and organ movements and changes. Since the reference standard contours are provided for the CT scan, this inherent imperfection forces the model to focus more on the CT modality, for which the reference standard contours are provided in this retrospective study. Furthermore, although a prediction can appear to be of high quality when overlayed on its corresponding MRI, when the model predicts using the MRI modality only, the imperfect registration of the MRI to the CT, which was primarily used to manually contour the OARs, negatively affects the segmentation metrics. Therefore, these scores do not represent the ability of the model to precisely predict OAR contours from the MR scans. 

Our experiments demonstrated that a good-quality registration, combined with a MD during the training, can help a model use information from different modalities. In particular, the MD augmentation had a considerable effect on the ability of the model to predict from the MRI modality. While training without MD yielded slightly higher overall performance, it compromised the model’s robustness to missing modalities. Evaluating these findings in the context of other cancer sites, such as abdominal cancer, where organ movements are frequent, is suggested for future research.

Our experiments also showed that utilizing both modalities, rather than just one, significantly improves the performance of the segmentation model for certain soft tissue structures. This finding aligns with the established use of MRI to support manual segmentation of soft tissues on CT during dataset creation~\cite{podobnik_han-seg_2023}. However, it was empirically observed that lower quality registration reduces the performance gap between models trained with CT only and those trained with both CT and MRI. 



The stochastic nature of the rigid registration optimization makes it computationally intensive, accounting for a substantial part of the prediction time. Future research could investigate deep learning–based registration methods as a promising alternative to accelerate this step and improve registration quality, which would likely enhance the overall performance of the proposed pipeline. The proposed pipeline can also predict only from a single modality. Therefore, if time is a constraint, users may choose to predict from only one modality to expedite the predictions; however, this might come at the cost of reduced precision. Additionally, as seen in Section~\ref{sect:predtime}, predicting contours at the resolution used to train the model, as done by~\textcite{wasserthal_totalsegmentator_2023}, would avoid resampling predicted logits and save prediction time. 

Generating missing modalities from existing ones with deep-learning~\parencite{8494797,Li_Yu_Wang_Heng_2020,liu2020head,jiang2020self,lei2020ct,jiang2018tumor,jue2019integrating,yang2020synthetic} could address the challenge of missing modalities and prevent the need for registration; however, this approach may come at the cost of a more complex, computationally intensive, and less explainable pipeline. Errors made by such models may also negatively affect the final predicted contours. Nevertheless, considering the pace of innovation in generative AI, incorporating these approaches is suggested for future research.

\section{Conclusions}
The proposed pipeline can effectively segment a wide range of OAR in H\&N cancer patients from CT scans and MRI. It achieved the best mean DS and mean HD among all participants of the Head and Neck Organ-At-Risk CT \& MR Segmentation Challenge, establishing a new state-of-the-art. The pipeline can be used interchangeably with CT or MR scans, but works best with both modalities provided simultaneously. This pipeline can accelerate the H\&N OAR segmentation process, circumventing the tedious task of manual segmentation while eliminating inter-observer variability.

\section{Acknowledgments}
This work was supported by the Canada Research Chair Program (grant \#252136), Catalyst Grant from the University of Calgary, and NSERC Discovery Grant (RGPIN-2024-04966). This work was partly supported by Mitacs through the Mitacs Accelerate program. Computing resources were provided by the University of British Columbia Cloud Innovation Centre, powered by Amazon Web Services (AWS). This research was also partly enabled by support provided by Calcul Quebec \url{https://www.calculquebec.ca/en/}, the Digital Research Alliance of Canada \url{https://alliancecan.ca/}, and the University of Calgary Advanced Research Computing (ARC) Cluster \url{https://rcs.ucalgary.ca/RCS_Home_Page}. Additionally, this research was made possible by generous support from Google’s exploreCSR program.

\newpage

\printbibliography

@article{argiris_head_2008,
	title = {Head and neck cancer},
	volume = {371},
	issn = {1474-547X},
	doi = {10.1016/S0140-6736(08)60728-X},
	language = {eng},
	number = {9625},
	journal = {The Lancet},
	author = {Argiris, Athanassios and Karamouzis, Michalis V. and Raben, David and Ferris, Robert L.},
	month = may,
	year = {2008},
	pmid = {18486742},
	pmcid = {PMC7720415},
	pages = {1695--1709},
}

@article{caudell_future_2017,
	title = {The future of personalised radiotherapy for head and neck cancer},
	volume = {18},
	issn = {1474-5488},
	doi = {10.1016/S1470-2045(17)30252-8},
	language = {eng},
	number = {5},
	journal = {The Lancet. Oncology},
	author = {Caudell, Jimmy J. and Torres-Roca, Javier F. and Gillies, Robert J. and Enderling, Heiko and Kim, Sungjune and Rishi, Anupam and Moros, Eduardo G. and Harrison, Louis B.},
	month = may,
	year = {2017},
	pmid = {28456586},
	pmcid = {PMC7771279},
	pages = {e266--e273},
}

@article{wong_comparing_2020,
	title = {Comparing deep learning-based auto-segmentation of organs at risk and clinical target volumes to expert inter-observer variability in radiotherapy planning},
	volume = {144},
	issn = {1879-0887},
	doi = {10.1016/j.radonc.2019.10.019},
	language = {eng},
	journal = {Radiotherapy and Oncology: Journal of the European Society for Therapeutic Radiology and Oncology},
	author = {Wong, Jordan and Fong, Allan and McVicar, Nevin and Smith, Sally and Giambattista, Joshua and Wells, Derek and Kolbeck, Carter and Giambattista, Jonathan and Gondara, Lovedeep and Alexander, Abraham},
	month = mar,
	year = {2020},
	pmid = {31812930},
	pages = {152--158},
}

@article{podobnik_han-seg_2023,
	title = {{HaN}-{Seg}: {The} head and neck organ-at-risk {CT} and {MR} segmentation dataset},
	volume = {50},
	issn = {2473-4209},
	shorttitle = {{HaN}-{Seg}},
	doi = {10.1002/mp.16197},
	language = {eng},
	number = {3},
	journal = {Medical Physics},
	author = {Podobnik, Gašper and Strojan, Primož and Peterlin, Primož and Ibragimov, Bulat and Vrtovec, Tomaž},
	month = mar,
	year = {2023},
	pmid = {36594372},
	pages = {1917--1927},
}

@article{lowekamp_design_2013,
	title = {The {Design} of {SimpleITK}},
	volume = {7},
	issn = {1662-5196},
	doi = {10.3389/fninf.2013.00045},
	language = {eng},
	journal = {Frontiers in Neuroinformatics},
	author = {Lowekamp, Bradley C. and Chen, David T. and Ibáñez, Luis and Blezek, Daniel},
	year = {2013},
	pmid = {24416015},
	pmcid = {PMC3874546},
	pages = {45},
}

@inproceedings{mattes_nonrigid_2001,
author = {Mattes, David and Haynor, David R. and Vesselle, Hubert and Lewellyn, Thomas K. and Eubank, William},
title = {{Nonrigid multimodality image registration}},
volume = {4322},
booktitle = {Medical Imaging 2001: Image Processing},
editor = {Milan Sonka and Kenneth M. Hanson},
organization = {International Society for Optics and Photonics},
publisher = {SPIE},
pages = {1609 -- 1620},
year = {2001},
doi = {10.1117/12.431046},
URL = {https://doi.org/10.1117/12.431046}
}

@article{mattes_pet-ct_2003,
	title = {{PET}-{CT} image registration in the chest using free-form deformations},
	volume = {22},
	issn = {0278-0062},
	doi = {10.1109/TMI.2003.809072},
	language = {eng},
	number = {1},
	journal = {IEEE Transactions on Medical Imaging},
	author = {Mattes, David and Haynor, David R. and Vesselle, Hubert and Lewellen, Thomas K. and Eubank, William},
	month = jan,
	year = {2003},
	pmid = {12703765},
	pages = {120--128},
}

@article{isensee_nnu-net_2021,
	title = {{nnU}-{Net}: a self-configuring method for deep learning-based biomedical image segmentation},
	volume = {18},
	copyright = {2020 The Author(s), under exclusive licence to Springer Nature America, Inc.},
	issn = {1548-7105},
	shorttitle = {{nnU}-{Net}},
	url = {https://www.nature.com/articles/s41592-020-01008-z},
	doi = {10.1038/s41592-020-01008-z},
	language = {en},
	number = {2},
	urldate = {2023-02-09},
	journal = {Nature Methods},
	author = {Isensee, Fabian and Jaeger, Paul F. and Kohl, Simon A. A. and Petersen, Jens and Maier-Hein, Klaus H.},
	month = feb,
	year = {2021},
	pages = {203--211},
}

@article{verhaart_relevance_2014,
	title = {The relevance of {MRI} for patient modeling in head and neck hyperthermia treatment planning: {A} comparison of {CT} and {CT}‐{MRI} based tissue segmentation on simulated temperature},
	volume = {41},
	issn = {0094-2405, 2473-4209},
	shorttitle = {The relevance of {MRI} for patient modeling in head and neck hyperthermia treatment planning},
	url = {https://aapm.onlinelibrary.wiley.com/doi/10.1118/1.4901270},
	doi = {10.1118/1.4901270},
	language = {en},
	number = {12},
	urldate = {2024-04-09},
	journal = {Medical Physics},
	author = {Verhaart, René F. and Fortunati, Valerio and Verduijn, Gerda M. and Van Der Lugt, Aad and Van Walsum, Theo and Veenland, Jifke F. and Paulides, Margarethus M.},
	month = dec,
	year = {2014},
	pages = {123302},
}

@article{khoo_comparison_1999,
	title = {Comparison of {MRI} with {CT} for the radiotherapy planning of prostate cancer: a feasibility study.},
	volume = {72},
	issn = {1748-880X, 0007-1285},
	shorttitle = {Comparison of {MRI} with {CT} for the radiotherapy planning of prostate cancer},
	url = {https://academic.oup.com/bjr/article/72/858/590-7/7320889},
	doi = {10.1259/bjr.72.858.10560342},
	language = {en},
	number = {858},
	urldate = {2024-04-09},
	journal = {The British Journal of Radiology},
	author = {Khoo, V S and Padhani, A R and Tanner, S F and Finnigan, D J and Leach, M O and Dearnaley, D P},
	month = jun,
	year = {1999},
	pages = {590--597},
}

@ARTICLE{8494797,
  author={Huo, Yuankai and Xu, Zhoubing and Moon, Hyeonsoo and Bao, Shunxing and Assad, Albert and Moyo, Tamara K. and Savona, Michael R. and Abramson, Richard G. and Landman, Bennett A.},
  journal={IEEE Transactions on Medical Imaging}, 
  title={SynSeg-Net: Synthetic Segmentation Without Target Modality Ground Truth}, 
  year={2019},
  volume={38},
  number={4},
  pages={1016-1025},
  doi={10.1109/TMI.2018.2876633}}

@article{Li_Yu_Wang_Heng_2020, title={Towards Cross-Modality Medical Image Segmentation with Online Mutual Knowledge Distillation}, volume={34}, url={https://ojs.aaai.org/index.php/AAAI/article/view/5421}, DOI={10.1609/aaai.v34i01.5421}, abstractNote={&lt;p&gt;The success of deep convolutional neural networks is partially attributed to the massive amount of annotated training data. However, in practice, medical data annotations are usually expensive and time-consuming to be obtained. Considering multi-modality data with the same anatomic structures are widely available in clinic routine, in this paper, we aim to exploit the prior knowledge (&lt;em&gt;e.g.&lt;/em&gt;, shape priors) learned from one modality (&lt;em&gt;aka.&lt;/em&gt;, assistant modality) to improve the segmentation performance on another modality (&lt;em&gt;aka.&lt;/em&gt;, target modality) to make up annotation scarcity. To alleviate the learning difficulties caused by modality-specific appearance discrepancy, we first present an Image Alignment Module (IAM) to narrow the appearance gap between assistant and target modality data. We then propose a novel Mutual Knowledge Distillation (MKD) scheme to thoroughly exploit the modality-shared knowledge to facilitate the target-modality segmentation. To be specific, we formulate our framework as an integration of two individual segmentors. Each segmentor not only explicitly extracts one modality knowledge from corresponding annotations, but also implicitly explores another modality knowledge from its counterpart in mutual-guided manner. The ensemble of two segmentors would further integrate the knowledge from both modalities and generate reliable segmentation results on target modality. Experimental results on the public multi-class cardiac segmentation data, &lt;em&gt;i.e.&lt;/em&gt;, MM-WHS 2017, show that our method achieves large improvements on CT segmentation by utilizing additional MRI data and outperforms other state-of-the-art multi-modality learning methods.&lt;/p&gt;}, number={01}, journal={Proceedings of the AAAI Conference on Artificial Intelligence}, author={Li, Kang and Yu, Lequan and Wang, Shujun and Heng, Pheng-Ann}, year={2020}, month=apr, pages={775-783} }

@article{liu2020head,
  title={Head and neck multi-organ auto-segmentation on CT images aided by synthetic MRI},
  author={Liu, Yingzi and Lei, Yang and Fu, Yabo and Wang, Tonghe and Zhou, Jun and Jiang, Xiaojun and McDonald, Mark and Beitler, Jonathan J and Curran, Walter J and Liu, Tian and others},
  journal={Medical Physics},
  volume={47},
  number={9},
  pages={4294--4302},
  year={2020},
  publisher={Wiley Online Library}
}

@article{jiang2020self,
  title={Self-derived organ attention for unpaired CT-MRI deep domain adaptation based MRI segmentation},
  author={Jiang, Jue and Hu, Yu-Chi and Tyagi, Neelam and Wang, Chuang and Lee, Nancy and Deasy, Joseph O and Sean, Berry and Veeraraghavan, Harini},
  journal={Physics in Medicine \& Biology},
  volume={65},
  number={20},
  pages={205001},
  year={2020},
  publisher={IOP Publishing}
}

@article{lei2020ct,
  title={CT prostate segmentation based on synthetic MRI-aided deep attention fully convolution network},
  author={Lei, Yang and Dong, Xue and Tian, Zhen and Liu, Yingzi and Tian, Sibo and Wang, Tonghe and Jiang, Xiaojun and Patel, Pretesh and Jani, Ashesh B and Mao, Hui and others},
  journal={Medical Physics},
  volume={47},
  number={2},
  pages={530--540},
  year={2020},
  publisher={Wiley Online Library}
}

@inproceedings{jiang2018tumor,
  title={Tumor-aware, adversarial domain adaptation from CT to MRI for lung cancer segmentation},
  author={Jiang, Jue and Hu, Yu-Chi and Tyagi, Neelam and Zhang, Pengpeng and Rimner, Andreas and Mageras, Gig S and Deasy, Joseph O and Veeraraghavan, Harini},
  booktitle={Medical Image Computing and Computer Assisted Intervention--MICCAI 2018: 21st International Conference, Granada, Spain, September 16-20, 2018, Proceedings, Part II 11},
  pages={777--785},
  year={2018},
  organization={Springer}
}

@inproceedings{jue2019integrating,
  title={Integrating cross-modality hallucinated MRI with CT to aid mediastinal lung tumor segmentation},
  author={Jue, Jiang and Jason, Hu and Neelam, Tyagi and Andreas, Rimner and Sean, Berry L and Joseph, Deasy O and Harini, Veeraraghavan},
  booktitle={MMedical Image Computing and Computer Assisted Intervention--MICCAI 2019: 22nd International Conference, Shenzhen, China, October 13--17, 2019, Proceedings, Part VI 22},
  pages={221--229},
  year={2019},
  organization={Springer}
}

@article{yang2020synthetic,
  title={Synthetic MRI-aided multi-organ CT segmentation for head and neck radiotherapy treatment planning},
  author={Yang, X and Lei, Y and Liu, Y and Wang, T and Zhou, J and Jiang, X and McDonald, MW and Beitler, JJ and Curran, WJ and Liu, T},
  journal={International Journal of Radiation Oncology, Biology, Physics},
  volume={108},
  number={3},
  pages={e341},
  year={2020},
  publisher={Elsevier}
}

@inproceedings{podobnik2023multimodal,
  title={Multimodal CT and MR segmentation of head and neck organs-at-risk},
  author={Podobnik, Ga{\v{s}}per and Strojan, Primo{\v{z}} and Peterlin, Primo{\v{z}} and Ibragimov, Bulat and Vrtovec, Toma{\v{z}}},
  booktitle={International Conference on Medical Image Computing and Computer-Assisted Intervention},
  pages={745--755},
  year={2023},
  organization={Springer}
}

@ARTICLE{8933421,
  author={Yang, Wanqi and Shi, Yinghuan and Park, Sang Hyun and Yang, Ming and Gao, Yang and Shen, Dinggang},
  journal={{IEEE} Journal of Biomedical and Health Informatics}, 
  title={An Effective MR-Guided {CT} Network Training for Segmenting Prostate in {CT} Images}, 
  year={2020},
  volume={24},
  number={8},
  pages={2278-2291},
  doi={10.1109/JBHI.2019.2960153}
}

@INPROCEEDINGS{8354170,

  author={Valindria, Vanya V. and Pawlowski, Nick and Rajchl, Martin and Lavdas, Ioannis and Aboagye, Eric O. and Rockall, Andrea G. and Rueckert, Daniel and Glocker, Ben},

  booktitle={2018 IEEE Winter Conference on Applications of Computer Vision (WACV)}, 

  title={Multi-modal Learning from Unpaired Images: Application to Multi-organ Segmentation in CT and MRI}, 

  year={2018},

  volume={},

  number={},

  pages={547-556},

  doi={10.1109/WACV.2018.00066}}

@ARTICLE{8764342,

  author={Dou, Qi and Ouyang, Cheng and Chen, Cheng and Chen, Hao and Glocker, Ben and Zhuang, Xiahai and Heng, Pheng-Ann},

  journal={IEEE Access}, 

  title={PnP-AdaNet: Plug-and-Play Adversarial Domain Adaptation Network at Unpaired Cross-Modality Cardiac Segmentation}, 

  year={2019},

  volume={7},

  number={},

  pages={99065-99076},

  doi={10.1109/ACCESS.2019.2929258}}

@ARTICLE{7169562,

  author={Neverova, Natalia and Wolf, Christian and Taylor, Graham and Nebout, Florian},

  journal={IEEE Transactions on Pattern Analysis and Machine Intelligence}, 

  title={ModDrop: Adaptive Multi-Modal Gesture Recognition}, 

  year={2016},

  volume={38},

  number={8},

  pages={1692-1706},

  doi={10.1109/TPAMI.2015.2461544}}

@InProceedings{10.1007/978-3-319-55050-3_8,
author="Li, Xiaomeng
and Dou, Qi
and Chen, Hao
and Fu, Chi-Wing
and Heng, Pheng-Ann",
editor="Yao, Jianhua
and Vrtovec, Toma{\v{z}}
and Zheng, Guoyan
and Frangi, Alejandro
and Glocker, Ben
and Li, Shuo",
title="Multi-scale and Modality Dropout Learning for Intervertebral Disc Localization and Segmentation",
booktitle="Computational Methods and Clinical Applications for Spine Imaging",
year="2016",
publisher="Springer International Publishing",
address="Cham",
pages="85--91",
isbn="978-3-319-55050-3"
}

@InProceedings{10.1007/978-3-031-16443-9_43,
author="Liu, Han
and Fan, Yubo
and Li, Hao
and Wang, Jiacheng
and Hu, Dewei
and Cui, Can
and Lee, Ho Hin
and Zhang, Huahong
and Oguz, Ipek",
editor="Wang, Linwei
and Dou, Qi
and Fletcher, P. Thomas
and Speidel, Stefanie
and Li, Shuo",
title="ModDrop++: A Dynamic Filter Network with Intra-subject Co-training for Multiple Sclerosis Lesion Segmentation with Missing Modalities",
booktitle="Medical Image Computing and Computer Assisted Intervention -- MICCAI 2022",
year="2022",
publisher="Springer Nature Switzerland",
address="Cham",
pages="444--453",
isbn="978-3-031-16443-9"
}

@ARTICLE{5338015,
  author={Klein, Stefan and Staring, Marius and Murphy, Keelin and Viergever, Max A. and Pluim, Josien P. W.},
  journal={IEEE Transactions on Medical Imaging}, 
  title={elastix: A Toolbox for Intensity-Based Medical Image Registration}, 
  year={2010},
  volume={29},
  number={1},
  pages={196-205},
  doi={10.1109/TMI.2009.2035616}}

@article{regparam,
  author = {Sara Leibfarth and David Mönnich and Stefan Welz and Christine Siegel and Nina Schwenzer and Holger Schmidt and Daniel Zips and Daniela Thorwarth},
  title = {A strategy for multimodal deformable image registration to integrate PET/MR into radiotherapy treatment planning},
  journal = {Acta Oncologica},
  volume = {52},
  number = {7},
  pages = {1353--1359},
  year = {2013},
  publisher = {Taylor \& Francis},
  doi = {10.3109/0284186X.2013.813964},
  note ={PMID: 23879651},
  URL = { https://doi.org/10.3109/0284186X.2013.813964},
  eprint = {https://doi.org/10.3109/0284186X.2013.813964}
}

@article{FEDOROV20121323,
title = {3D Slicer as an image computing platform for the Quantitative Imaging Network},
journal = {Magnetic Resonance Imaging},
volume = {30},
number = {9},
pages = {1323-1341},
year = {2012},
note = {Quantitative Imaging in Cancer},
issn = {0730-725X},
doi = {https://doi.org/10.1016/j.mri.2012.05.001},
url = {https://www.sciencedirect.com/science/article/pii/S0730725X12001816},
author = {Andriy Fedorov and Reinhard Beichel and Jayashree Kalpathy-Cramer and Julien Finet and Jean-Christophe Fillion-Robin and Sonia Pujol and Christian Bauer and Dominique Jennings and Fiona Fennessy and Milan Sonka and John Buatti and Stephen Aylward and James V. Miller and Steve Pieper and Ron Kikinis}
}

@article{podobnik_han-seg_2024,
	title = {{HaN}-{Seg}: {The} head and neck organ-at-risk {CT} and {MR} segmentation challenge},
	volume = {198},
	issn = {01678140},
	shorttitle = {{HaN}-{Seg}},
	url = {https://linkinghub.elsevier.com/retrieve/pii/S0167814024006807},
	doi = {10.1016/j.radonc.2024.110410},
	language = {en},
	urldate = {2024-07-30},
	journal = {Radiotherapy and Oncology},
	author = {Podobnik, Gašper and Ibragimov, Bulat and Tappeiner, Elias and Lee, Chanwoong and Kim, Jin Sung and Mesbah, Zacharia and Modzelewski, Romain and Ma, Yihao and Yang, Fan and Rudecki, Mikołaj and Wodziński, Marek and Peterlin, Primož and Strojan, Primož and Vrtovec, Tomaž},
	month = sep,
	year = {2024},
	pages = {110410},
}

@Article{ijerph19159057,
AUTHOR = {D’Aviero, Andrea and Re, Alessia and Catucci, Francesco and Piccari, Danila and Votta, Claudio and Piro, Domenico and Piras, Antonio and Di Dio, Carmela and Iezzi, Martina and Preziosi, Francesco and Menna, Sebastiano and Quaranta, Flaviovincenzo and Boschetti, Althea and Marras, Marco and Miccichè, Francesco and Gallus, Roberto and Indovina, Luca and Bussu, Francesco and Valentini, Vincenzo and Cusumano, Davide and Mattiucci, Gian Carlo},
TITLE = {Clinical Validation of a Deep-Learning Segmentation Software in Head and Neck: An Early Analysis in a Developing Radiation Oncology Center},
JOURNAL = {International Journal of Environmental Research and Public Health},
VOLUME = {19},
YEAR = {2022},
NUMBER = {15},
ARTICLE-NUMBER = {9057},
URL = {https://www.mdpi.com/1660-4601/19/15/9057},
PubMedID = {35897425},
ISSN = {1660-4601},
DOI = {10.3390/ijerph19159057}
}

@article{chen_validation_2024,
	title = {Validation of a deep-learning segmentation model for adult and pediatric head and neck radiotherapy in different patient positions},
	volume = {29},
	issn = {24056316},
	url = {https://linkinghub.elsevier.com/retrieve/pii/S2405631623001185},
	doi = {10.1016/j.phro.2023.100527},
	language = {en},
	urldate = {2024-08-01},
	journal = {Physics and Imaging in Radiation Oncology},
	author = {Chen, Linda and Platzer, Patricia and Reschl, Christian and Schafasand, Mansure and Nachankar, Ankita and Lukas Hajdusich, Christoph and Kuess, Peter and Stock, Markus and Habraken, Steven and Carlino, Antonio},
	month = jan,
	year = {2024},
	pages = {100527},
}

@article{wasserthal_totalsegmentator_2023,
	title = {{TotalSegmentator}: {Robust} {Segmentation} of 104 {Anatomic} {Structures} in {CT} {Images}},
	volume = {5},
	issn = {2638-6100},
	shorttitle = {{TotalSegmentator}},
	url = {http://pubs.rsna.org/doi/10.1148/ryai.230024},
	doi = {10.1148/ryai.230024},
	language = {en},
	number = {5},
	urldate = {2024-08-03},
	journal = {Radiology: Artificial Intelligence},
	author = {Wasserthal, Jakob and Breit, Hanns-Christian and Meyer, Manfred T. and Pradella, Maurice and Hinck, Daniel and Sauter, Alexander W. and Heye, Tobias and Boll, Daniel T. and Cyriac, Joshy and Yang, Shan and Bach, Michael and Segeroth, Martin},
	month = sep,
	year = {2023},
	pages = {e230024},
}

@article{totalsegmentator_mri,
	title = {{TotalSegmentator} {MRI}: {Robust} {Sequence}-independent {Segmentation} of {Multiple} {Anatomic} {Structures} in {MRI}},
	volume = {314},
	issn = {0033-8419, 1527-1315},
	shorttitle = {{TotalSegmentator} {MRI}},
	url = {http://pubs.rsna.org/doi/10.1148/radiol.241613},
	doi = {10.1148/radiol.241613},
	language = {en},
	number = {2},
	urldate = {2025-05-08},
	journal = {Radiology},
	author = {Akinci D’Antonoli, Tugba and Berger, Lucas K. and Indrakanti, Ashraya K. and Vishwanathan, Nathan and Weiss, Jakob and Jung, Matthias and Berkarda, Zeynep and Rau, Alexander and Reisert, Marco and Küstner, Thomas and Walter, Alexandra and Merkle, Elmar M. and Boll, Daniel T. and Breit, Hanns-Christian and Nicoli, Andrew Phillip and Segeroth, Martin and Cyriac, Joshy and Yang, Shan and Wasserthal, Jakob},
	month = feb,
	year = {2025},
	pages = {e241613},
}

@article{isensee2024nnu,
  title={nnu-net revisited: A call for rigorous validation in 3d medical image segmentation},
  author={Isensee, Fabian and Wald, Tassilo and Ulrich, Constantin and Baumgartner, Michael and Roy, Saikat and Maier-Hein, Klaus and Jaeger, Paul F},
  journal={arXiv preprint arXiv:2404.09556},
  year={2024}
}

@article{Xu_2022,
doi = {10.1088/1361-6560/ac840f},
url = {https://dx.doi.org/10.1088/1361-6560/ac840f},
year = {2022},
month = {aug},
publisher = {IOP Publishing},
volume = {67},
number = {17},
pages = {17TR01},
author = {Jiangchang Xu and Bolun Zeng and Jan Egger and Chunliang Wang and Örjan Smedby and Xiaoyi Jiang and Xiaojun Chen},
title = {A review on AI-based medical image computing in head and neck surgery},
journal = {Physics in Medicine \& Biology}
}

@misc{glisrt_dataset,
	title = {Glioma {Image} {Segmentation} for {Radiotherapy}: {RT} targets, barriers to cancer spread, and organs at risk ({GLIS}-{RT})},
	copyright = {TCIA Limited Access License},
	shorttitle = {Glioma {Image} {Segmentation} for {Radiotherapy}},
	url = {https://www.cancerimagingarchive.net/collection/glis-rt/},
	doi = {10.7937/TCIA.T905-ZQ20},
	urldate = {2025-05-08},
	publisher = {The Cancer Imaging Archive},
	author = {Shusharina, Nadya and Bortfeld, Thomas},
	year = {2021},
}

@inproceedings{glisrt_paper,
	address = {Cham},
	title = {Cross-{Modality} {Brain} {Structures} {Image} {Segmentation} for the {Radiotherapy} {Target} {Definition} and {Plan} {Optimization}},
	isbn = {9783030718275},
	doi = {10.1007/978-3-030-71827-5_1},
	language = {en},
	booktitle = {Segmentation, {Classification}, and {Registration} of {Multi}-modality {Medical} {Imaging} {Data}},
	publisher = {Springer International Publishing},
	author = {Shusharina, Nadya and Bortfeld, Thomas and Cardenas, Carlos and De, Brian and Diao, Kevin and Hernandez, Soleil and Liu, Yufei and Maroongroge, Sean and Söderberg, Jonas and Soliman, Moaaz},
	editor = {Shusharina, Nadya and Heinrich, Mattias P. and Huang, Ruobing},
	year = {2021},
	pages = {3--15},
}

@misc{burdenko_dataset,
	title = {Burdenko's {Glioblastoma} {Progression} {Dataset} ({Burdenko}-{GBM}-{Progression})},
	copyright = {TCIA Limited Access License, Creative Commons Attribution 4.0 International},
	url = {https://www.cancerimagingarchive.net/collection/burdenko-gbm-progression/},
	doi = {10.7937/E1QP-D183},
	urldate = {2025-05-08},
	publisher = {The Cancer Imaging Archive},
	author = {Zolotova, Svetlana V. and Golanov, Andrey V. and Pronin, Igor N. and Dalechina, Alexandra V. and Nikolaeva, Anna A. and Belyashova, Alexandra S. and Usachev, Dmitry Y. and Kondrateva, Ekaterina A. and Druzhinina, Polina V. and Shirokikh, Boris N. and Saparov, Talgat N. and Belyaev, Mikhail G. and Kurmukov, Anvar I.},
	year = {2023},
}

@article{hnscc_paper,
	title = {Imaging and clinical data archive for head and neck squamous cell carcinoma patients treated with radiotherapy},
	volume = {5},
	copyright = {2018 The Author(s)},
	issn = {2052-4463},
	url = {https://www.nature.com/articles/sdata2018173},
	doi = {10.1038/sdata.2018.173},
	language = {en},
	number = {1},
	urldate = {2025-05-08},
	journal = {Scientific Data},
	author = {Grossberg, Aaron J. and Mohamed, Abdallah S. R. and Elhalawani, Hesham and Bennett, William C. and Smith, Kirk E. and Nolan, Tracy S. and Williams, Bowman and Chamchod, Sasikarn and Heukelom, Jolien and Kantor, Michael E. and Browne, Theodora and Hutcheson, Katherine A. and Gunn, G. Brandon and Garden, Adam S. and Morrison, William H. and Frank, Steven J. and Rosenthal, David I. and Freymann, John B. and Fuller, Clifton D.},
	month = sep,
	year = {2018},
	pages = {180173},
}

@misc{hnscc_dataset,
	title = {Data from {Head} and {Neck} {Cancer} {CT} {Atlas} ({Head}-{Neck}-{CT}-{Atlas})},
	copyright = {TCIA Limited Access License, Creative Commons Attribution 3.0 Unported},
	url = {https://www.cancerimagingarchive.net/analysis-result/head-neck-ct-atlas/},
	doi = {10.7937/K9/TCIA.2017.UMZ8DV6S},
	urldate = {2025-05-08},
	publisher = {The Cancer Imaging Archive},
	author = {Grossberg, Aaron and Mohamed, Abdallah and El Halawani, Hesham and Bennett, William and Smith, Kirk and Nolan, Tracy and Chamchod, Sasikarn and Kantor, Michael and Browne, Theodora and Hutcheson, Katherine and Gunn, Gary and Garden, Adam and Frank, Steven and Rosenthal, David and Freymann, John and Fuller, Clifton},
	collaborator = {{TCIA Team}},
	year = {2017},
}

@article{otsu_thresholding,
author = "R. Beare",
title = "Histogram-based thresholding - some missing methods",
howpublished = "\url{http://hdl.handle.net/10380/3279}",
year = 2011,
month = 06,
institution = "Monash University",
publisher = "The Insight Journal",
doi = "10.54294/efycla",
}

@article{han_contouring_guidelines,
	title = {{CT}-based delineation of organs at risk in the head and neck region: {DAHANCA}, {EORTC}, {GORTEC}, {HKNPCSG}, {NCIC} {CTG}, {NCRI}, {NRG} {Oncology} and {TROG} consensus guidelines},
	volume = {117},
	issn = {01678140},
	shorttitle = {{CT}-based delineation of organs at risk in the head and neck region},
	url = {https://linkinghub.elsevier.com/retrieve/pii/S0167814015004016},
	doi = {10.1016/j.radonc.2015.07.041},
	language = {en},
	number = {1},
	urldate = {2025-05-22},
	journal = {Radiotherapy and Oncology},
	author = {Brouwer, Charlotte L. and Steenbakkers, Roel J.H.M. and Bourhis, Jean and Budach, Wilfried and Grau, Cai and Grégoire, Vincent and Van Herk, Marcel and Lee, Anne and Maingon, Philippe and Nutting, Chris and O’Sullivan, Brian and Porceddu, Sandro V. and Rosenthal, David I. and Sijtsema, Nanna M. and Langendijk, Johannes A.},
	month = oct,
	year = {2015},
	pages = {83--90},
}

@article{he_multitrans_2023,
	title = {{MultiTrans}: {Multi}-scale feature fusion transformer with transfer learning strategy for multiple organs segmentation of head and neck {CT} images},
	volume = {18},
	issn = {2590-0935},
	shorttitle = {{MultiTrans}},
	url = {https://www.sciencedirect.com/science/article/pii/S2590093523000309},
	doi = {10.1016/j.medntd.2023.100235},
	urldate = {2025-06-06},
	journal = {Medicine in Novel Technology and Devices},
	author = {He, Yufang and Song, Fan and Wu, Wangjiang and Tian, Suqing and Zhang, Tianyi and Zhang, Shuming and Zhang, Peng and Ma, Chenbin and Feng, Youdan and Yang, Ruijie and Zhang, Guanglei},
	month = jun,
	year = {2023},
	pages = {100235},
}

@article{gao_focusnetv2_2021,
	title = {{FocusNetv2}: {Imbalanced} large and small organ segmentation with adversarial shape constraint for head and neck {CT} images},
	volume = {67},
	issn = {1361-8415},
	shorttitle = {{FocusNetv2}},
	url = {https://www.sciencedirect.com/science/article/pii/S136184152030195X},
	doi = {10.1016/j.media.2020.101831},
	urldate = {2025-06-06},
	journal = {Medical Image Analysis},
	author = {Gao, Yunhe and Huang, Rui and Yang, Yiwei and Zhang, Jie and Shao, Kainan and Tao, Changjuan and Chen, Yuanyuan and Metaxas, Dimitris N. and Li, Hongsheng and Chen, Ming},
	month = jan,
	year = {2021},
	pages = {101831},
}

@article{tappeiner_multi_organ_2019,
	title = {Multi-organ segmentation of the head and neck area: an efficient hierarchical neural networks approach},
	volume = {14},
	issn = {1861-6429},
	shorttitle = {Multi-organ segmentation of the head and neck area},
	url = {https://doi.org/10.1007/s11548-019-01922-4},
	doi = {10.1007/s11548-019-01922-4},
	language = {en},
	number = {5},
	urldate = {2025-06-06},
	journal = {International Journal of Computer Assisted Radiology and Surgery},
	author = {Tappeiner, Elias and Pröll, Samuel and Hönig, Markus and Raudaschl, Patrick F. and Zaffino, Paolo and Spadea, Maria F. and Sharp, Gregory C. and Schubert, Rainer and Fritscher, Karl},
	month = may,
	year = {2019},
	pages = {745--754},
}

@article{luan_accurate_2024,
	title = {Accurate and robust auto‐segmentation of head and neck organ‐at‐risks based on a novel {CNN} fine‐tuning workflow},
	volume = {25},
	issn = {1526-9914, 1526-9914},
	url = {https://aapm.onlinelibrary.wiley.com/doi/10.1002/acm2.14248},
	doi = {10.1002/acm2.14248},
	language = {en},
	number = {1},
	urldate = {2025-06-06},
	journal = {Journal of Applied Clinical Medical Physics},
	author = {Luan, Shunyao and Wu, Kun and Wu, Yuan and Zhu, Benpeng and Wei, Wei and Xue, Xudong},
	month = jan,
	year = {2024},
	pages = {e14248},
}

@article{singh_multi_organ_2024,
	title = {Multi-organ segmentation of organ-at-risk ({OAR}'s) of head and neck site using ensemble learning technique},
	volume = {30},
	issn = {1078-8174},
	url = {https://www.sciencedirect.com/science/article/pii/S1078817424000385},
	doi = {10.1016/j.radi.2024.02.001},
	number = {2},
	urldate = {2025-06-06},
	journal = {Radiography},
	author = {Singh, S. and Singh, B. K. and Kumar, A.},
	month = mar,
	year = {2024},
	pages = {673--680},
}

@article{tappeiner_tackling_2022,
	title = {Tackling the class imbalance problem of deep learning-based head and neck organ segmentation},
	volume = {17},
	issn = {1861-6429},
	url = {https://doi.org/10.1007/s11548-022-02649-5},
	doi = {10.1007/s11548-022-02649-5},
	language = {en},
	number = {11},
	urldate = {2025-06-06},
	journal = {International Journal of Computer Assisted Radiology and Surgery},
	author = {Tappeiner, Elias and Welk, Martin and Schubert, Rainer},
	month = nov,
	year = {2022},
	pages = {2103--2111},
}

@article{dicomrttool_paper,
title = {Simple Python Module for Conversions Between DICOM Images and Radiation Therapy Structures, Masks, and Prediction Arrays},
journal = {Practical Radiation Oncology},
volume = {11},
number = {3},
pages = {226-229},
year = {2021},
issn = {1879-8500},
doi = {https://doi.org/10.1016/j.prro.2021.02.003},
url = {https://www.sciencedirect.com/science/article/pii/S1879850021000485},
author = {Brian M. Anderson and Kareem A. Wahid and Kristy K. Brock},
}
\newpage

\end{document}


\maketitle




\appendix
\section{Registration robustness}

To evaluate the impact of the stochastic aspect of the registration process on the downstream segmentation, a robustness experiment with repeated preprocessing, registration and inference on the same patient was conducted. Contours were predicted ten times with our end-to-end pipeline for the patients of the internal testing set of the first fold. Consequently, registration was run ten times for each patient. MRI scans were registered to the CT scan but the CT scan was not used for predictions. Predictions were made from the MRI scan only, using the model trained with MD, to focus on the impact of the registration. Ten Dice Similarity scores (DS) and Hausdorff Distances (HD) were obtained for each OAR of each patient. Absolute error between the DS and HD of the first repetition and the nine other repetitions were computed and analyzed.

\begin{figure}[h!] 
  \centering
 \includegraphics[width=0.95\textwidth]{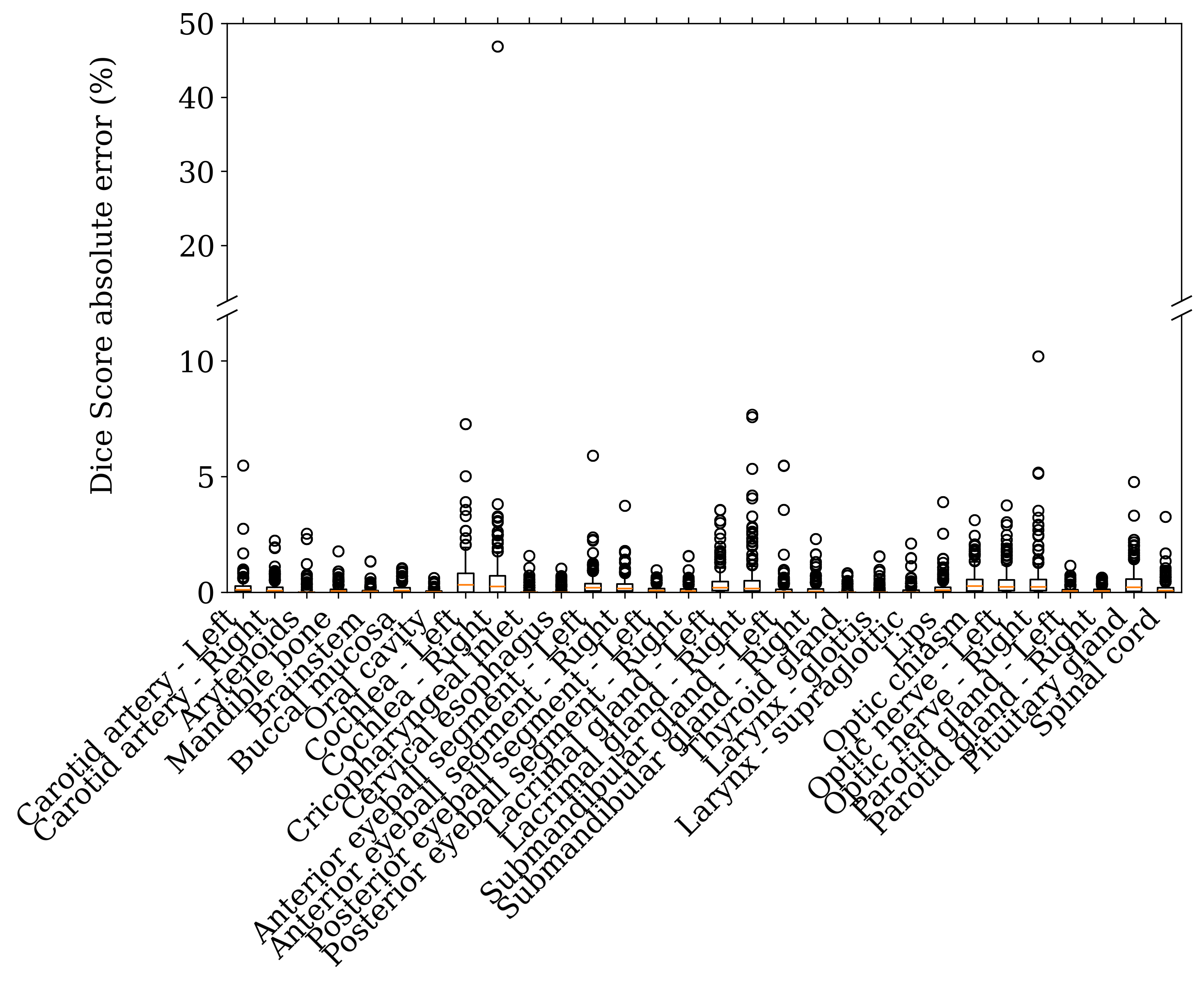}
  \caption{Absolute errors in Dice Score between the first repetition and nine other repetitions of our pipeline inference with MRI only.}
  \label{fig:robust_ds}
\end{figure}
\begin{figure}[h!] 
  \centering
 \includegraphics[width=0.95\textwidth]{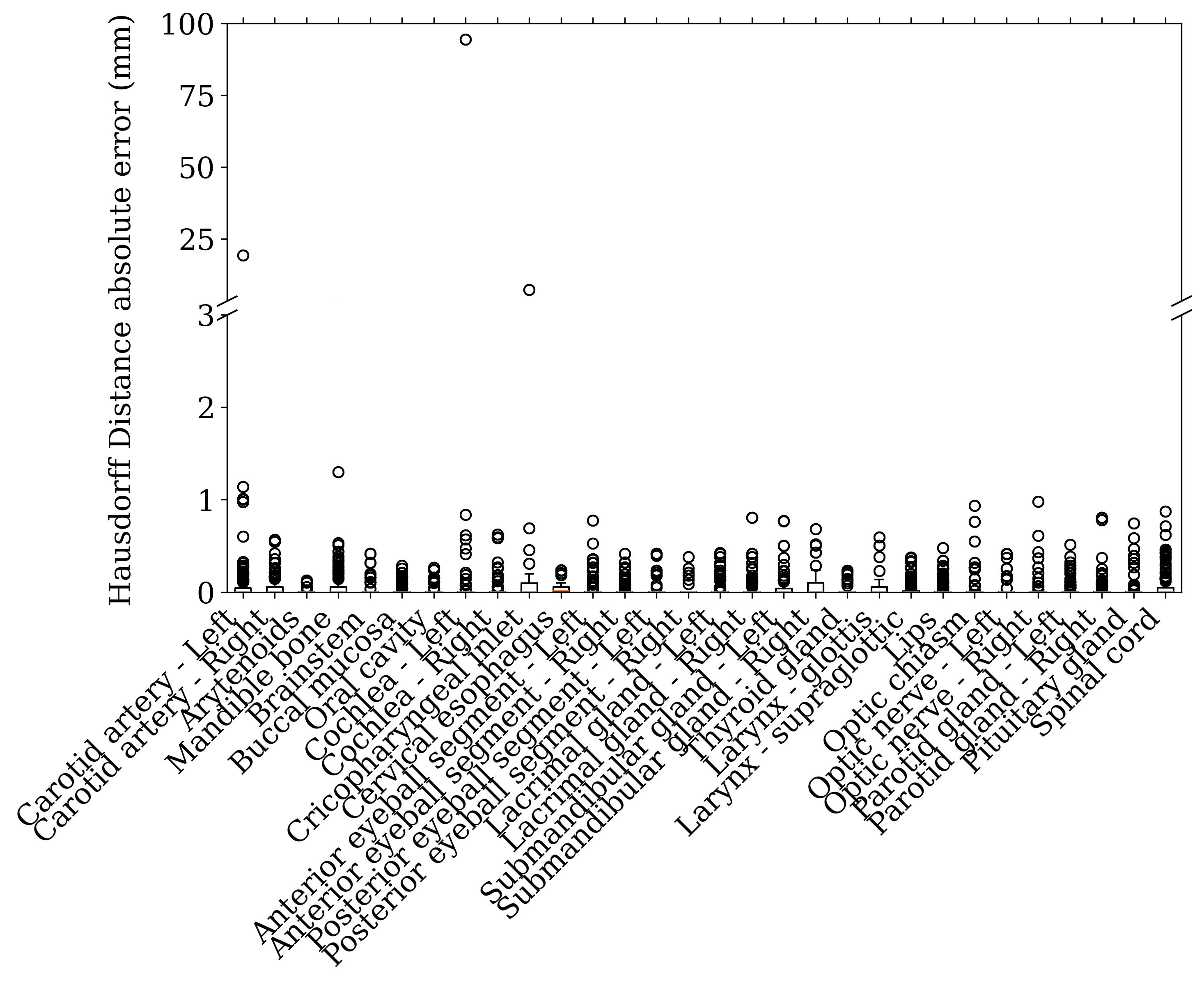}
  \caption{Absolute errors in Hausdorff Distance between the first repetition and nine other repetitions of our pipeline inference with MRI only.}
  \label{fig:robust_hd}
\end{figure}
The average absolute error in DS across multiple inference repetitions of the end-to-end MRI-only pipeline on the same patient, for all OARs and patients in our first fold internal testing set was 0.24\% ± 0.82\%, with a range of 0.00 to 46.90\%. For the HD, the average absolute error across repetitions was 0.06 mm ± 1.46 mm, spanning from 0.00 to 94.38 mm. Figure~\ref{fig:robust_ds}, and \ref{fig:robust_hd} show details of these absolute errors for each of the OAR contoured. These results demonstrate the robustness of the registration pipeline.

\newpage

\section{Multi-step training for dataset expansion}
\begin{figure}[h!] 
  \centering
 \includegraphics[width=0.95\textwidth]{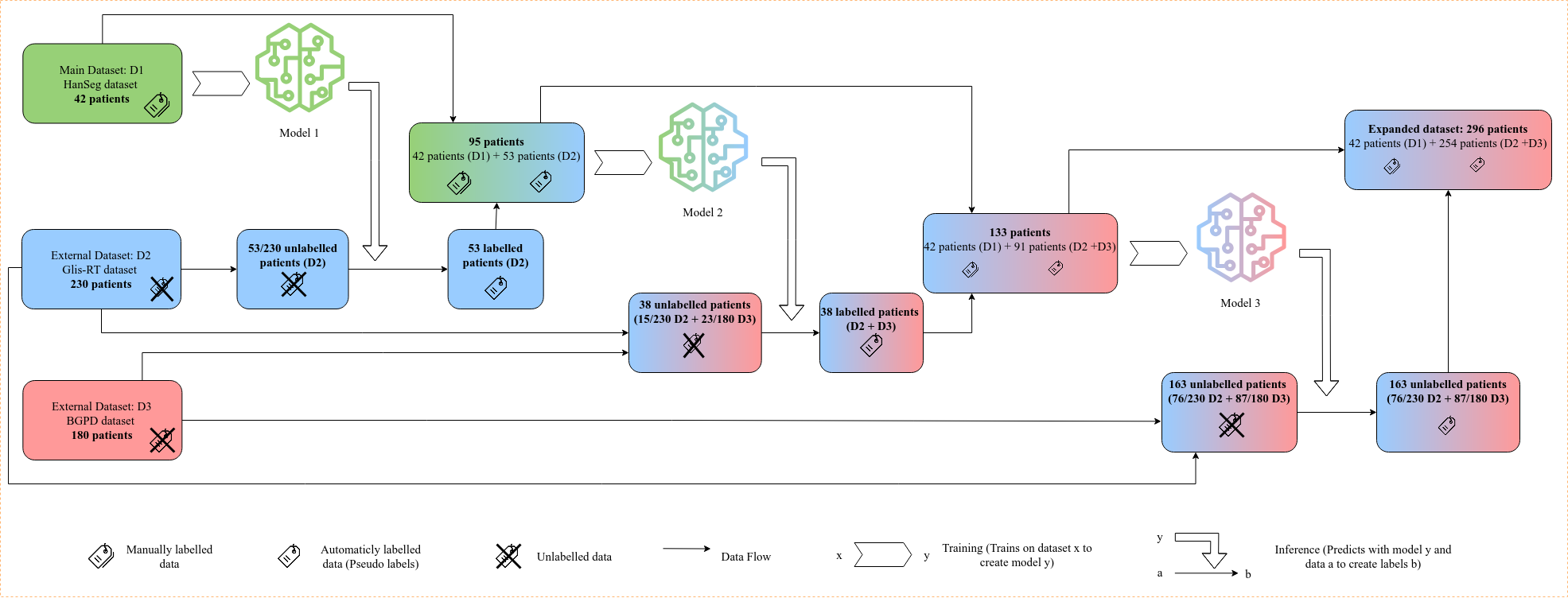}
  \caption{Detailed multi-step training process to expand our original dataset. This figure extends Figure 2 from the main paper with detailed number of patients used at the different stage of the dataset creation.}
  \label{fig:multistep_training}
\end{figure}
Figure~\ref{fig:multistep_training} illustrates the sequential training process implemented for dataset expansion in this study. Initially, pseudo-contours were generated for the Glis-RT dataset, followed by the BGPD dataset. Among the 410 patients examined by a clinician, 254 pseudo-contours were validated and subsequently used to expand the training dataset.

\section{Hounsfield Unit distributions}

\begin{figure}[h!] 
  \centering
 \includegraphics[width=0.95\textwidth]{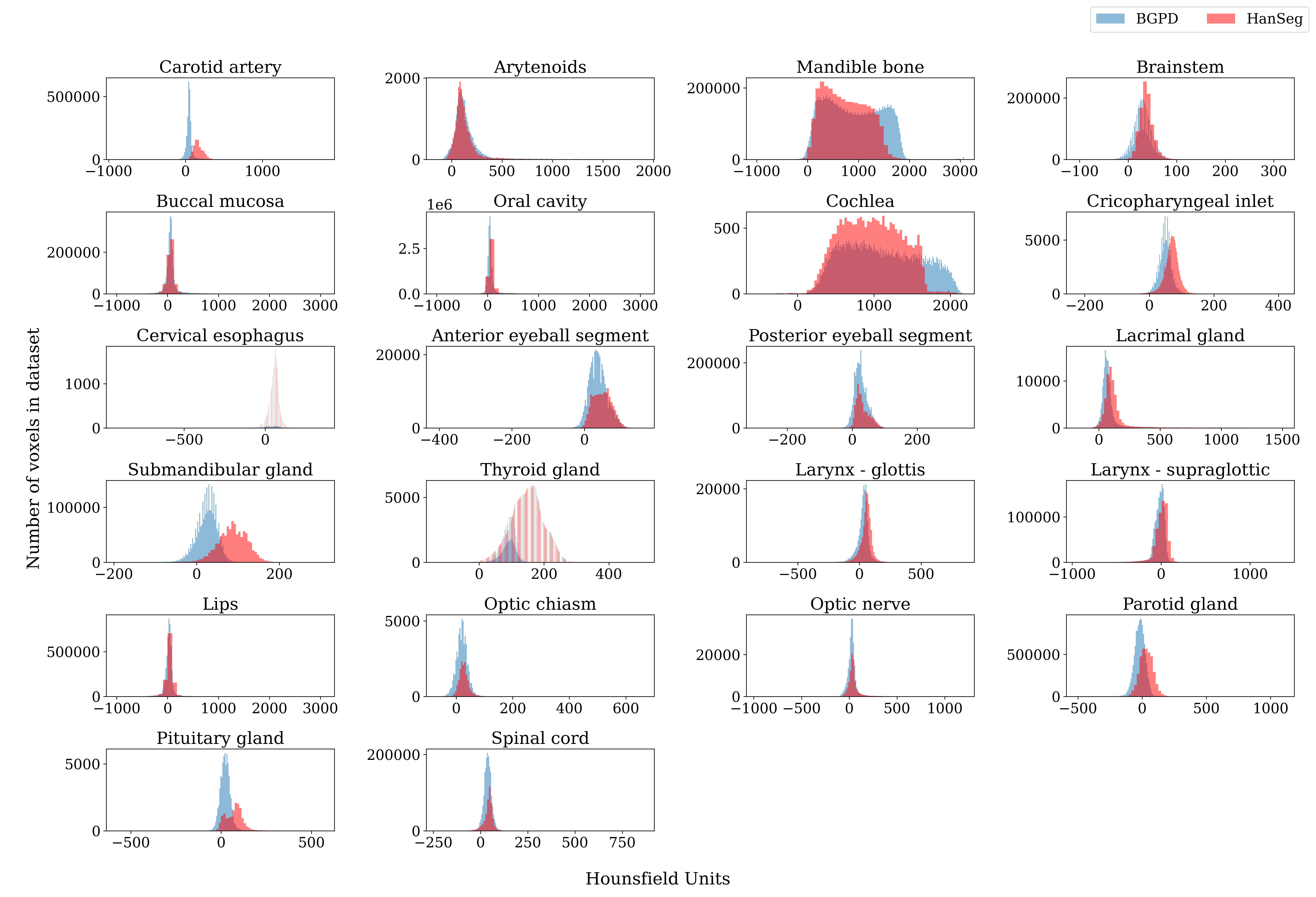}
  \caption{Hounsfield Unit distributions for the different OARs in BGPD and HaN-Seg dataset.}
  \label{fig:HU_bgpd}
\end{figure}

\begin{figure}[h!] 
  \centering
 \includegraphics[width=0.95\textwidth]{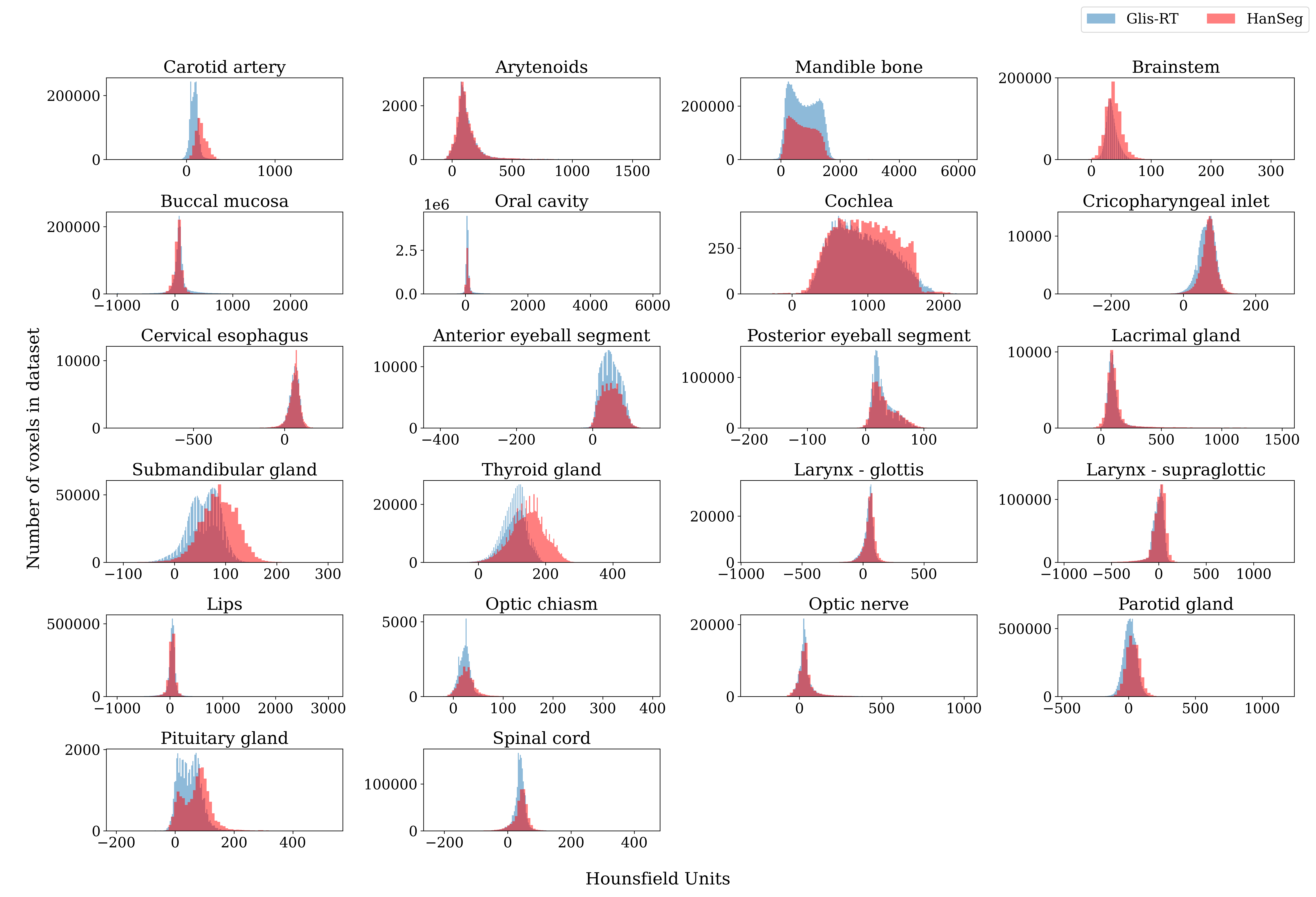}
  \caption{Hounsfield Unit distributions for the different OARs in Glis-RT and HaN-Seg dataset.}
  \label{fig:HU_glisrt}
\end{figure}

Figures~\ref{fig:HU_bgpd} and \ref{fig:HU_glisrt} show the distributions of Hounsfield Units (HU) for the different datasets used during training including the H\&N OARs CT and MR segmentation (HaN-Seg) dataset~\parencite{podobnik_han-seg_2023}  and two datasets from The Cancer Imaging Archive (TCIA) : the Glioma Image Segmentation for Radiotherapy dataset (Glis-RT)~\cite{glisrt_dataset, glisrt_paper} and Burdenko's Glioblastoma Progression Dataset (BGPD)~\cite{burdenko_dataset}. HU were obtained by masking the CT scans with HaN-Seg manual contours and BGPD and Glis-RT pseudo contours used for the final model training.  
For most organs at risk (OARs), the distribution of values shows substantial overlap between the HaN-Seg and TCIA datasets. The largest discrepancies are observed for the carotid arteries, the thyroid and submandibular glands.
\begin{figure}
    \centering
      \subfloat[]{\includegraphics[width=0.49\textwidth]{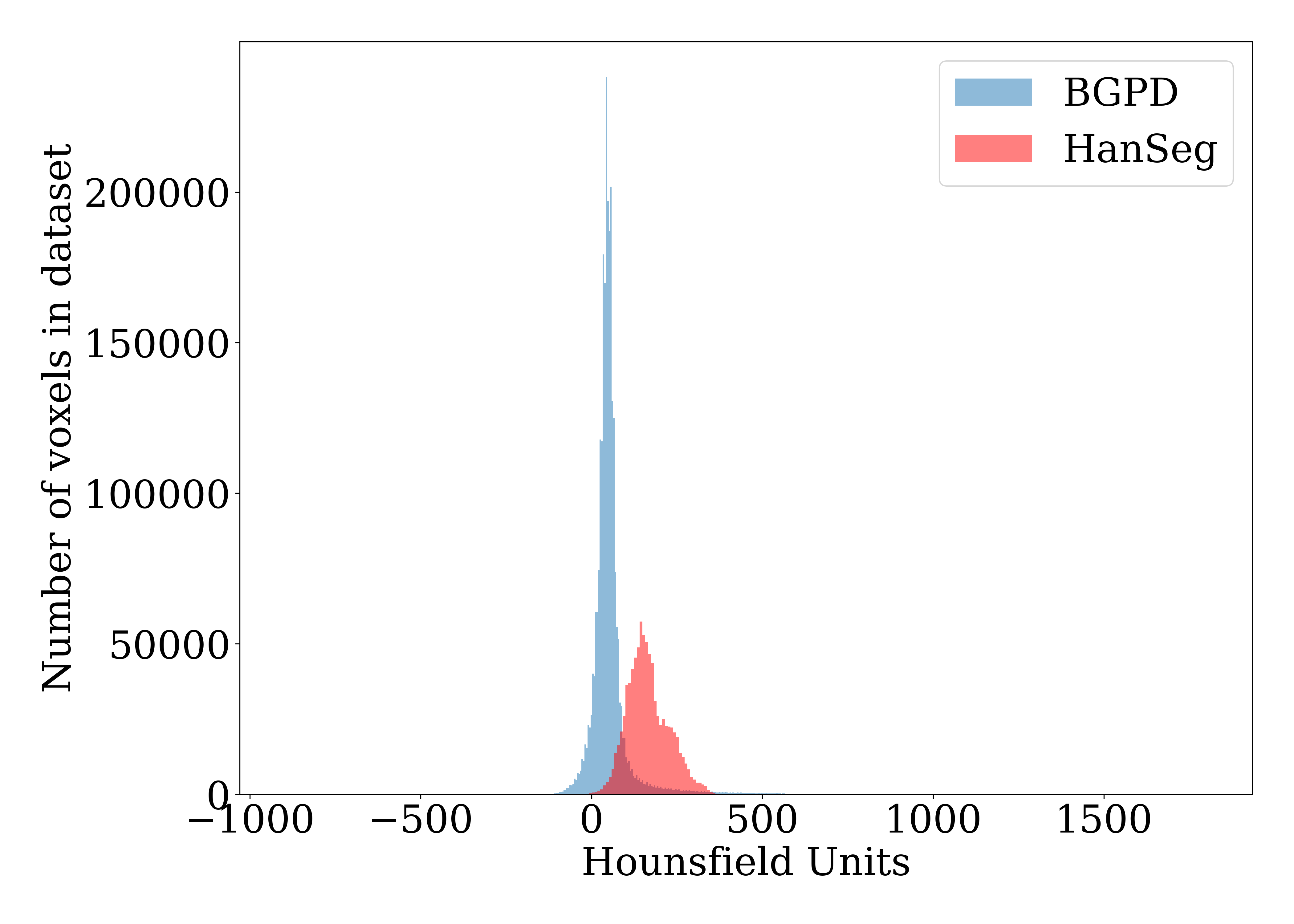}\label{fig:f1}}
   \hfill
   \subfloat[]{\includegraphics[width=0.49\textwidth]{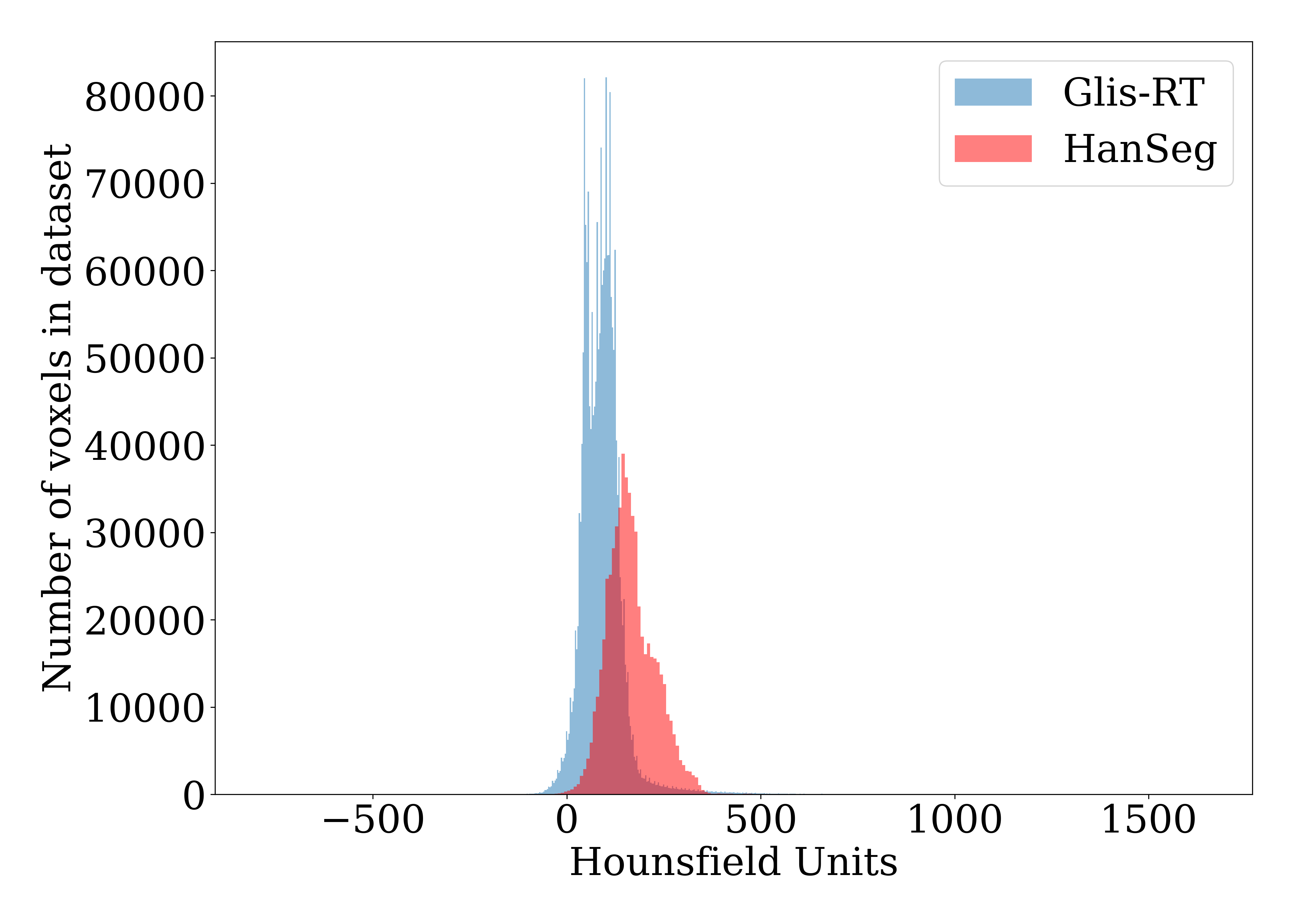}\label{fig:f2}}
   \caption{Hounsfield Unit distribution of the manual carotid contour for the HaN-Seg challenge dataset and of the validated pseudo contours for patients in the BGPD and Glis-RT datasets. It can be observed that the distribution means if the TCIA datasets are different from the HaN-Seg dataset. }  
    \label{Fig:hu_dist_carotid}
\end{figure}
%
%

During the training dataset expansion process, the pseudo-contours generated were accurate for most OARs; however, the model did not generalize well for the carotid arteries when applied to new datasets. This limitation can be attributed to differences in Hounsfield Unit (HU) distributions, as illustrated in Figure~\ref{Fig:hu_dist_carotid}, which may result from variations in the calibrations of CT scanners, particularly for soft tissues. 
Although the HU distributions across datasets appeared visually similar, statistical analysis (Kolmogorov–Smirnov (KS) test, $\alpha$ = 0.5) revealed significant differences for most OARs between the HaN-Seg dataset and the BCGP or Glis-RT datasets. Notably, the carotid arteries exhibited the largest divergence, with the highest KS statistic. These discrepancies hindered the model's ability to accurately segment the carotid arteries when it had not been trained on patients from the same dataset.

The prediction strategy was adjusted for the carotid arteries to address this issue and improve contour quality. Instead of selecting the class with the highest softmax probability as done in the nnU-Net pipeline, a threshold-based approach was used, where a voxel was assigned a class when its corresponding logit exceeded a defined threshold. For the first Glis-RT review session, the threshold was empirically set to 0.001, while for the first BGPD review session, it was set to 0.01. In subsequent review sessions, the conventional softmax-based criterion was reinstated for the carotid arteries, as the model had then been trained on CT scans from these datasets and was thus more robust to Hounsfield Unit variability.

\newpage

\section{Detailed Challenge Submission Results}\label{app:sub}

\begin{table}[h!]
    \scriptsize
    \centering
    
    \begin{tabular}{|P{10em}|P{6.5em}|P{6.5em}||P{10em}|P{6.5em}|P{6.5em}|}
    \hline
    \textbf{Organ at risk}&\textbf{Dice Score} (\%) $\uparrow$& \textbf{Hausdorff Distance} (mm) $\downarrow$ & \textbf{Organ at risk} & \textbf{Dice Score} (\%) $\uparrow$& \textbf{Hausdorff Distance} (mm)$\downarrow$\tabularnewline 
    \hline\hline
    Anterior eyeball segment - Left & $79.96 \pm 7.34$ [61.22 88.69] & $2.26 \pm 0.80$ [1.22 4.00]  & Lips & $74.43 \pm 9.13$ [60.12 91.21] & $6.21 \pm 3.39$ [1.61 12.60] \\ \hline 
Anterior eyeball segment - Right & $80.88 \pm 5.00$ [68.93 88.95] & $1.98 \pm 0.61$ [0.67 3.24]  & Mandible bone & $94.86 \pm 1.84$ [89.91 96.95] & $1.30 \pm 0.89$ [0.51 4.00] \\ \hline 
Arytenoids & $64.94 \pm 13.66$ [32.34 80.38] & $3.38 \pm 2.73$ [1.21 12.25]  & Optic nerve - Left & $71.81 \pm 8.05$ [56.65 85.90] & $2.27 \pm 0.94$ [0.56 3.89] \\ \hline 
Brainstem & $86.64 \pm 2.57$ [82.57 91.57] & $4.72 \pm 1.36$ [2.16 6.07]  & Optic nerve - Right & $74.54 \pm 5.58$ [65.80 83.76] & $2.41 \pm 1.15$ [0.94 4.72] \\ \hline 
Buccal mucosa & $71.91 \pm 7.52$ [55.46 81.48] & $5.47 \pm 2.68$ [2.35 13.06]  & Optic chiasm & $47.05 \pm 11.19$ [26.44 62.34] & $4.20 \pm 1.94$ [1.84 9.91] \\ \hline 
Carotid artery - Left & $83.69 \pm 4.70$ [74.02 90.95] & $3.46 \pm 4.17$ [0.62 14.55]  & Oral cavity & $90.11 \pm 4.20$ [78.31 94.30] & $5.16 \pm 2.25$ [2.09 9.91] \\ \hline 
Carotid artery - Right & $85.80 \pm 2.69$ [80.18 89.73] & $1.88 \pm 1.95$ [0.54 7.60]  & Parotid gland - Left & $87.31 \pm 2.38$ [81.91 91.79] & $4.68 \pm 2.63$ [2.65 12.88] \\ \hline 
Cervical esophagus & $60.98 \pm 12.28$ [35.07 84.83] & $8.32 \pm 4.21$ [2.00 18.16]  & Parotid gland - Right & $85.12 \pm 3.70$ [77.49 89.00] & $5.30 \pm 2.27$ [3.23 11.62] \\ \hline 
Cochlea - Left & $74.40 \pm 9.58$ [58.91 92.37] & $1.69 \pm 0.80$ [0.60 3.00]  & Pituitary gland & $71.64 \pm 11.95$ [45.23 92.81] & $2.13 \pm 0.92$ [0.68 4.00] \\ \hline 
Cochlea - Right & $72.12 \pm 9.10$ [53.70 84.05] & $2.07 \pm 0.65$ [1.20 3.00]  & Posterior eyeball segment - Left & $93.04 \pm 1.64$ [89.39 95.74] & $1.64 \pm 0.38$ [0.87 2.00] \\ \hline 
Cricopharyngeal inlet & $65.87 \pm 8.45$ [46.98 76.49] & $5.94 \pm 2.60$ [2.65 12.02]  & Posterior eyeball segment - Right & $92.26 \pm 1.55$ [88.68 94.60] & $1.78 \pm 0.59$ [0.87 3.00] \\ \hline 
Lacrimal gland - Left & $65.59 \pm 11.43$ [40.36 80.84] & $3.05 \pm 1.21$ [1.52 6.02]  & Spinal cord & $83.08 \pm 2.65$ [78.50 86.34] & $2.17 \pm 1.25$ [1.27 6.00] \\ \hline 
Lacrimal gland - Right & $64.98 \pm 13.47$ [34.92 80.57] & $3.15 \pm 1.63$ [1.34 6.18]  & Submandibular gland - Left & $85.08 \pm 8.97$ [55.44 92.44] & $3.65 \pm 2.23$ [1.52 10.88] \\ \hline 
Larynx - glottis & $76.54 \pm 7.44$ [60.56 85.33] & $3.08 \pm 1.40$ [1.52 6.00]  & Submandibular gland - Right & $85.09 \pm 7.22$ [67.77 91.97] & $3.89 \pm 2.87$ [1.27 12.04] \\ \hline 
Larynx - supraglottic & $81.47 \pm 6.39$ [69.07 89.01] & $3.42 \pm 1.55$ [1.98 7.26]  & Thyroid gland & $89.74 \pm 3.62$ [81.12 94.45] & $2.41 \pm 1.71$ [0.51 6.17] \\ \hline

    \end{tabular}
    \caption{Mean Dice scores obtained on the test patients of the \href{https://han-seg2023.grand-challenge.org/}{Head and Neck Organ-At-Risk CT \& MR Segmentation Challenge}. These scores were obtained by submitting to the challenge platform and are available on \href{https://han-seg2023.grand-challenge.org/evaluation/6845bb71-cf52-4d8b-a87e-4aa754fff658/}{the challenge website}. The reported measures are in the form of mean $\pm$ standard deviation [minimum maximum]. $\uparrow$ indicates that higher is best and $\downarrow$ indicates smaller is best. }
    \label{table:dice_han_challenge}
\end{table}
\begin{figure}[h!]
    \centering
    \includegraphics[width=0.7\textwidth]{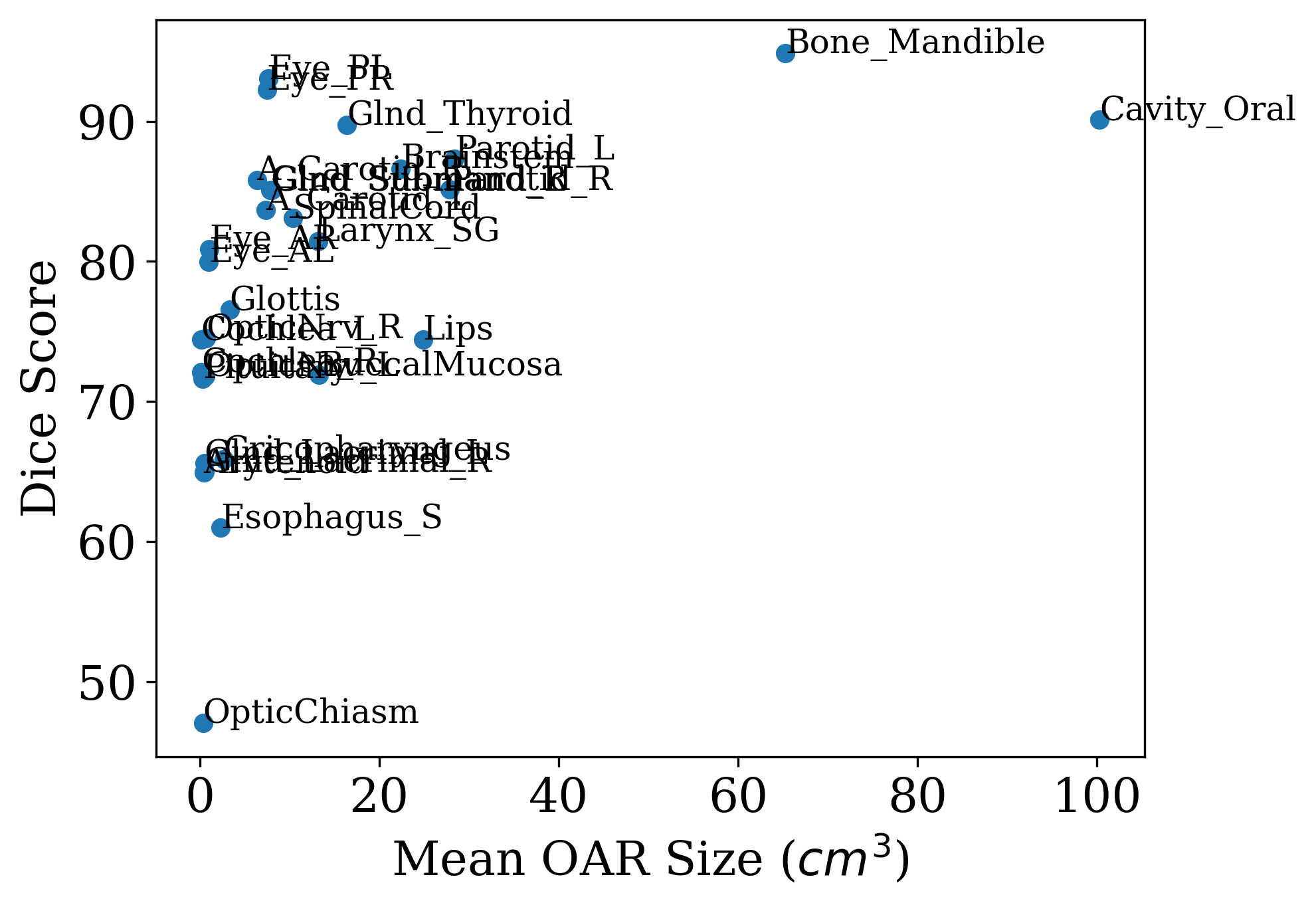}
    \caption{Mean Dice scores obtained on the test patients of the \href{https://han-seg2023.grand-challenge.org/}{Head and Neck Organ-At-Risk CT \& MR Segmentation Challenge} against the average size of the organs in the HaN-Seg training dataset. These scores were obtained by submitting to the challenge platform and are available on \href{https://han-seg2023.grand-challenge.org/evaluation/6845bb71-cf52-4d8b-a87e-4aa754fff658/}{the challenge website}.}
    \label{fig:dice_size}
\end{figure}
Table~\ref{table:dice_han_challenge} and Figure~\ref{fig:dice_size} show the different performances of our pipeline across OARs on the held-out internal testing set of the Head and Neck Organ-At-Risk CT \& MR Segmentation Challenge. It can be observed that the bigger is the OAR, the better is the DS. The optic chiasm exhibited the lowest DS, with an average value of 47.05\%. This low score can be attributed to the thinness of the optic chiasm, which can result in its omission between slices when the interslice distance is substantial. Additionally, the optic chiasm displays better contrast on MRI scans, making the ground truth in the dataset challenge highly dependent on quality of the original registration method used to contour this organ-at-risk (OAR).

\newpage

\section{Custom Standardization for the challenge}

\begin{figure}[h!] 
  \centering
 \includegraphics[width=0.80\textwidth]{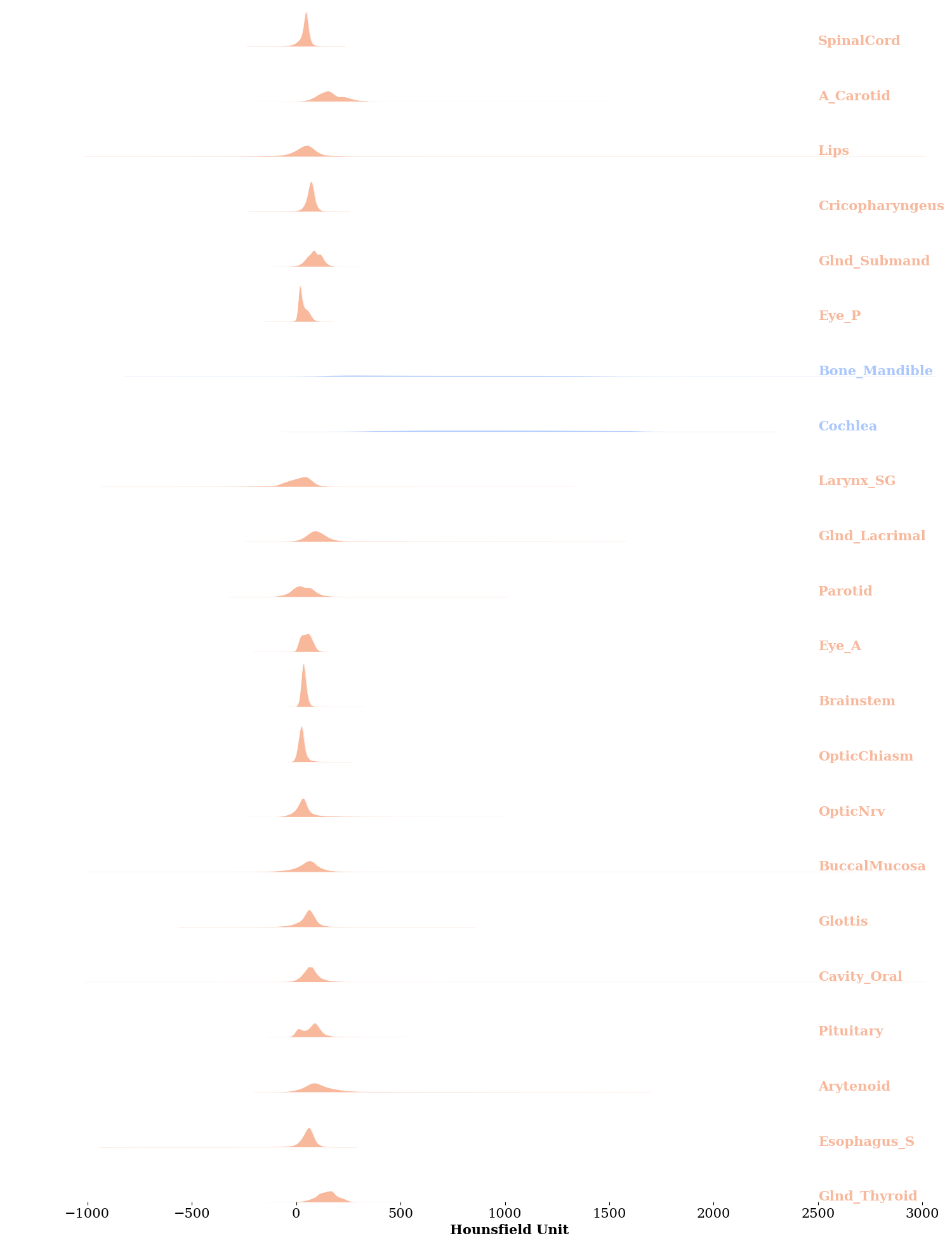}
  \caption{The distribution of Hounsfield Units for all CT scans provided in the Head and Neck Organ-At-Risk CT \& MR Segmentation Challenge training dataset is presented in the form of kernel density estimates. Organs highlighted in blue represent those emphasized by our bone tissue standardization, while those highlighted in orange are emphasized by our soft tissue standardization.}
  \label{fig:HU_per_oar}
\end{figure}

A custom preprocessing was developed and used for submission to the HaN-Seg challenge. Two separate volumes were produced from the CT scan data: one focusing on soft tissue and the other on bone tissue. The soft tissue volume was standardized to the Hounsfield Unit (HU) range of -500 to 500, using the mean and standard deviation of the voxel values within this range in the HaN-Seg dataset. Likewise, the bony tissue volume was standardized using a similar process with the HU range of 0 to 3000. Figure~\ref{fig:HU_per_oar} illustrates the context for this particular standardization scheme, where most of the OAR HU distribution fall in the range [-500, 500]. This method differs from the nnU-Net CT standardization, which calculates the mean and standard deviation across all foreground CT voxels, corresponding to voxels contained in any contour, without differentiating between soft tissue and bone tissue. When data was preprocessed this way, if Modality Dropout was used during the model training, this augmentation would randomly either zero out both CT channels, zero out the MRI channel or keep both modalities.  

The MRI volume was also standardized in a custom way, where both the mean and standard deviation were calculated using only the voxels in the MRI field of view registered to CT voxels within the aforementioned range for soft tissue. The nnU-Net MRI standardization would consider all voxels from a patient MRI scan to compute the mean and standard deviation, even voxels in the CT field of view but not in the MRI field of view, representing background air instead of body tissues.

These mean and standard deviation values were only computed on the HaN-Seg challenge dataset and not on the 296 patients dataset. 


To assess the generalizability of the preprocessing method, a model was trained using this approach on the expanded dataset (n = 296) and evaluated on the internal internal testing set of fold 0. Table~\ref{table:std_train_infer} shows the average performance of models trained and inferred on different inputs. According to the corrected Wilcoxon signed-rank test, no significant differences can be observed at a 95\% confidence level for any OAR contours between a model trained with CT only with nnU-Net standardization or ours, as well as for a model trained with MRI only. Table~\ref{table:sig_ctmri_nnunsetstd_ourstd} shows performance of a model trained with CT and MRI scans, for OAR for which differences are significant. No significant differences are observed for HD in this case. On average a model trained with CT and MRI using our custom standardization scheme is slightly better for DS with 90.20 against 90.16 for nnU-Net standardization, but performs worst considering HD with 1.78 mm against  1.55 mm.
\begin{table}[h!]
\scriptsize
\centering
\begin{tabular}{|P{8em}|P{6em}|P{6em}|P{7em}|P{6em}|P{6em}|P{7em}|}
\cline{1-7}
\multirow{3}{*}{\textbf{Metric}}& \multicolumn{6}{c|}{\textbf{Inputs for training and inference}}  \\
\cline{2-7} & \multicolumn{3}{c|}{\textbf{nnU-Net standardization}}  &\multicolumn{3}{c|}{\textbf{Our standardization}}  \\
\cline{2-7} & \textbf{CT only} & \textbf{MRI only (MRI FOV)} & \textbf{CT and MRI (with MD)} & \textbf{CT only} & \textbf{MRI only (MRI FOV)} & \textbf{CT and MRI (with MD)}\\ \hline

\textbf{Dice Score (\%) $\uparrow$} & 90.06 $\pm$ 5.35 [73.78 97.75]& 69.16 $\pm$ 16.3 [26.56 93.88]& 90.16 $\pm$ 5.41 [73.19 97.59] & 90.01 $\pm$ 5.26 [74.14 97.71]& 68.81 $\pm$ 16.66 [24.82 93.93]& 90.20 $\pm$ 5.37 [73.72 97.66]\\ \hline
\textbf{Hausdorff Distance (mm) $\downarrow$}  & 1.59 $\pm$ 0.71 [1.02 3.81]& 4.41 $\pm$ 3.21 [1.45 16.15]&1.55 $\pm$ 0.64 [1.03 3.89]  &1.93 $\pm$ 1.01 [1.02 4.82] & 4.62 $\pm$ 3.57 [1.46 18.36]& 1.78 $\pm$ 0.91 [1.02 3.93] \\ \hline

\end{tabular}
\caption{Segmentation metrics obtained by comparing predictions of our pipeline trained with different inputs standardized with nnU-Net strategy or ours. Results were first averaged for each OAR over all cases and finally over all organs. Results are presented as: mean across patients and OARs $\pm$ standard deviation among the mean metrics of each OAR [minimum OAR metric averaged over cases    and maximum]. $\uparrow$ indicates that higher is best and $\downarrow$ indicates smaller is best. }
    \label{table:std_train_infer}
\end{table}
\begin{table}[h!]
    \scriptsize
    \centering
    \begin{tabular}{|P{6em}|P{11em}|P{8em}|P{8em}|P{7em}|P{6em}|}
    \hline
    \multirow{3}{*}{\textbf{Metric}} & \multirow{3}{*}{\textbf{OAR}} & \multicolumn{2}{c|}{\textbf{Model performance}} & \multirow{3}{*}{\begin{tabular}[c]{@{}c@{}}\textbf{Paired}\\ \textbf{differences}\end{tabular}} & \multirow{3}{*}{\begin{tabular}[c]{@{}c@{}}\textbf{Corrected}\\ \textbf{p-value}\end{tabular}} \tabularnewline 
    \cline{3-4}
    && \textbf{nnU-Net standardization} & \textbf{Custom standardization}  && \tabularnewline 
    \hline\hline
\multirow{25}{*}{\makecell[c]{Dice Score\\ (\%) $\uparrow$}} & Carotid artery - Left (n=60) &  $91.27 \pm 4.49$ [77.82 96.77] & $91.82 \pm 3.78$ [80.47 97.29] & $-0.55 \pm 1.58$ [-6.42 3.65] & $4.27E-02$ \\ \cline{2-6} 
 & Carotid artery - Right (n=60) &  $91.79 \pm 3.82$ [77.40 96.87] & $92.30 \pm 3.41$ [82.78 97.28] & $-0.51 \pm 1.45$ [-8.68 1.76] & $2.02E-02$ \\ \cline{2-6} 
 & Mandible bone (n=60) &  $97.59 \pm 1.53$ [90.87 99.09] & $97.66 \pm 1.52$ [91.82 99.20] & $-0.07 \pm 0.27$ [-0.95 1.01] & $2.99E-02$ \\ \cline{2-6} 
 & Brainstem (n=60) &  $95.70 \pm 2.63$ [87.94 98.61] & $96.13 \pm 2.86$ [86.51 98.79] & $-0.43 \pm 0.88$ [-3.48 2.30] & $1.41E-03$ \\ \cline{2-6} 
 & Oral cavity (n=60) &  $96.78 \pm 2.78$ [86.02 98.82] & $96.75 \pm 3.53$ [76.69 99.04] & $0.03 \pm 1.26$ [-1.27 9.33] & $1.10E-02$ \\ \cline{2-6} 
 & Cochlea - Right (n=60) &  $86.82 \pm 15.77$ [0.00 100.00] & $87.80 \pm 15.87$ [0.00 98.18] & $-0.99 \pm 2.65$ [-6.97 4.42] & $4.54E-02$ \\ \cline{2-6} 
 & Posterior eyeball segment - Right (n=60) &  $96.51 \pm 1.71$ [90.15 98.23] & $96.63 \pm 1.70$ [90.04 98.45] & $-0.12 \pm 0.32$ [-1.05 0.57] & $4.04E-02$ \\ \cline{2-6} 
 & Submandibular gland - Left (n=59) &  $94.63 \pm 4.34$ [73.42 97.96] & $94.89 \pm 4.39$ [73.63 98.32] & $-0.26 \pm 0.66$ [-2.59 1.41] & $1.10E-02$ \\ \cline{2-6} 
 & Submandibular gland - Right (n=60) &  $94.50 \pm 5.39$ [62.30 97.91] & $93.43 \pm 13.15$ [0.00 98.14] & $1.07 \pm 10.83$ [-6.76 83.87] & $2.10E-02$ \\ \cline{2-6} 
 & Larynx - supraglottic (n=59) &  $93.12 \pm 4.39$ [77.49 98.04] & $93.63 \pm 4.43$ [77.66 97.85] & $-0.51 \pm 0.97$ [-5.10 1.15] & $1.41E-03$ \\ \cline{2-6} 
 & Optic chiasm (n=59) &  $80.20 \pm 8.80$ [46.68 92.79] & $81.81 \pm 9.24$ [54.57 94.55] & $-1.61 \pm 3.81$ [-12.74 7.16] & $1.10E-02$ \\ \cline{2-6} 
 & Parotid gland - Left (n=60) &  $94.97 \pm 5.33$ [64.22 98.51] & $95.28 \pm 5.31$ [64.64 98.77] & $-0.31 \pm 0.66$ [-3.14 1.20] & $1.41E-03$ \\ \cline{2-6} 
 & Parotid gland - Right (n=60) &  $95.14 \pm 4.46$ [74.85 98.52] & $95.44 \pm 4.22$ [78.58 98.68] & $-0.30 \pm 0.83$ [-3.94 1.68] & $1.49E-02$ \\
 \hline
    \end{tabular}
    \caption{Average performance for significantly different OAR performances at 95\% confidence level according to the corrected Wilcoxon signed-rank test between a model traine with nnU-Net standardization and a model trained with our custom standardization on CT and MRI scans. Results are presented as mean $\pm$ standard deviation [minimum maximum].}
    \label{table:sig_ctmri_nnunsetstd_ourstd}
\end{table}
%
%

To assess the impact of the field of view (FOV) on the performance metric, we trained different models on inputs data cropped to different FOVs. Table~\ref{table:std_inference} shows the performance of models trained using CT and MRI scans and inferred on specific inputs. According to the corrected Wilcoxon signed-rank test, no significant differences can be observed at a 95\% confidence level for any OAR contours when considering a model trained on MRI FOV and inferred using MRI only between nnU-Net and the proposed standardization. 

For models trained on the CT FOV and inferred with CT only, the only significant difference is observed for the DS of the larynx supraglottic which obtained an average DS of $93.07 \pm 4.51$ [75.89 98.07] with nnU-Net standardization against $93.61 \pm 4.47$ [77.11 97.83] for ours. For the MRI only prediction of a model trained on the CT FOV, the only significant difference is observed for the mandible bone in both DS and HD. Our custom standardization obtained an average DS of $64.66 \pm 20.90$ [28.24 90.27] against $63.58 \pm 21.37$ [26.65 90.27] for nnU-Net standaradization, and an average HD of $23.80 \pm 30.33$ [1.00 100.81] against $27.84 \pm 33.37$ [1.41 103.98].
\begin{table}[h!]
\tiny
\centering
\begin{tabular}{|P{14ex}|P{16ex}|P{16ex}|P{16ex}|P{17ex}|P{16ex}|P{16ex}|P{16ex}|P{17ex}|}
\cline{1-9}
\multirow{4}{*}{\textbf{Metric}}& \multicolumn{8}{c|}{\textbf{Inputs for inference of a model trained with CT and MRI of different FOV using MD }}  \\

\cline{2-9} & \multicolumn{4}{c|}{\textbf{nnU-Net standardization}}  &\multicolumn{4}{c|}{\textbf{Our standardization}}  \\ 

\cline{2-9} &\multicolumn{3}{c|}{\textbf{CT FOV}} & \textbf{MRI FOV}&\multicolumn{3}{c|}{\textbf{CT FOV}} & \textbf{MRI FOV} \\

\cline{2-9} & \textbf{CT only} & \textbf{MRI only} & \textbf{CT and MRI} & \textbf{MRI only} &\textbf{CT only} & \textbf{MRI only}  & \textbf{CT and MRI}&\textbf{MRI only}\\ \hline

\textbf{Dice Score (\%) $\uparrow$} &90.06 $\pm$ 5.41 [73.22 97.61] & 57.0 $\pm$ 26.49 [5.84 92.22]& 90.16 $\pm$ 5.41 [73.19 97.59]& 69.39 $\pm$ 16.03 [23.52 93.69]& 90.09 $\pm$ 5.34 [73.63 97.67]& 57.21 $\pm$ 26.53 [6.08 92.16]&90.2 $\pm$ 5.37 [73.72 97.66]& 69.71 $\pm$ 15.58 [25.38 93.9]\\ \hline
\textbf{Hausdorff Distance (mm) $\downarrow$}  & 1.58 $\pm$ 0.66 [1.04 3.99]& 9.09 $\pm$ 9.51 [1.47 36.11]& 1.55 $\pm$ 0.64 [1.03 3.89] & 4.64 $\pm$ 3.85 [1.44 21.35]&1.75 $\pm$ 0.92 [1.02 4.07]& 8.54 $\pm$ 8.85 [1.4 33.1]& 1.78 $\pm$ 0.91 [1.02 3.93] & 4.74 $\pm$ 3.55 [1.45 18.1]\\ \hline

\end{tabular}
\caption{Segmentation metrics obtained by comparing predictions of our pipeline trained with CT and MRI inputs standardized with nnU-Net strategy or ours. Results were first averaged for each OAR over all cases and finally over all organs. Results are presented as: mean across patients and OARs $\pm$ standard deviation among the mean metrics of each OAR [minimum OAR metric averaged over cases    and maximum]. $\uparrow$ indicates that higher is best and $\downarrow$ indicates smaller is best. }
    \label{table:std_inference}
\end{table}

Although the results above do not reveal a substantial difference between the two standardization methods, the custom standardization yielded our best performance in the HaN-Seg challenge. We hypothesize that this improved performance is attributable to the fact that all MRI scans from the HaN-Seg were acquired using the same scanner, making a dataset-specific standardization more effective than a patient-specific approach. In contrast, the extended dataset of 296 patients used for our ablation study includes MRI scans from multiple scanners, thereby diminishing the influence of the HaN-Seg-specific standardization on validation performance across this larger, heterogeneous dataset.

\newpage

\section{Pipeline performance}
\subsection{TCIA datasets}

\begin{figure}[h!] 
  \centering
 \includegraphics[width=0.95\textwidth]{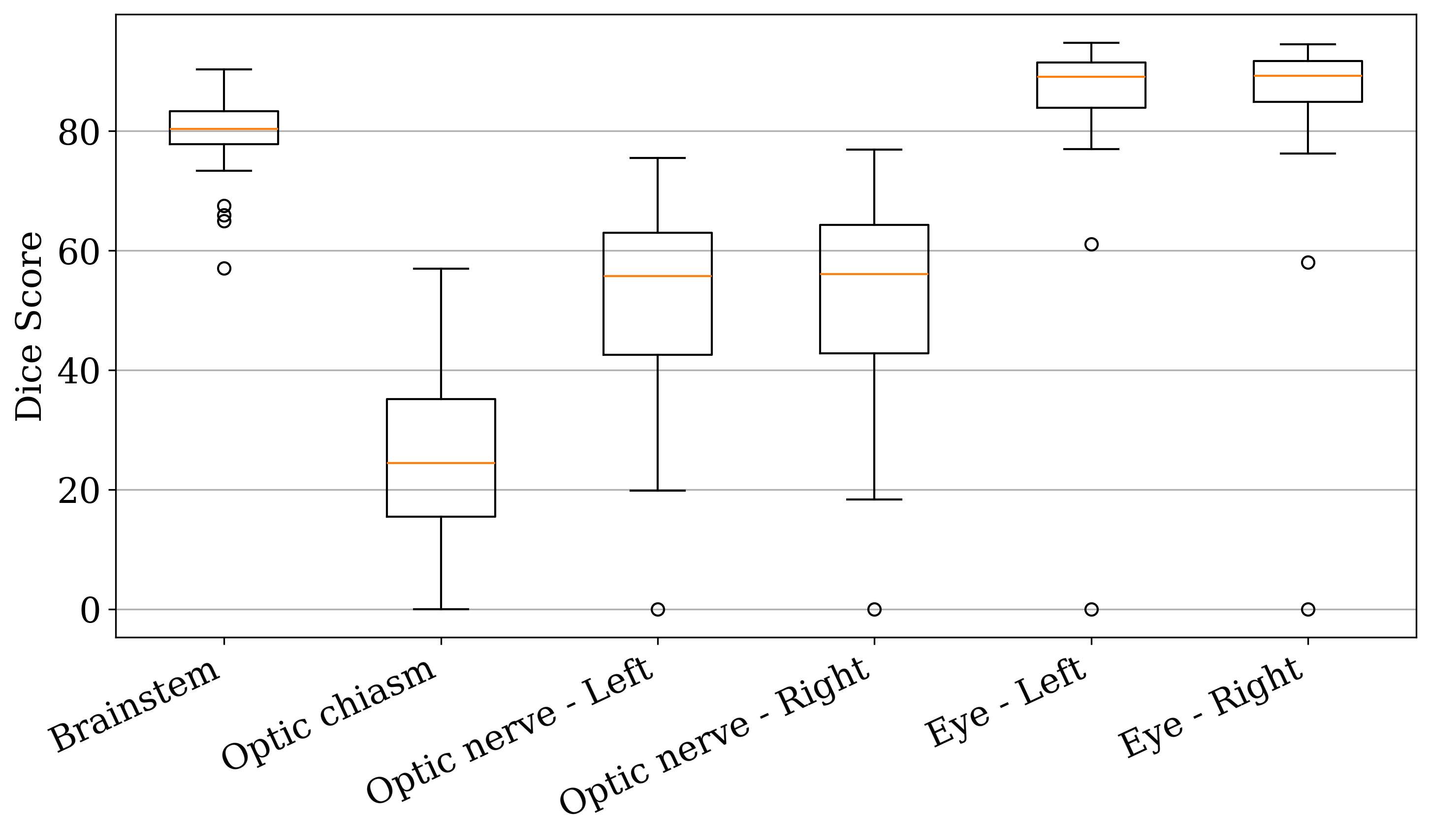}
  \caption{Agreement between our pipeline predictions and manual contours available in BGPD dataset.}
  \label{fig:bgpd_perf}
\end{figure}

After training, our pipeline predictions were compared to manual contours provided in different TCIA datasets: BGPD, GLis-RT and HNSCC. Contours were inferred for all scans which contained manual contours. Table~\ref{table:bgpd_perf} and Figure~\ref{fig:bgpd_perf} present the agreement between our predicted contours and BGPD manual contours.
\begin{table}[h!]
    \scriptsize
    \centering
    \begin{tabular}{|c|c|c|}
    \hline
    \textbf{OAR} (n=176) & \textbf{Mean Dice Score} $\uparrow$& \textbf{Mean Hausdorff Distance} $\downarrow$ \tabularnewline 
    \hline\hline
Brainstem & $80.21 \pm 4.38$ [57.06 90.30] & $7.28 \pm 2.99$ [3.00 25.81] \\ \hline 
Optic chiasm & $25.81 \pm 12.70$ [0.00 56.93] & $17.00 \pm 5.82$ [4.00 35.35] \\ \hline 
Optic nerve - Left & $52.76 \pm 13.46$ [0.00 75.49] & $7.22 \pm 18.49$ [2.24 235.04] \\ \hline 
Optic nerve - Right & $53.21 \pm 13.49$ [0.00 76.87] & $5.63 \pm 6.66$ [2.00 86.07] \\ \hline 
Eye - Left & $87.35 \pm 8.11$ [0.00 94.75] & $3.16 \pm 8.02$ [1.00 108.08] \\ \hline 
Eye - Right & $87.64 \pm 8.11$ [0.00 94.48] & $3.15 \pm 8.10$ [1.00 109.08] \\ \hline 
    \end{tabular}
    \caption{Agreement between our pipeline predictions and manual contours available in BGPD dataset. Results are presented as: mean across patients $\pm$ standard deviation [minimum maximum]. $\uparrow$ indicates that higher is best and $\downarrow$ indicates smaller is best.}
    \label{table:bgpd_perf}
\end{table}
%
%
It can be seen that the smallest agreement is found for the optic structures. Notably, in one patient case, a manual contouring error resulted in the left and right optic nerves and eye structures being swapped, leading to a DS of 0.00\% for these structures.
%
%
%
%

\begin{figure}[h!] 
  \centering
 \includegraphics[width=0.95\textwidth]{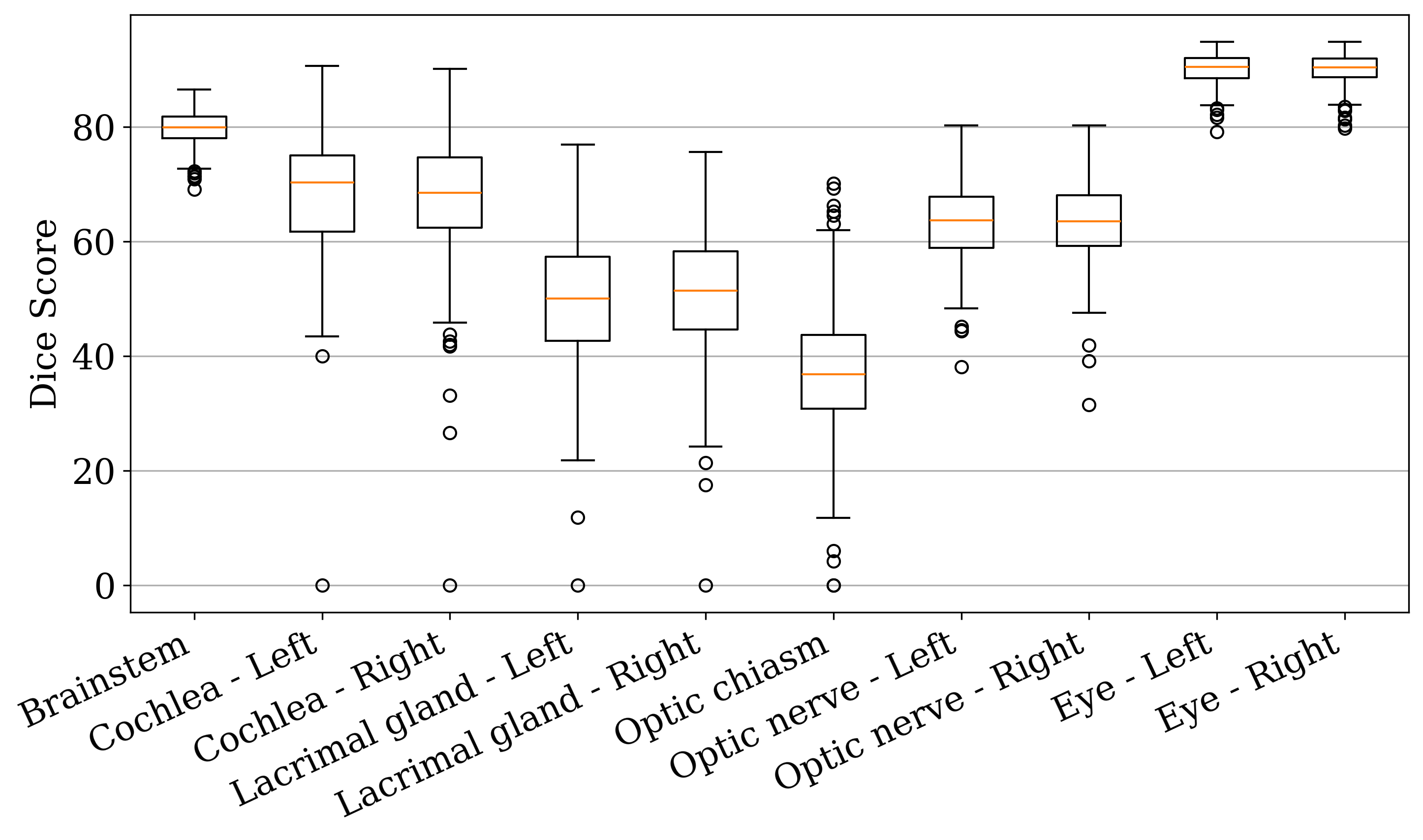}
  \caption{Agreement between our pipeline predictions and manual contours available in Glis-RT dataset. }
  \label{fig:glisrt_perf}
\end{figure}

Table~\ref{table:glisrt_perf} and Figure~\ref{fig:glisrt_perf} summarize the agreement of our pipeline predictions with Glis-RT manual contours. The same disagreement in optic structures can be observed, with lowest agreement being for the optic chiasm, the lacrimal glands and optic nerves. 
\begin{table}[h!]
    \scriptsize
    \centering
    \begin{tabular}{|c|c|c|}
    \hline
    \textbf{OAR} (n=230) & \textbf{Mean Dice Score} $\uparrow$& \textbf{Mean Hausdorff Distance} $\downarrow$ \tabularnewline 
    \hline\hline
Brainstem & $79.72 \pm 3.01$ [69.06 86.56] & $5.49 \pm 1.84$ [2.83 14.18] \\ \hline 
Cochlea - Left & $68.46 \pm 10.55$ [0.00 90.67] & $1.86 \pm 0.79$ [1.00 7.33] \\ \hline 
Cochlea - Right & $67.65 \pm 10.93$ [0.00 90.18] & $1.97 \pm 1.06$ [1.00 9.35] \\ \hline 
Lacrimal gland - Left & $49.75 \pm 10.90$ [0.00 76.87] & $6.33 \pm 6.35$ [1.41 94.05] \\ \hline 
Lacrimal gland - Right & $50.72 \pm 11.16$ [0.00 75.62] & $6.51 \pm 6.23$ [2.00 92.16] \\ \hline 
Optic chiasm & $37.12 \pm 11.13$ [0.00 70.09] & $9.70 \pm 4.21$ [2.24 27.09] \\ \hline 
Optic nerve - Left & $63.03 \pm 6.83$ [38.14 80.28] & $3.14 \pm 1.37$ [1.41 13.01] \\ \hline 
Optic nerve - Right & $63.33 \pm 6.79$ [31.46 80.27] & $3.04 \pm 1.79$ [1.41 22.85] \\ \hline 
Eye - Left & $89.95 \pm 2.59$ [79.15 94.85] & $1.72 \pm 0.54$ [1.00 3.61] \\ \hline 
Eye - Right & $89.93 \pm 2.66$ [79.77 94.87] & $1.68 \pm 0.50$ [1.00 3.00] \\ \hline
    \end{tabular}
    \caption{Agreement between our pipeline predictions and manual contours available in Glis-RT dataset. Results are presented as: mean across patients $\pm$ standard deviation [minimum maximum]. $\uparrow$ indicates that higher is best and $\downarrow$ indicates smaller is best.}
    \label{table:glisrt_perf}
\end{table}
%
%

\begin{figure}[h!] 
  \centering
 \includegraphics[width=0.95\textwidth]{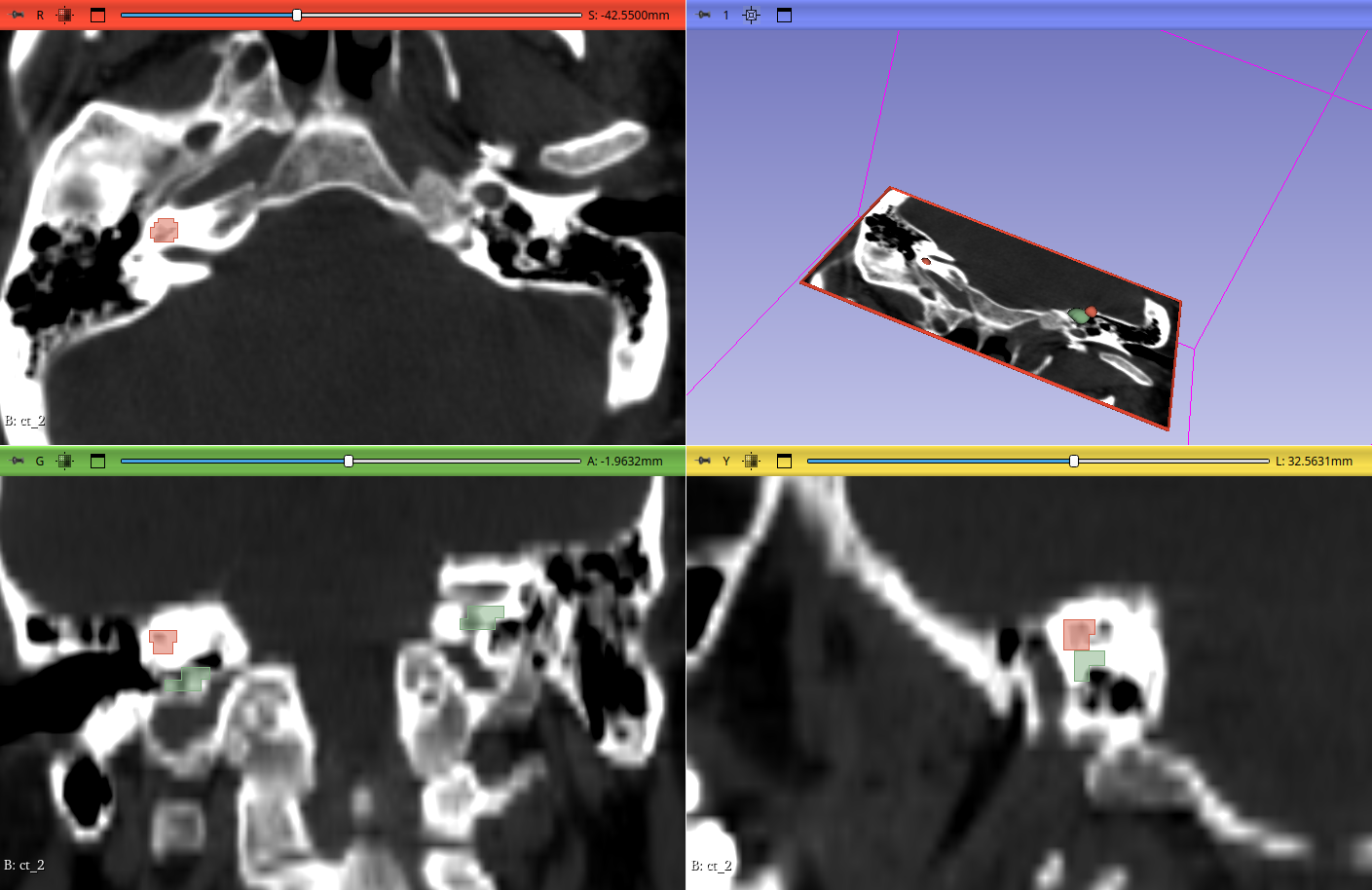}
  \caption{CT scan of a patient in Glis-RT dataset with manual contours of the cochleae shown in green along with our automated predictions shown in red.}
  \label{fig:cochlea_glisrt}
\end{figure}

\begin{figure}[h!] 
  \centering
  \subfloat[CT scan overlayed with contours.]{\includegraphics[width=0.49\textwidth]{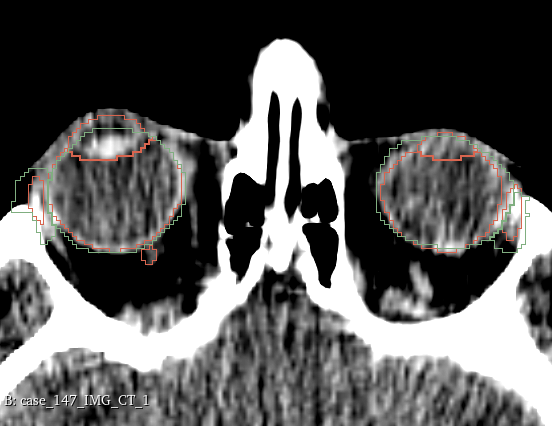}\label{fig:f1}}
  \hfill
  \subfloat[3D contour view.]{\includegraphics[width=0.49\textwidth]{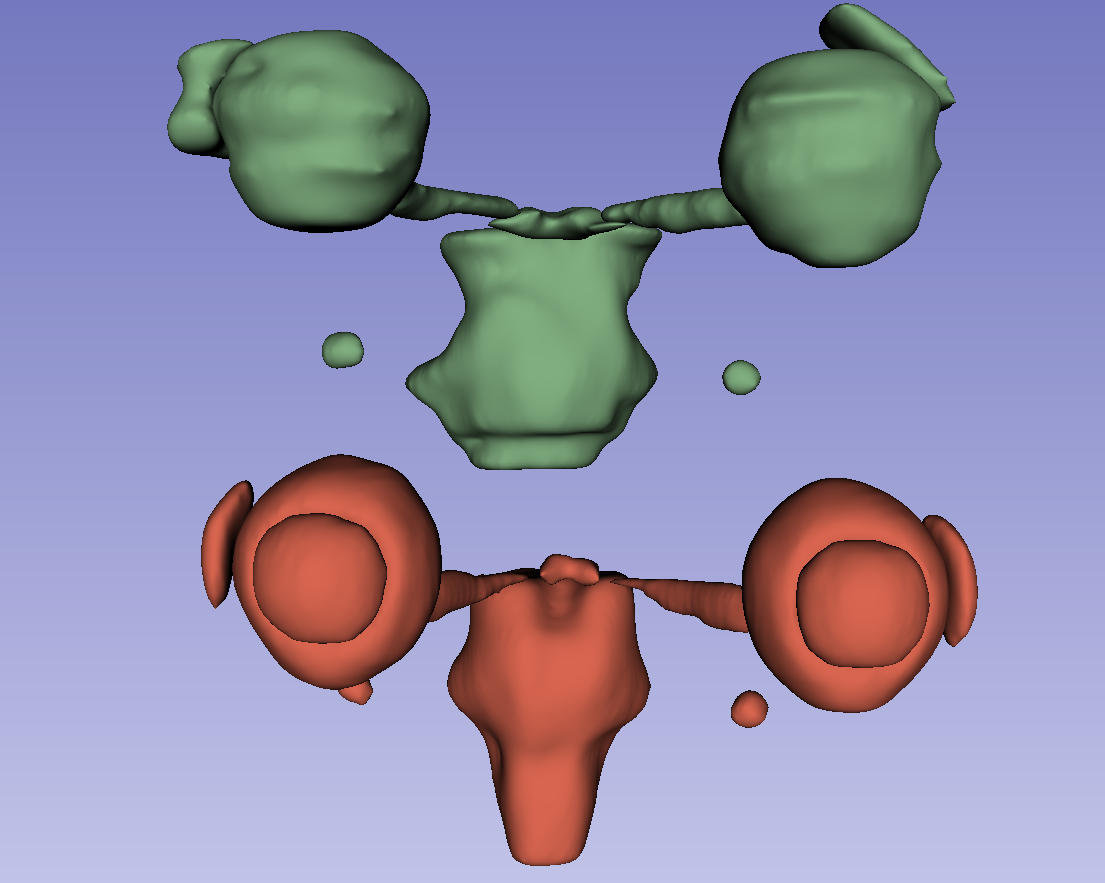}\label{fig:f2}}
  \caption{CT scan of a patient in Glis-RT dataset with manual contours of the optic structures shown in green along with our automated predictions shown in red. A vertical shift was applied to the automated contours in the 3D view for visual purposes.}
  \label{fig:lac_gland_glisrt}
\end{figure}

Similar to BGPD dataset, in one patient case, a manual contouring error resulted in swapped left and right, leading to a DS of 0.00\% for lacrimal glands. For another patient, a DS of 0.00\% for the cochleae was explained by the manual ground truth contour being misaligned with the structure.
Figure~\ref{fig:cochlea_glisrt} shows our pipeline prediction of the cochleae along with manual contours for this specific patient.
Figure~\ref{fig:lac_gland_glisrt} illustrates a case for which the DS between our pipeline prediction and the manual contour was 11.80\% and 54.33\% for the left and right lacrimal glands, respectively. It can be seen that our prediction is more faithful to the glands seen on the CT scan.

%
%
\begin{table}[h!]
    \scriptsize
    \centering
    \begin{tabular}{|c|c|c|}
    \hline
    \textbf{OAR} & \textbf{Mean Dice Score} $\uparrow$& \textbf{Mean Hausdorff Distance} $\downarrow$ \tabularnewline 
    \hline\hline
Mandible bone  (n=105) & $81.12 \pm 11.72$ [0.00 92.38] & $8.19 \pm 19.20$ [1.00 195.07] \\ \hline 
Brainstem  (n=179) & $66.39 \pm 18.09$ [0.00 85.66] & $5.73 \pm 3.54$ [2.24 34.06] \\ \hline 
Parotid gland - Left  (n=175) & $78.42 \pm 9.45$ [17.71 90.10] & $5.52 \pm 2.86$ [2.24 19.10] \\ \hline 
Parotid gland - Right  (n=180) & $78.69 \pm 10.70$ [0.00 88.91] & $6.93 \pm 13.43$ [2.00 151.04] \\ \hline 
Spinal cord  (n=187) & $59.24 \pm 9.55$ [14.20 84.48] & $27.37 \pm 13.30$ [2.83 59.91] \\ \hline 
Oral cavity  (n=51) & $37.55 \pm 19.74$ [0.00 69.98] & $24.92 \pm 10.70$ [9.11 56.24] \\ \hline 
Cochlea - Left  (n=58) & $25.13 \pm 18.24$ [0.00 86.96] & $7.28 \pm 3.87$ [1.00 23.11] \\ \hline 
Cochlea - Right  (n=59) & $24.81 \pm 18.00$ [0.00 77.08] & $7.07 \pm 3.65$ [1.00 22.37] \\ \hline 
Optic chiasm  (n=15) & $20.02 \pm 7.69$ [4.99 36.41] & $9.13 \pm 2.85$ [4.58 14.87] \\ \hline 
Optic nerve - Left  (n=13) & $52.22 \pm 21.37$ [0.00 70.34] & $8.37 \pm 12.21$ [2.00 47.71] \\ \hline 
Optic nerve - Right  (n=13) & $50.51 \pm 20.06$ [0.00 68.24] & $8.10 \pm 12.44$ [1.96 49.28] \\ \hline 
Eye - Left  (n=10) & $87.24 \pm 3.09$ [79.65 91.95] & $2.07 \pm 0.57$ [1.41 3.46] \\ \hline 
Eye - Right  (n=10) & $87.00 \pm 3.65$ [78.16 90.81] & $1.99 \pm 0.73$ [1.41 3.61] \\ \hline 
Submandibular gland - Right  (n=11) & $67.12 \pm 23.33$ [0.00 88.29] & $6.15 \pm 6.01$ [1.41 24.54] \\ \hline 
Submandibular gland - Left  (n=11) & $74.31 \pm 11.02$ [52.40 86.59] & $4.04 \pm 1.82$ [2.24 8.00] \\ \hline 
    \end{tabular}
    \caption{Agreement between our pipeline predictions and manual contours available in HNSCC dataset. Results are presented as: mean across patients $\pm$ standard deviation [minimum maximum]. $\uparrow$ indicates that higher is best and $\downarrow$ indicates smaller is best.}
    \label{table:hnscc_perf}
\end{table}

\begin{figure}[h!] 
  \centering
 \includegraphics[width=0.95\textwidth]{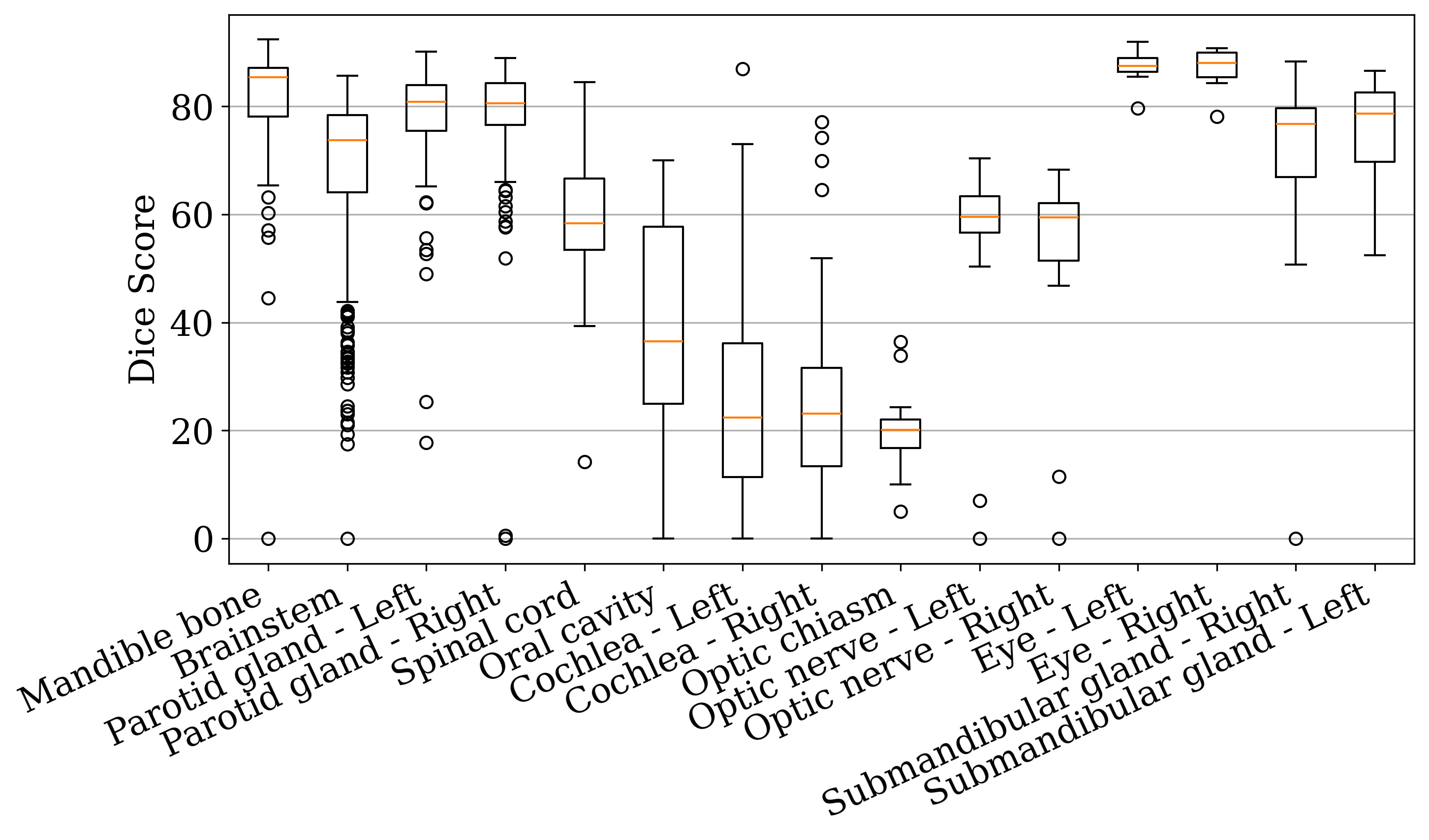}
  \caption{Agreement between our pipeline predictions and manual contours available in HNSCC dataset. }
  \label{fig:hnscc_perf}
\end{figure}

\begin{figure}[h!] 
  \centering
  \subfloat[CT scan overlayed with contours optic chiasm.]{\includegraphics[width=0.49\textwidth]{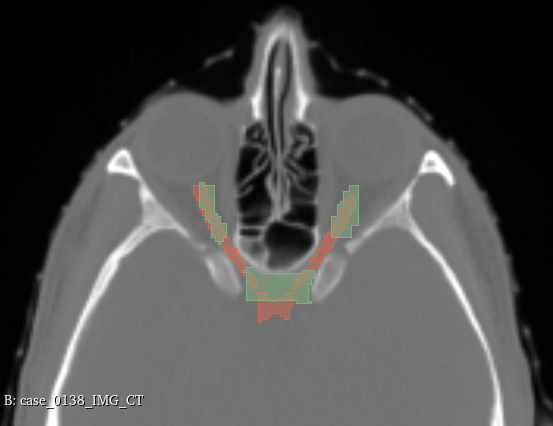}\label{fig:f1}}
  \hfill
  \subfloat[oral cavity.]{\includegraphics[width=0.49\textwidth]{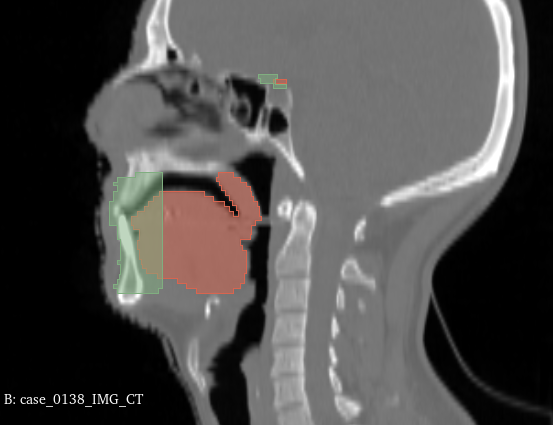}\label{fig:f2}}
  \caption{CT scan of a patient in the HNSCC dataset with manual contours of the optic chiasm, nerves and the oral cavity shown in green along with our automated predictions shown in red.}
  \label{fig:ocav_ochias_hnscc}
\end{figure}

\begin{figure}[h!] 
  \centering
 \includegraphics[width=0.5\textwidth]{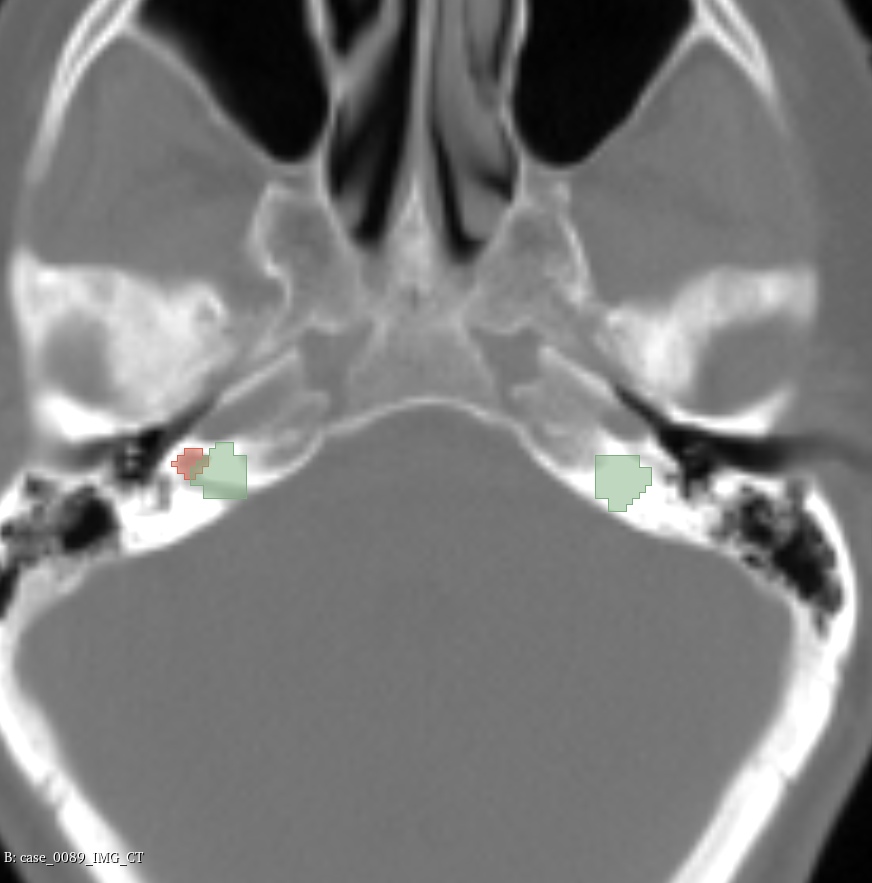}
  \caption{CT scan of a patient in the HNSCC dataset with manual contours of the cochleae shown in green along with our automated predictions shown in red.}
  \label{fig:cochlea_hnscc}
\end{figure}

\begin{figure}[h!] 
  \centering
 \includegraphics[width=0.5\textwidth]{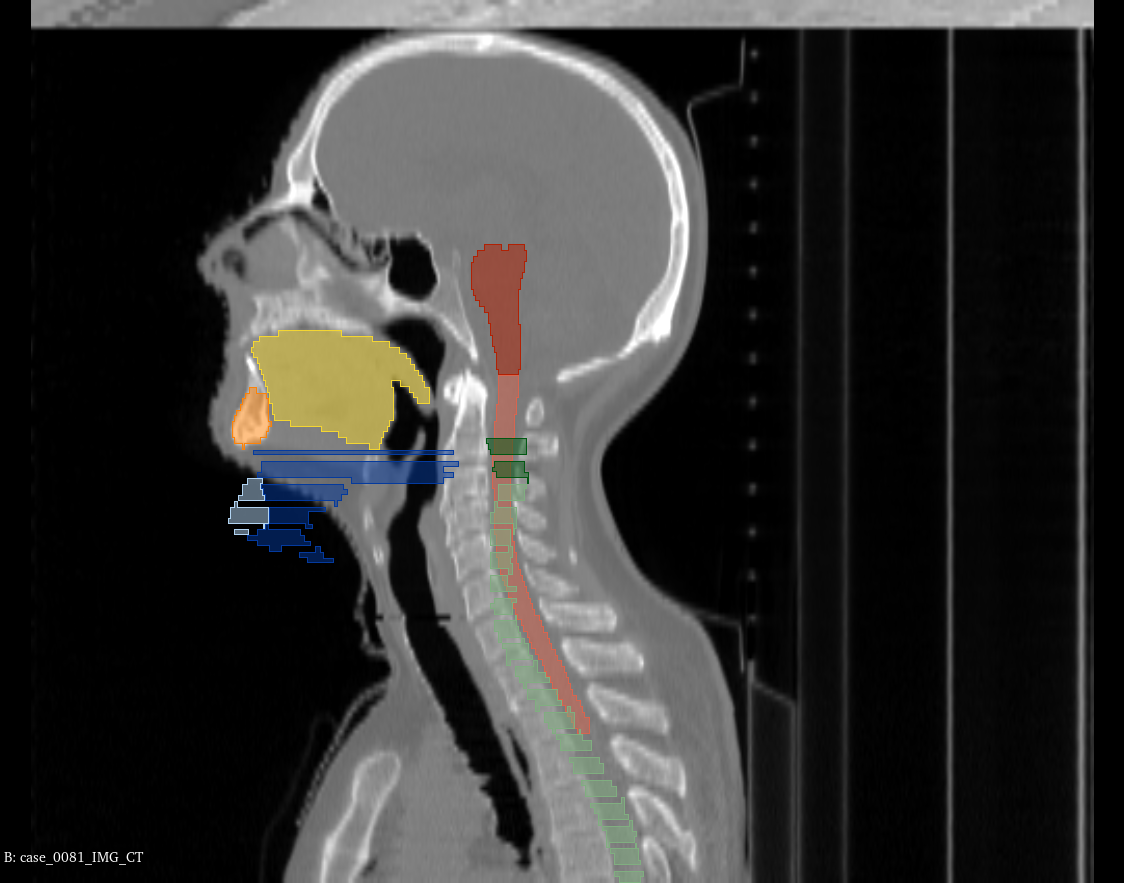}
  \caption{CT scan of a patient from the HNSCC dataset, displaying manual contours alongside automated predictions. Manual segmentations are shown in dark green for the brainstem, light green for the spinal cord, dark blue for the oral cavity, and light blue for the mandible bone. Automated segmentations are represented in dark red for the brainstem, light red for the spinal cord, yellow for the oral cavity, and orange for the mandible bone. }
  \label{fig:wrong_patient_hnscc}
\end{figure}

Table~\ref{table:hnscc_perf} and Figure~\ref{fig:hnscc_perf} summarize the agreement between our pipeline predictions and HNSCC manual contours. OAR with the worst agreement are the optic chiasm, the cochleae and the oral cavity.
Figure~\ref{fig:ocav_ochias_hnscc} displays predictions of our pipeline and manual ground truth contours for a patient whose DSs were 4.99, 58.44, 61.22 and 15.42\% for the optic chiasm, the left and right optic nerves and the oral cavity, respectively. It can be seen that the low DS for the oral cavity is not due to a poor prediction from our pipeline but rather due to a partial manual contour.
Figure~\ref{fig:cochlea_hnscc} shows a patient whose DSs were 0.00\% and 5.19\% for the left and right cochleae, respectively. The manual contour appears to be misaligned with the structures.
Figure~\ref{fig:wrong_patient_hnscc} shows a patient whose DSs were 0.00, 0.00, 0.00 0.57 and 14.20\% for the mandible bone, the brainstem, the oral cavity, the right parotid gland and the spinal cord, respectively. The poor reconstruction quality of the CT scan appears to have caused a systematic shift in all manually delineated structures.
%
%

Different part of the esophagus, spinal cord, and larynx structures are contoured in HNSCC and in our training dataset and thus these contours agreement are not shown.  

Overall, the weak agreement between our pipeline predictions and some OARs in TCIA datasets does not seem to be related to poor quality predictions of our pipeline but rather to different contouring guidelines for some OARs, e.g., the optic chiasm and optic nerves, or approximate manual contouring in TCIA datasets.

\subsection{Commercial software}

\begin{figure}[h!] 
  \centering
 \includegraphics[width=0.95\textwidth]{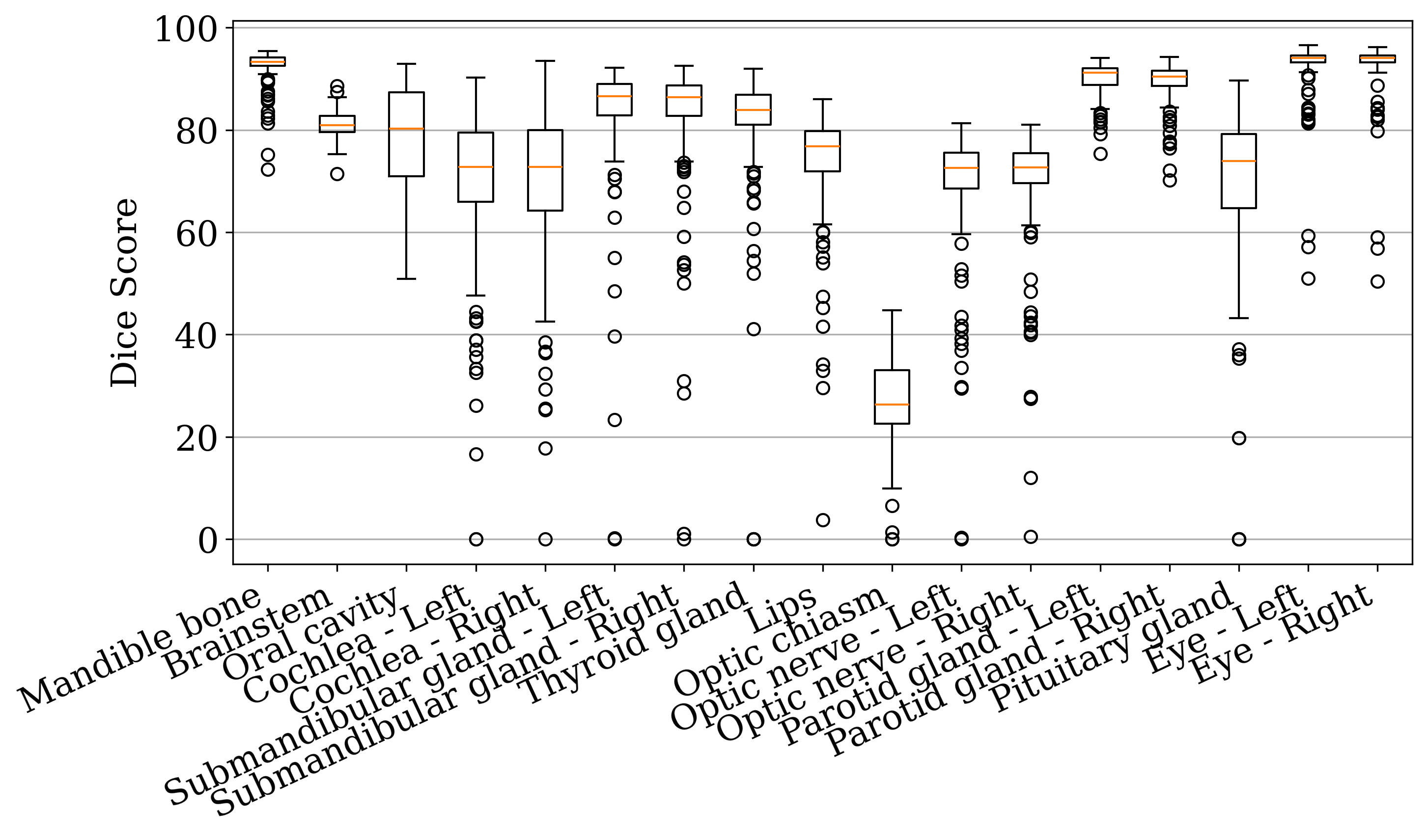}
  \caption{Dice score obtained when comparing our pipeline predictions and Limbus AI predictions on the HNSCC CT scans.}
  \label{fig:limbus_boxplots}
\end{figure}

\begin{figure}
    \centering
    \includegraphics[width=0.5\textwidth]{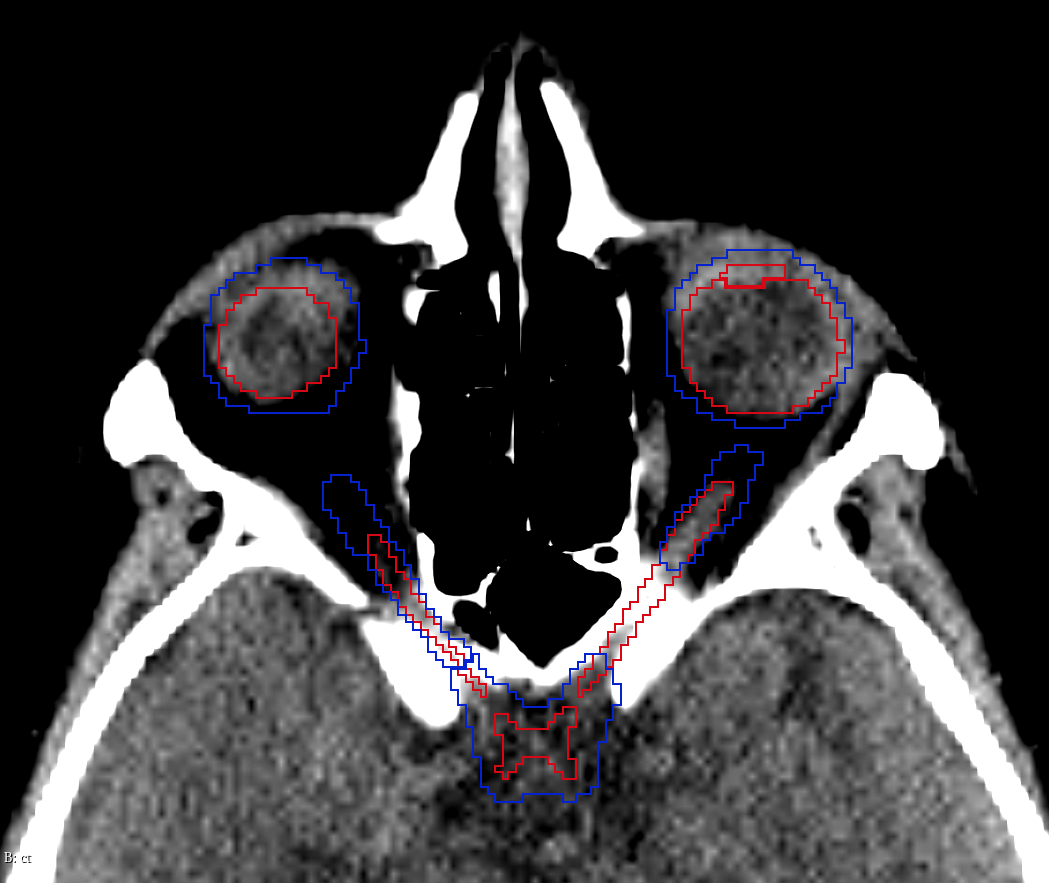}
    \caption{Limbus AI prediction (blue) and the prediction of our pipeline using the CT only (red) overlayed on the CT scan of a patient in HNSCC dataset. Contours of the eyes, optic nerves and optic chiasm can be observed.}
    \label{fig:limbus_ours_chiasm}
\end{figure}

Table~\ref{table:limbusai_ours} and Figure~\ref{fig:limbus_boxplots} present the agreement between Limbus AI predictions and our pipeline predictions on 17 different structures. A strong agreement can be observed for most of the OARs compared, except for the optic chiasm, whose average DS is 27.12\%. 
\begin{table}[h!]
    \scriptsize
    \centering
    \begin{tabular}{|c|c|c|}
    \hline
    \textbf{OAR} & \textbf{Mean Dice Score} $\uparrow$& \textbf{Mean Hausdorff Distance} $\downarrow$ \tabularnewline 
    \hline\hline
Mandible bone  (n=188) & $92.70 \pm 2.97$ [72.23 95.41] & $5.10 \pm 26.10$ [1.00 194.95] \\ \hline 
Brainstem  (n=188) & $81.07 \pm 2.38$ [71.41 88.63] & $3.41 \pm 0.93$ [1.73 10.34] \\ \hline 
Oral cavity  (n=188) & $78.64 \pm 9.50$ [50.85 92.95] & $7.89 \pm 3.89$ [2.00 21.40] \\ \hline 
Cochlea - Left  (n=188) & $70.02 \pm 14.43$ [0.00 90.24] & $1.21 \pm 0.35$ [1.00 3.16] \\ \hline 
Cochlea - Right  (n=187) & $70.43 \pm 14.45$ [0.00 93.48] & $1.25 \pm 0.35$ [1.00 3.22] \\ \hline 
Submandibular gland - Left  (n=188) & $83.53 \pm 11.97$ [0.00 92.16] & $3.04 \pm 2.79$ [1.41 32.89] \\ \hline 
Submandibular gland - Right  (n=188) & $83.16 \pm 12.43$ [0.00 92.52] & $3.31 \pm 2.95$ [1.41 29.07] \\ \hline 
Thyroid gland  (n=188) & $80.64 \pm 14.96$ [0.00 91.98] & $3.50 \pm 5.88$ [1.00 57.07] \\ \hline 
Lips  (n=188) & $74.08 \pm 10.29$ [3.80 86.00] & $4.68 \pm 4.14$ [2.00 42.84] \\ \hline 
Optic chiasm  (n=187) & $27.12 \pm 8.56$ [0.00 44.79] & $7.44 \pm 1.62$ [4.24 13.64] \\ \hline 
Optic nerve - Left  (n=187) & $69.68 \pm 11.64$ [0.00 81.29] & $3.31 \pm 3.73$ [1.00 16.81] \\ \hline 
Optic nerve - Right  (n=187) & $70.13 \pm 10.98$ [0.49 81.03] & $2.14 \pm 2.45$ [1.00 18.24] \\ \hline 
Parotid gland - Left  (n=188) & $90.12 \pm 3.00$ [75.32 94.08] & $2.67 \pm 1.62$ [1.41 11.36] \\ \hline 
Parotid gland - Right  (n=188) & $89.43 \pm 3.62$ [70.16 94.23] & $3.57 \pm 10.89$ [1.41 150.71] \\ \hline 
Pituitary gland  (n=187) & $69.80 \pm 16.59$ [0.00 89.61] & $1.71 \pm 1.00$ [1.00 12.53] \\ \hline 
Eye - Left  (n=187) & $92.69 \pm 5.46$ [51.02 96.52] & $1.14 \pm 0.61$ [1.00 7.40] \\ \hline 
Eye - Right  (n=187) & $92.78 \pm 5.46$ [50.37 96.14] & $1.11 \pm 0.39$ [1.00 4.00] \\ \hline 
    \end{tabular}
    \caption{Comparison between our pipeline predictions and Limbus AI predictions on the HNSCC dataset}
    \label{table:limbusai_ours}
\end{table}
Figure~\ref{fig:limbus_ours_chiasm} shows the Limbus AI prediction along with our prediction for a patient whose optic chiasm DS was 22.71\%, optic nerves DS were 41.82\% and 41.87\%, and eyes DS were 82.80\% and 81.31\%. The low DS value for the optic chiasm seems to be attributed to different contouring styles.

\begin{figure}
    \centering
    \subfloat[Slide 1]{\includegraphics[width=0.49\textwidth]{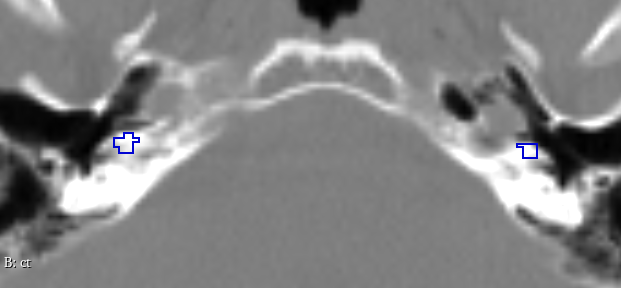}\label{fig:f1}}
    \hfill
    \subfloat[Slide 2]{\includegraphics[width=0.49\textwidth]{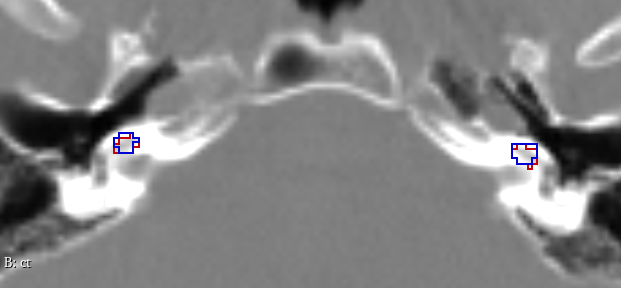}\label{fig:f2}}
    \hfill 
    \subfloat[Slide 3]{\includegraphics[width=0.49\textwidth]{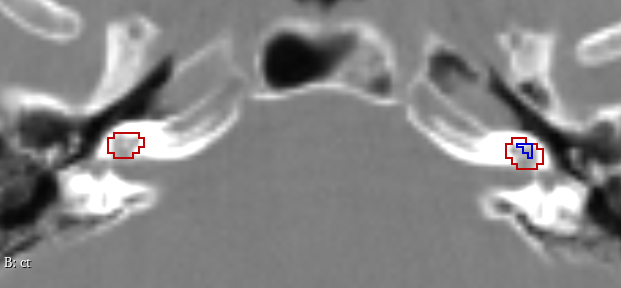}\label{fig:f3}}
    \hfill
    \subfloat[Slide 4]{\includegraphics[width=0.49\textwidth]{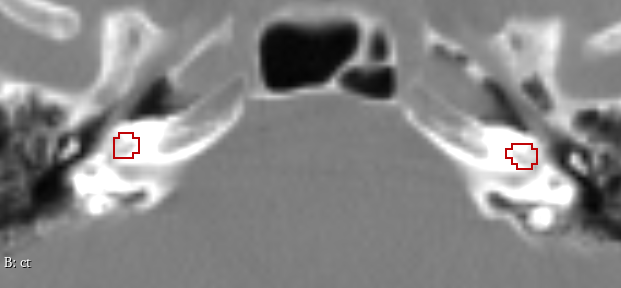}\label{fig:f4}}
  \caption{Limbus AI prediction of the cochleae (blue) and the prediction of our pipeline using the CT only (red) overlayed on the CT scan of a patient in HNSCC dataset. All slices containing cochleae predictions of either softare are shown, ordered from lower to upper body.}
  
    \label{fig:cochlea_limbus_ours}
\end{figure}

Figure~\ref{fig:cochlea_limbus_ours} shows the extent of Limbus AI and our pipeline predictions for the cochleae. Limbus AI contour appears to be imprecise and not exactly aligned with the true structure.

Overall, a strong agreement is observed between the two softwares, with the exception of the optic chiasm contour, whose contour is in general thinner with our software. Consequently, these contours were chosen to compare performances obtained from a model trained on the HaN-Seg challenge training dataset (n=42) and our created expanded dataset (n=296). Table~\ref{table:small_vs_big_dataset_limbus} shows the performance of models trained on HaN-Seg and our expanded dataset, on OARs whose performance is significantly different at a 95\% confidence level according to the corrected Wilcoxon signed-rank test.

\begin{table}[h!]
    \scriptsize
    \centering
    \begin{tabular}{|P{9em}|P{11em}|P{8em}|P{8em}|P{7em}|P{6em}|}
    \hline
    \multirow{4}{*}{\textbf{Metric}} & \multirow{4}{*}{\textbf{OAR}} & \multicolumn{2}{c|}{\textbf{Model performance}}  & \multirow{4}{*}{\begin{tabular}[c]{@{}c@{}}\textbf{Paired}\\ \textbf{differences}\end{tabular}} & \multirow{4}{*}{{\begin{tabular}[c]{@{}c@{}}\textbf{Corrected}\\ \textbf{p-value}\end{tabular}}} \tabularnewline 
    \cline{3-4}
    && \textbf{Trained on the HaN-Seg dataset} & \textbf{Trained on the expanded dataset} && \tabularnewline 
    \hline\hline
\multirow{21}{*}{Dice Score (\%) $\uparrow$} & Oral cavity (n=188) &  $78.58 \pm 9.48$ [52.14 92.95] & $77.25 \pm 10.23$ [49.83 92.57] & $1.33 \pm 2.09$ [-6.24 9.53] & $1.22E-16$ \\ \cline{2-6} 
 & Cochlea - Left (n=188) &  $70.18 \pm 14.48$ [0.00 90.24] & $68.12 \pm 13.55$ [0.00 87.04] & $2.06 \pm 7.95$ [-20.91 30.22] & $1.76E-04$ \\ \cline{2-6} 
 & Cochlea - Right (n=187) &  $70.43 \pm 14.46$ [0.00 93.48] & $66.46 \pm 14.11$ [0.00 94.12] & $3.97 \pm 7.62$ [-27.02 23.89] & $5.73E-11$ \\ \cline{2-6} 
 & Submandibular gland - Left (n=188) &  $83.20 \pm 13.30$ [0.00 92.16] & $82.18 \pm 12.59$ [0.00 92.40] & $1.02 \pm 5.87$ [-61.39 22.22] & $5.38E-10$ \\ \cline{2-6} 
 & Submandibular gland - Right (n=188) &  $83.04 \pm 12.63$ [0.00 92.52] & $81.82 \pm 14.67$ [0.00 91.94] & $1.22 \pm 4.30$ [-15.36 27.01] & $2.60E-08$ \\ \cline{2-6} 
 & Lips (n=188) &  $74.05 \pm 10.27$ [3.80 86.01] & $73.26 \pm 8.93$ [15.62 85.60] & $0.79 \pm 4.13$ [-26.25 8.78] & $2.05E-08$ \\ \cline{2-6} 
 & Optic nerve - Left (n=187) &  $69.68 \pm 11.63$ [0.00 81.29] & $69.06 \pm 11.90$ [0.00 81.01] & $0.62 \pm 3.20$ [-7.62 10.43] & $3.53E-02$ \\ \cline{2-6} 
 & Parotid gland - Left (n=188) &  $90.10 \pm 3.03$ [75.32 94.08] & $88.62 \pm 3.80$ [65.96 93.29] & $1.48 \pm 1.64$ [-2.58 15.25] & $1.79E-29$ \\ \cline{2-6} 
 & Parotid gland - Right (n=188) &  $89.41 \pm 3.62$ [70.49 94.23] & $88.02 \pm 4.44$ [65.00 93.34] & $1.39 \pm 1.69$ [-1.53 12.53] & $6.56E-26$ \\ \cline{2-6} 
 & Eye - Left (n=187) &  $92.69 \pm 5.46$ [51.02 96.46] & $92.25 \pm 5.49$ [49.36 95.87] & $0.43 \pm 0.62$ [-0.86 2.32] & $1.17E-14$ \\ \cline{2-6} 
 & Eye - Right (n=187) &  $92.77 \pm 5.46$ [50.37 96.20] & $92.38 \pm 5.44$ [49.10 96.16] & $0.39 \pm 0.61$ [-1.36 1.98] & $2.03E-13$ \\ 
 \hline \hline
 \multirow{21}{*}{\makecell[c]{Hausdorff Distance \\ (mm) $\downarrow$}} & Oral cavity (n=188) &  $9.10 \pm 16.97$ [2.00 234.97] & $8.35 \pm 4.15$ [2.00 19.24] & $0.75 \pm 16.81$ [-4.58 230.07] & $7.52E-09$ \\ \cline{2-6} 
 & Cochlea - Left (n=188) &  $1.21 \pm 0.35$ [1.00 3.16] & $1.31 \pm 0.37$ [1.00 3.16] & $-0.10 \pm 0.20$ [-0.90 0.55] & $1.67E-09$ \\ \cline{2-6} 
 & Cochlea - Right (n=187) &  $1.24 \pm 0.35$ [1.00 3.22] & $1.37 \pm 0.40$ [1.00 3.16] & $-0.12 \pm 0.24$ [-1.16 0.56] & $1.18E-09$ \\ \cline{2-6} 
 & Lips (n=188) &  $4.69 \pm 4.14$ [2.00 42.84] & $4.56 \pm 2.90$ [2.24 23.35] & $0.14 \pm 1.95$ [-1.27 21.11] & $1.56E-02$ \\ \cline{2-6} 
 & Optic nerve - Left (n=187) &  $3.33 \pm 3.73$ [1.00 16.73] & $4.12 \pm 4.44$ [1.00 20.98] & $-0.80 \pm 1.64$ [-10.59 3.56] & $1.18E-09$ \\ \cline{2-6} 
 & Optic nerve - Right (n=187) &  $2.13 \pm 2.44$ [1.00 18.22] & $2.63 \pm 3.17$ [1.00 19.59] & $-0.51 \pm 1.21$ [-7.82 2.15] & $1.10E-08$ \\ \cline{2-6} 
 & Parotid gland - Left (n=188) &  $4.10 \pm 19.68$ [1.41 272.27] & $4.24 \pm 12.91$ [1.41 176.95] & $-0.14 \pm 7.25$ [-12.52 95.32] & $1.67E-09$ \\ \cline{2-6} 
 & Parotid gland - Right (n=188) &  $4.23 \pm 13.83$ [1.41 150.71] & $3.33 \pm 2.67$ [1.41 19.43] & $0.89 \pm 14.00$ [-15.31 149.30] & $2.19E-06$ \\ \cline{2-6} 
 & Pituitary gland (n=187) &  $1.70 \pm 1.00$ [1.00 12.53] & $1.64 \pm 1.08$ [1.00 13.42] & $0.07 \pm 0.36$ [-1.94 1.00] & $1.47E-03$ \\ \cline{2-6} 
 & Eye - Left (n=187) &  $1.14 \pm 0.61$ [1.00 7.40] & $1.16 \pm 0.64$ [1.00 7.95] & $-0.02 \pm 0.12$ [-1.00 0.41] & $2.09E-02$ \\ \cline{2-6} 
 & Eye - Right (n=187) &  $1.11 \pm 0.39$ [1.00 4.00] & $1.14 \pm 0.41$ [1.00 4.00] & $-0.03 \pm 0.13$ [-1.00 0.41] & $2.28E-03$ \\
 \hline
    \end{tabular}
    \caption{Average performance for significantly different OAR performances at 95\% confidence level according to the corrected Wilcoxon signed-rank test between predictions of models trained on the HaN-Seg training dataset only and on the expanded dataset. Results are presented as mean $\pm$ standard deviation [minimum maximum].}
    \label{table:small_vs_big_dataset_limbus}
\end{table}

\newpage

\section{Detailed OAR performance}
\begin{table}[h!]
    \tiny
    \centering   
    \begin{tabular}{|P{8em}|P{6em}|P{6em}|P{6em}|P{6em}|P{6em}|P{6em}|}
    \cline{1-7}
    \multirow{2}{*}{ \textbf{OAR}}& \multicolumn{2}{c|}{\textbf{MRI inference}} & \multicolumn{2}{c|}{\textbf{CT inference}} & \multicolumn{2}{c|}{\textbf{CT + MRI inference}} \\ 
    \cline{2-7}
     &\textbf{Dice Score} (\%) $\uparrow$& \textbf{Hausdorff Distance} (mm) $\downarrow$ &\textbf{Dice Score} (\%) $\uparrow$& \textbf{Hausdorff Distance} (mm) $\downarrow$&\textbf{Dice Score} (\%) $\uparrow$& \textbf{Hausdorff Distance} (mm) $\downarrow$ \\ \hline 
    Carotid artery - Left & $69.05 \pm 15.74$ [7.55 87.63] & $5.29 \pm 4.73$ [1.00 25.16]  & $90.65 \pm 5.24$ [72.10 96.42] & $1.46 \pm 1.57$ [1.00 12.45] & $90.98 \pm 4.59$ [74.49 96.06] & $1.37 \pm 1.46$ [1.00 12.08] \\ \hline 
Carotid artery - Right & $70.48 \pm 13.27$ [33.98 87.67] & $4.53 \pm 3.32$ [1.41 15.39]  & $91.00 \pm 4.73$ [71.96 96.86] & $1.31 \pm 0.85$ [1.00 6.50] & $91.24 \pm 4.15$ [76.34 96.80] & $1.25 \pm 0.64$ [1.00 5.10] \\ \hline 
Arytenoids & $26.18 \pm 19.35$ [0.00 50.23] & $9.05 \pm 11.31$ [2.24 43.21]  & $75.68 \pm 8.85$ [64.68 88.33] & $1.46 \pm 0.57$ [1.00 2.45] & $75.58 \pm 9.25$ [62.79 88.39] & $1.44 \pm 0.58$ [1.00 2.45] \\ \hline 
Mandible bone & $81.48 \pm 10.58$ [32.96 90.96] & $7.31 \pm 11.85$ [1.41 59.03]  & $97.45 \pm 1.36$ [92.75 99.01] & $0.99 \pm 0.14$ [0.00 1.41] & $97.44 \pm 1.38$ [92.66 98.97] & $1.00 \pm 0.15$ [0.00 1.41] \\ \hline 
Brainstem & \textbf{93.64}$ \pm 4.82$ [68.78 98.35] & $1.54 \pm 0.85$ [1.00 4.58]  & \textbf{95.36}$ \pm 2.93$ [85.62 98.56] & $1.34 \pm 0.60$ [1.00 3.32] & $95.96 \pm 2.65$ [87.61 98.75] & $1.22 \pm 0.49$ [1.00 3.16] \\ \hline 
Buccal mucosa & $73.34 \pm 11.60$ [37.19 88.78] & $3.87 \pm 1.98$ [1.00 10.29]  & $90.15 \pm 7.24$ [58.90 96.69] & $1.74 \pm 1.50$ [1.00 8.06] & $90.20 \pm 7.20$ [59.67 96.82] & $1.74 \pm 1.48$ [1.00 8.00] \\ \hline 
Oral cavity & $88.62 \pm 7.45$ [48.88 95.58] & $4.32 \pm 2.67$ [1.00 18.63]  & $96.81 \pm 2.86$ [86.56 99.25] & $1.51 \pm 1.13$ [0.00 5.39] & $96.82 \pm 2.89$ [86.36 99.28] & $1.51 \pm 1.16$ [0.00 5.48] \\ \hline 
Cochlea - Left & $69.94 \pm 18.01$ [0.00 90.91] & $1.63 \pm 1.11$ [1.00 7.24]  & $88.11 \pm 6.71$ [72.97 100.00] & $1.01 \pm 0.34$ [0.00 2.24] & $88.12 \pm 6.57$ [75.00 100.00] & $1.01 \pm 0.31$ [0.00 2.45] \\ \hline 
Cochlea - Right & $73.58 \pm 17.91$ [0.00 92.86] & $1.53 \pm 1.01$ [1.00 6.48]  & $88.21 \pm 11.05$ [17.43 99.54] & $1.09 \pm 0.79$ [0.00 6.71] & $88.01 \pm 11.16$ [15.96 99.54] & $1.09 \pm 0.79$ [0.00 6.71] \\ \hline 
Cricopharyngeal inlet & $38.88 \pm 27.36$ [0.07 73.69] & $6.05 \pm 3.85$ [2.00 14.46]  & $80.77 \pm 10.18$ [62.89 96.33] & $1.65 \pm 0.54$ [1.00 2.45] & $80.72 \pm 10.12$ [62.39 96.37] & $1.70 \pm 0.61$ [1.00 2.83] \\ \hline 
Cervical esophagus & $58.16 \pm 12.61$ [40.77 75.86] & $4.70 \pm 2.38$ [1.41 8.06]  & $78.11 \pm 9.83$ [64.62 95.73] & $2.50 \pm 1.32$ [1.00 4.90] & $78.84 \pm 9.95$ [64.29 95.63] & $2.56 \pm 1.40$ [1.00 5.00] \\ \hline 
Anterior eyeball segment - Left & $76.24 \pm 12.81$ [15.75 90.85] & $2.12 \pm 1.14$ [1.00 5.83]  & $91.46 \pm 5.74$ [68.29 96.73] & $1.12 \pm 0.33$ [1.00 3.00] & $91.75 \pm 5.63$ [68.73 97.00] & $1.12 \pm 0.34$ [1.00 3.00] \\ \hline 
Anterior eyeball segment - Right & $76.67 \pm 13.19$ [11.27 93.12] & $2.18 \pm 1.24$ [1.00 6.90]  & $91.61 \pm 5.89$ [61.85 97.26] & $1.15 \pm 0.46$ [1.00 3.61] & $91.89 \pm 5.74$ [62.31 96.76] & $1.13 \pm 0.46$ [1.00 3.61] \\ \hline 
Posterior eyeball segment - Left & $90.16 \pm 6.66$ [50.42 95.49] & $1.66 \pm 0.71$ [1.00 4.24]  & $96.55 \pm 1.61$ [91.87 98.68] & $1.05 \pm 0.25$ [1.00 2.83] & $96.62 \pm 1.60$ [91.89 98.84] & $1.06 \pm 0.29$ [1.00 3.16] \\ \hline 
Posterior eyeball segment - Right & $90.22 \pm 6.41$ [51.85 96.26] & $1.64 \pm 0.85$ [1.00 5.48]  & $96.53 \pm 1.64$ [90.54 98.50] & $1.05 \pm 0.20$ [1.00 2.24] & $96.60 \pm 1.62$ [90.67 98.56] & $1.04 \pm 0.19$ [1.00 2.24] \\ \hline 
Lacrimal gland - Left & $74.34 \pm 14.49$ [12.67 90.36] & $1.84 \pm 0.97$ [1.00 5.74]  & $87.81 \pm 7.04$ [65.80 96.22] & $1.19 \pm 0.50$ [1.00 3.61] & $88.29 \pm 7.03$ [66.37 95.32] & $1.19 \pm 0.49$ [1.00 3.16] \\ \hline 
Lacrimal gland - Right & $75.01 \pm 14.30$ [0.56 91.69] & $1.90 \pm 1.10$ [1.00 7.02]  & $87.15 \pm 9.08$ [54.49 95.80] & $1.36 \pm 1.00$ [1.00 7.28] & $87.48 \pm 8.64$ [54.97 95.57] & $1.32 \pm 0.93$ [1.00 7.07] \\ \hline 
Submandibular gland - Left & $62.92 \pm 21.18$ [0.00 89.21] & $6.60 \pm 3.66$ [2.24 14.37]  & $90.51 \pm 9.93$ [42.11 98.11] & $1.69 \pm 1.07$ [1.00 5.00] & $90.51 \pm 10.41$ [38.30 98.08] & $1.65 \pm 1.08$ [1.00 5.00] \\ \hline 
Submandibular gland - Right & $62.59 \pm 21.91$ [0.00 86.89] & $9.29 \pm 13.98$ [2.00 88.24]  & $89.05 \pm 13.80$ [21.21 97.21] & $1.80 \pm 1.32$ [1.00 7.00] & $89.25 \pm 13.06$ [24.56 97.44] & $1.76 \pm 1.30$ [1.00 7.00] \\ \hline 
Thyroid gland & $49.07 \pm 31.73$ [0.00 82.81] & $18.01 \pm 21.44$ [2.24 60.63]  & $88.29 \pm 6.50$ [71.60 96.26] & $1.81 \pm 1.63$ [1.00 6.63] & $87.94 \pm 6.76$ [70.89 96.34] & $1.86 \pm 1.62$ [1.00 6.63] \\ \hline 
Larynx - glottis & $44.10 \pm 18.21$ [5.65 68.71] & $5.60 \pm 3.71$ [1.73 14.00]  & $85.27 \pm 7.31$ [73.07 95.03] & $1.39 \pm 0.50$ [1.00 2.45] & $85.19 \pm 7.17$ [73.34 94.25] & $1.37 \pm 0.47$ [1.00 2.24] \\ \hline 
Larynx - supraglottic & $49.57 \pm 22.49$ [0.00 77.26] & $7.37 \pm 4.84$ [2.24 19.10]  & $87.16 \pm 17.29$ [0.00 96.65] & $1.77 \pm 1.56$ [1.00 7.66] & $87.08 \pm 17.28$ [0.00 96.41] & $1.79 \pm 1.62$ [1.00 7.87] \\ \hline 
Lips & $75.44 \pm 11.98$ [23.48 90.28] & $3.90 \pm 2.16$ [1.00 12.41]  & $92.07 \pm 7.00$ [53.12 97.42] & $1.47 \pm 1.34$ [1.00 10.49] & $92.10 \pm 6.88$ [54.09 97.45] & $1.48 \pm 1.34$ [1.00 10.19] \\ \hline 
Optic chiasm & $70.67 \pm 17.06$ [0.00 92.70] & $1.64 \pm 0.84$ [1.00 5.83]  & $79.34 \pm 11.00$ [45.73 93.90] & $1.26 \pm 0.46$ [1.00 3.00] & $80.47 \pm 10.40$ [45.93 94.61] & $1.19 \pm 0.43$ [1.00 3.00] \\ \hline 
Optic nerve - Left & $66.84 \pm 16.31$ [2.53 88.47] & $1.79 \pm 0.79$ [1.00 4.00]  & $87.14 \pm 7.08$ [65.43 96.40] & $1.11 \pm 0.39$ [1.00 3.61] & $86.96 \pm 7.15$ [65.44 96.16] & $1.11 \pm 0.39$ [1.00 3.61] \\ \hline 
Optic nerve - Right & $70.05 \pm 15.76$ [16.29 87.59] & $1.90 \pm 1.89$ [1.00 14.99]  & $88.35 \pm 7.00$ [66.67 97.04] & $1.10 \pm 0.36$ [0.00 2.45] & $88.27 \pm 7.17$ [65.61 97.33] & $1.09 \pm 0.40$ [0.00 2.61] \\ \hline 
Parotid gland - Left & $82.57 \pm 8.80$ [53.02 92.82] & $5.41 \pm 3.25$ [1.41 19.80]  & $95.06 \pm 4.54$ [77.09 98.73] & $1.91 \pm 2.32$ [1.00 17.23] & $95.20 \pm 4.38$ [77.33 98.82] & $1.70 \pm 1.34$ [1.00 7.87] \\ \hline 
Parotid gland - Right & $84.14 \pm 6.28$ [64.63 92.04] & $5.72 \pm 3.62$ [1.41 21.00]  & $95.22 \pm 4.09$ [81.53 98.36] & $1.95 \pm 2.86$ [1.00 22.14] & $95.30 \pm 3.99$ [82.35 98.43] & $1.88 \pm 2.85$ [1.00 22.29] \\ \hline 
Pituitary gland & $77.71 \pm 14.69$ [19.64 94.50] & $1.48 \pm 0.85$ [1.00 6.15]  & $88.51 \pm 10.39$ [31.85 97.45] & $1.09 \pm 0.36$ [0.00 2.92] & $88.49 \pm 10.06$ [34.53 96.50] & $1.09 \pm 0.36$ [0.00 2.83] \\ \hline 
Spinal cord & $75.66 \pm 15.08$ [25.37 94.85] & $7.07 \pm 22.67$ [1.00 179.64]  & $92.80 \pm 5.67$ [74.71 98.59] & $1.27 \pm 0.59$ [1.00 3.16] & $93.43 \pm 5.23$ [76.34 98.52] & $1.23 \pm 0.50$ [1.00 2.83] \\ \hline

    \end{tabular}
    \caption{Segmentation performance of our pipeline trained with Modality Dropout, predicting with MRI only, CT only and CT and MRI scans. Predictions are made with our fold 0 model and evaluated on our internal internal testing set, cropped to the MRI field of view. Results were first averaged for each OAR over all cases and finally over all organs. Results are presented as: mean across patients and OARs $\pm$ standard deviation among the mean metrics of each OAR [minimum OAR metric averaged over cases   and maximum]. $\uparrow$ indicates that higher is best and $\downarrow$ indicates smaller is best.}
    \label{table:mri_vs_ct}
\end{table}
Table~\ref{table:mri_vs_ct} presents the breakdown of segmentation performance for each OAR when predicting from either CT or MRI with a model trained with MD, on scans cropped to the field of view of the MRI scan. It can be seen that overall the model predicts better from the CT scan. Even structures that are well defined on the MRI do not result in a better segmentation performance when predicting with the MRI. The brainstem segmentation performance between the MRI only predictions and the CT only prediction is close. The optic nerves for example than can be easily delineated from an MRI scan have better segmentation metrics when predicting with the CT scan. This is attributed to the fact that contours were made primarily on the CT scan, with only a support from the MRI scan for soft tissues. 

\newpage

\section{Modality Dropout} \label{app:modadrop}
\subsection{MRI FOV}
Table~\ref{table:modalitydropout_mrifov} shows the segmentation performance of the developed pipeline trained with and without the MD augmentation on the MRI and CT scans cropped to the MRI field of view, and with ground truth segmentation masks also cropped to the field of view of the MRI scans. Contours were predicted using only CT, using only MRI, and inferring using both CT and MRI jointly.

\begin{table}[h!]
    \scriptsize
    \centering   
    \begin{tabular}{|P{8em}|P{6em}|P{6em}|P{7em}|P{6em}|P{6em}|P{7em}|}
    \cline{1-7}
    \multirow{2}{*}{ \textbf{Metric}}& \multicolumn{3}{c|}{\textbf{With Modality Dropout}} & \multicolumn{3}{c|}{\textbf{Without Modality Dropout}}\tabularnewline 
    \cline{2-7}
     & \textbf{CT only}&\textbf{MRI only} & \textbf{CT + MRI}&  \textbf{CT only}&\textbf{MRI only} & \textbf{CT + MRI}\tabularnewline 
    \hline\hline
    
    \textbf{Dice Score (\%) $\uparrow$} & 88.73 $\pm$ 6.22 [73.17 97.43] & 69.39 $\pm$ 16.03 [23.52 93.69] & 88.89 $\pm$ 6.15 [73.85 97.4] & 86.22 $\pm$ 8.79 [60.55 97.25]& 0.65 $\pm$ 3.5 [0.0 19.51]& 88.48 $\pm$ 7.0 [67.65 97.45]\tabularnewline \hline
    \textbf{Hausdorff Distance (mm) $\downarrow$} & 1.58 $\pm$ 0.63 [0.97 4.14] & 4.64 $\pm$ 3.85 [1.44 21.35]& 1.55 $\pm$ 0.57 [0.97 3.75] & 2.06 $\pm$ 1.24 [1.0 6.65] & 138.1 $\pm$ 71.64 [35.75 268.35]& 1.74 $\pm$ 1.07 [0.98 5.84]\tabularnewline \hline

    \end{tabular}
    \caption{
    Average segmentation metrics obtained by our fold 0 model on the patients of our fold 0 internal internal testing set. Results were first averaged for each OAR over all cases and finally over all OARs. Results are presented as: mean across patients and OARs $\pm$ standard deviation among the mean metrics of each OAR [minimum OAR metric averaged over cases  and maximum]. $\uparrow$ indicates that higher is best and $\downarrow$ indicates smaller is best.}
    \label{table:modalitydropout_mrifov}
\end{table}

Similar conclusions about the impact of MD augmentation to the full field of view experiments can be drawn for this table. 
It can be also observed that difference in performance of a model trained without the MD comparing an inference with CT only and one with CT and MRI is bigger using a smaller filed of view. DS goes from 86.22\% for CT only to 88.48\% for CT and MRI prediction with MRI field of view against 89.85\% to 90.24\% for the full field of view. It can be explained by the fact that for a model trained on MRI field of view, the MRI contains body information everywhere. When using the CT FOV, the MR scans also contain some background air voxels and thus the model naturally relies more on the CT scan, which contains more reliable information.

\subsection{Significant differences}

\subsubsection{CT only predictions vs predictions with CT and MRI}

%
%
\begin{table}[h!]
    \scriptsize
    \centering
    \begin{tabular}{|P{10em}|P{9em}|P{7em}|P{7em}|P{7em}|P{6em}|}
    \hline
    \multirow{4}{*}{\textbf{Metric}} & \multirow{4}{*}{\textbf{OAR}} & \multicolumn{2}{c|}{\textbf{Model performance}} & \multirow{4}{*}{\begin{tabular}[c]{@{}c@{}}\textbf{Paired}\\ \textbf{differences}\end{tabular}} & \multirow{4}{*}{\begin{tabular}[c]{@{}c@{}}\textbf{Corrected}\\ \textbf{p-value}\end{tabular}} \tabularnewline 
    \cline{3-4}
    && \textbf{CT only predictions} & \textbf{Predictions with MRI and CT}  & & \tabularnewline 
    \hline\hline
\multirow{3}{*}{Dice Score (\%) $\uparrow$} & Brainstem (n=60) &  $95.10 \pm 3.11$ [84.38 98.59] & $95.70 \pm 2.63$ [87.94 98.61] & $-0.60 \pm 1.18$ [-3.67 3.72] & $2.78E-03$ \\ \cline{2-6} 
 & Spinal cord (n=60) &  $93.60 \pm 5.68$ [74.25 97.90] & $93.97 \pm 5.18$ [76.09 98.23] & $-0.37 \pm 0.89$ [-4.47 1.70] & $1.85E-02$ \\
 \hline \hline
 Hausdorff Distance (mm) $\downarrow$ & Brainstem (n=60) &  $1.36 \pm 0.63$ [1.00 3.32] & $1.23 \pm 0.48$ [1.00 3.00] & $0.13 \pm 0.27$ [-0.41 1.32] & $2.20E-02$ \\

 \hline
    \end{tabular}
    \caption{Average performance for significantly different OAR performances at 95\% confidence level according to the corrected Wilcoxon signed-rank test between CT only prediction and predictions with CT and MRI scans with a model trained with MD. Results are presented as mean $\pm$ standard deviation [minimum maximum].}
    \label{table:ct_vs_ctmri_md}
\end{table}
Table~\ref{table:ct_vs_ctmri_md} shows performance of a model trained with MD and inferred using CT scans only and MRI and CT scans, for OARs whose performance were significantly different at a 95\% confidence level according to the corrected Wilcoxon signed-rank test. 

\begin{table}[h!]
    \scriptsize
    \centering
    \begin{tabular}{|P{10em}|P{11em}|P{7em}|P{7em}|P{7em}|P{6em}|}
    \hline
    \multirow{4}{*}{\textbf{Metric}} & \multirow{4}{*}{\textbf{OAR}} & \multicolumn{2}{c|}{\textbf{Model performance}} & \multirow{4}{*}{\begin{tabular}[c]{@{}c@{}}\textbf{Paired}\\ \textbf{differences}\end{tabular}} & \multirow{4}{*}{\begin{tabular}[c]{@{}c@{}}\textbf{Corrected}\\ \textbf{p-value}\end{tabular}}\tabularnewline 
    \cline{3-4}
    && \textbf{CT only predictions} & \textbf{Predictions with MRI and CT} & & \tabularnewline 
    \hline\hline
\multirow{25}{*}{Dice Score (\%) $\uparrow$} & Carotid artery - Left (n=60) &  $90.89 \pm 5.19$ [75.21 97.32] & $91.77 \pm 4.44$ [76.18 97.34] & $-0.88 \pm 2.02$ [-10.53 2.47] & $2.46E-05$ \\ \cline{2-6} 
 & Brainstem (n=60) &  $94.01 \pm 3.55$ [78.94 98.05] & $95.76 \pm 2.88$ [86.43 98.84] & $-1.75 \pm 1.91$ [-8.08 5.36] & $1.59E-07$ \\ \cline{2-6} 
 & Anterior eyeball segment - Left (n=60) &  $90.55 \pm 5.83$ [67.25 95.96] & $91.41 \pm 5.82$ [68.55 97.35] & $-0.86 \pm 1.95$ [-5.04 5.45] & $7.11E-04$ \\ \cline{2-6} 
 & Anterior eyeball segment - Right (n=60) &  $90.88 \pm 5.83$ [62.87 97.14] & $91.61 \pm 5.95$ [60.55 97.32] & $-0.73 \pm 1.99$ [-7.01 3.49] & $2.60E-02$ \\ \cline{2-6} 
 & Posterior eyeball segment - Left (n=60) &  $96.16 \pm 2.07$ [85.37 98.56] & $96.50 \pm 2.07$ [85.45 98.51] & $-0.34 \pm 0.47$ [-1.63 0.78] & $2.46E-05$ \\ \cline{2-6} 
 & Posterior eyeball segment - Right (n=60) &  $96.29 \pm 1.76$ [90.28 98.39] & $96.51 \pm 1.80$ [89.80 98.46] & $-0.22 \pm 0.41$ [-1.63 0.73] & $4.04E-04$ \\ \cline{2-6} 
 & Lacrimal gland - Left (n=60) &  $86.69 \pm 9.74$ [33.33 96.22] & $87.68 \pm 9.29$ [33.33 95.51] & $-0.99 \pm 2.36$ [-9.06 3.17] & $9.49E-03$ \\ \cline{2-6} 
 & Lacrimal gland - Right (n=60) &  $85.98 \pm 10.21$ [40.78 94.75] & $87.25 \pm 9.95$ [40.78 96.20] & $-1.28 \pm 2.32$ [-10.14 5.07] & $2.93E-04$ \\ \cline{2-6} 
 & Larynx - glottis (n=50) &  $90.28 \pm 5.62$ [71.32 96.50] & $90.04 \pm 6.04$ [70.07 96.51] & $0.24 \pm 0.81$ [-1.44 4.10] & $4.90E-02$ \\ \cline{2-6} 
 & Optic chiasm (n=59) &  $77.93 \pm 11.38$ [41.50 93.73] & $81.61 \pm 9.48$ [48.08 92.59] & $-3.68 \pm 5.73$ [-18.12 15.42] & $2.68E-05$ \\ \cline{2-6} 
 & Parotid gland - Left (n=60) &  $94.85 \pm 5.59$ [64.76 98.32] & $95.19 \pm 5.22$ [64.89 98.49] & $-0.34 \pm 0.82$ [-4.66 0.58] & $4.04E-04$ \\ \cline{2-6} 
 & Parotid gland - Right (n=60) &  $95.25 \pm 4.29$ [77.77 98.53] & $95.45 \pm 4.05$ [80.78 98.50] & $-0.19 \pm 0.52$ [-3.01 0.62] & $1.09E-02$ \\ \cline{2-6} 
 & Spinal cord (n=60) &  $93.60 \pm 5.59$ [73.97 98.01] & $94.40 \pm 4.76$ [76.81 98.00] & $-0.80 \pm 1.45$ [-6.84 1.42] & $4.04E-04$ \\
 \hline \hline
 \multirow{11}{*}{\makecell[c]{Hausdorff Distance\\ (mm) $\downarrow$}} & Carotid artery - Left (n=60) &  $1.48 \pm 1.35$ [1.00 10.49] & $1.25 \pm 0.58$ [1.00 4.90] & $0.23 \pm 0.85$ [-0.32 5.59] & $3.99E-02$ \\ \cline{2-6} 
 & Brainstem (n=60) &  $1.60 \pm 0.86$ [1.00 5.00] & $1.21 \pm 0.44$ [1.00 3.16] & $0.39 \pm 0.63$ [-0.41 3.27] & $1.62E-04$ \\ \cline{2-6} 
 & Lacrimal gland - Right (n=60) &  $1.43 \pm 0.93$ [1.00 5.66] & $1.31 \pm 0.70$ [1.00 3.73] & $0.12 \pm 0.35$ [0.00 2.05] & $3.22E-02$ \\ \cline{2-6} 
 & Optic chiasm (n=59) &  $1.35 \pm 0.55$ [1.00 3.61] & $1.18 \pm 0.44$ [1.00 3.21] & $0.17 \pm 0.39$ [-0.97 1.37] & $2.87E-02$ \\ \cline{2-6} 
 & Parotid gland - Left (n=60) &  $2.03 \pm 3.43$ [1.00 26.98] & $1.79 \pm 2.25$ [1.00 17.06] & $0.24 \pm 1.28$ [-0.45 9.92] & $3.99E-02$ \\ \cline{2-6} 
 & Spinal cord (n=60) &  $2.56 \pm 8.48$ [1.00 66.71] & $2.47 \pm 8.54$ [1.00 67.08] & $0.09 \pm 0.26$ [-0.37 1.27] & $4.76E-02$ \\

 \hline
    \end{tabular}
    \caption{Average performance for significantly different OAR performances at 95\% confidence level according to the corrected Wilcoxon signed-rank test between CT only prediction and predictions with CT and MRI scans with a model trained without MD. Results are presented as mean $\pm$ standard deviation [minimum maximum].}
    \label{table:ct_vs_ctmri_nomd}
\end{table}
Table~\ref{table:ct_vs_ctmri_nomd} shows performance of a model trained without MD and inferred using CT scans only and MRI and CT scans, for OARs whose performance were significantly different at a 95\% confidence level according to the corrected Wilcoxon signed-rank test. 

\subsubsection{CT only predictions}
\begin{table}[h!]
    \scriptsize
    \centering
    \begin{tabular}{|P{10em}|P{11em}|P{7em}|P{7em}|P{7em}|P{6em}|}
    \hline
    \multirow{4}{*}{\textbf{Metric}} & \multirow{4}{*}{\textbf{OAR}} & \multicolumn{2}{c|}{\textbf{Model performance}} & \multirow{4}{*}{\begin{tabular}[c]{@{}c@{}}\textbf{Paired}\\ \textbf{differences}\end{tabular}} & \multirow{4}{*}{\begin{tabular}[c]{@{}c@{}}\textbf{Corrected}\\ \textbf{p-value}\end{tabular}}\tabularnewline 
    \cline{3-4}
    && \textbf{With Modality Dropout} & \textbf{Without Modality Dropout}  && \tabularnewline 
    \hline\hline
\multirow{21}{*}{Dice Score (\%) $\uparrow$} & Mandible bone (n=60) &  $97.61 \pm 1.52$ [90.75 99.12] & $97.72 \pm 1.41$ [92.09 99.21] & $-0.12 \pm 0.25$ [-1.34 0.53] & $1.01E-03$ \\ \cline{2-6} 
 & Brainstem (n=60) &  $95.10 \pm 3.11$ [84.38 98.59] & $94.01 \pm 3.55$ [78.94 98.05] & $1.09 \pm 1.30$ [-2.68 5.45] & $2.51E-06$ \\ \cline{2-6} 
 & Oral cavity (n=60) &  $96.77 \pm 2.78$ [86.26 98.95] & $96.97 \pm 2.66$ [86.61 99.14] & $-0.20 \pm 0.41$ [-1.83 0.58] & $4.99E-04$ \\ \cline{2-6} 
 & Anterior eyeball segment - Left (n=60) &  $91.41 \pm 5.92$ [65.73 96.86] & $90.55 \pm 5.83$ [67.25 95.96] & $0.86 \pm 1.78$ [-2.91 6.71] & $2.88E-03$ \\ \cline{2-6} 
 & Anterior eyeball segment - Right (n=60) &  $91.80 \pm 5.91$ [60.73 96.93] & $90.88 \pm 5.83$ [62.87 97.14] & $0.92 \pm 1.58$ [-2.14 5.80] & $4.49E-04$ \\ \cline{2-6} 
 & Posterior eyeball segment - Left (n=60) &  $96.39 \pm 1.94$ [87.52 98.51] & $96.16 \pm 2.07$ [85.37 98.56] & $0.22 \pm 0.47$ [-0.69 2.15] & $2.88E-03$ \\ \cline{2-6} 
 & Posterior eyeball segment - Right (n=60) &  $96.50 \pm 1.78$ [90.05 98.46] & $96.29 \pm 1.76$ [90.28 98.39] & $0.21 \pm 0.38$ [-0.67 1.08] & $5.96E-04$ \\ \cline{2-6} 
 & Submandibular gland - Left (n=59) &  $94.54 \pm 4.58$ [72.29 97.99] & $94.75 \pm 4.66$ [72.23 98.13] & $-0.21 \pm 0.64$ [-2.50 1.19] & $2.83E-02$ \\ \cline{2-6} 
 & Larynx - supraglottic (n=59) &  $93.07 \pm 4.51$ [75.89 98.07] & $93.48 \pm 4.10$ [78.68 97.79] & $-0.41 \pm 1.04$ [-5.91 1.14] & $1.66E-02$ \\ \cline{2-6} 
 & Optic chiasm (n=59) &  $79.67 \pm 9.50$ [50.27 95.40] & $77.93 \pm 11.38$ [41.50 93.73] & $1.74 \pm 5.01$ [-15.13 12.01] & $2.69E-02$ \\ \cline{2-6} 
 & Pituitary gland (n=60) &  $88.15 \pm 9.99$ [36.78 97.91] & $87.22 \pm 9.79$ [36.40 97.01] & $0.93 \pm 3.84$ [-10.33 17.80] & $3.65E-02$ \\  
\hline \hline
Hausdorff Distance (mm) $\downarrow$ & Brainstem (n=60) &  $1.36 \pm 0.63$ [1.00 3.32] & $1.60 \pm 0.86$ [1.00 5.00] & $-0.24 \pm 0.42$ [-1.68 1.08] & $1.98E-03$ \\ 
 \hline
    \end{tabular}
    \caption{Average performance for significantly different OAR performances at 95\% confidence level according to the corrected Wilcoxon signed-rank test between CT only prediction with model trained with and without MD. Results are presented as mean $\pm$ standard deviation [minimum maximum].}
    \label{table:ct_pred_md_vs_nomd}
\end{table}
Table~\ref{table:ct_pred_md_vs_nomd} shows performance of a model trained with and without MD and inferred using CT scans only, for OARs whose performance were significantly different at a 95\% confidence level according to the corrected Wilcoxon signed-rank test. 

\subsubsection{MRI only vs CT only}
\begin{table}[h!]
    \tiny
    \centering
    \begin{tabular}{|P{10em}|P{11em}|P{7em}|P{7em}|P{7em}|P{6em}|}
    \hline
    \multirow{2}{*}{\textbf{Metric}} & \multirow{2}{*}{\textbf{OAR}} & \multicolumn{2}{c|}{\textbf{Model performance}} & \multirow{2}{*}{\begin{tabular}[c]{@{}c@{}}\textbf{Paired}\\ \textbf{differences}\end{tabular}} & \multirow{2}{*}{\begin{tabular}[c]{@{}c@{}}\textbf{Corrected}\\ \textbf{p-value}\end{tabular}}\tabularnewline 
    \cline{3-4}
    && \textbf{CT only} & \textbf{MRI only*} &&  \tabularnewline 
    \hline\hline
    \multirow{57}{*}{Dice Score (\%) $\uparrow$} & Carotid artery - Left (n=60) &  $90.94 \pm 5.03$ [76.35 96.90] & $48.49 \pm 12.97$ [9.85 78.39] & $42.45 \pm 15.77$ [-1.91 81.13] & $4.20E-11$ \\ \cline{2-6} 
 & Carotid artery - Right (n=60) &  $91.60 \pm 4.50$ [72.30 97.05] & $50.14 \pm 10.35$ [24.46 76.24] & $41.46 \pm 12.93$ [2.31 71.58] & $4.20E-11$ \\ \cline{2-6} 
 & Arytenoids (n=50) &  $83.54 \pm 8.23$ [60.23 94.44] & $5.84 \pm 14.46$ [0.00 51.98] & $77.70 \pm 20.97$ [8.77 94.44] & $1.78E-14$ \\ \cline{2-6} 
 & Mandible bone (n=60) &  $97.61 \pm 1.52$ [90.75 99.12] & $63.58 \pm 21.37$ [26.65 90.27] & $34.02 \pm 22.05$ [4.57 72.36] & $4.20E-11$ \\ \cline{2-6} 
 & Buccal mucosa (n=60) &  $90.22 \pm 6.65$ [68.17 96.44] & $69.25 \pm 12.78$ [27.10 88.22] & $20.98 \pm 13.27$ [1.68 58.47] & $4.20E-11$ \\ \cline{2-6} 
 & Oral cavity (n=60) &  $96.77 \pm 2.78$ [86.26 98.95] & $76.96 \pm 17.05$ [34.34 94.61] & $19.81 \pm 17.61$ [1.25 64.34] & $4.20E-11$ \\ \cline{2-6} 
 & Cochlea - Left (n=60) &  $86.88 \pm 13.19$ [0.00 98.80] & $68.60 \pm 22.02$ [0.00 91.09] & $18.28 \pm 21.54$ [-7.21 95.96] & $2.23E-10$ \\ \cline{2-6} 
 & Cochlea - Right (n=60) &  $86.92 \pm 15.73$ [0.00 100.00] & $72.59 \pm 22.06$ [0.00 93.86] & $14.33 \pm 19.84$ [-4.45 97.14] & $1.35E-10$ \\ \cline{2-6} 
 & Cricopharyngeal inlet (n=50) &  $84.44 \pm 14.78$ [9.52 96.39] & $8.28 \pm 19.52$ [0.00 69.97] & $76.16 \pm 27.41$ [5.04 96.39] & $1.78E-14$ \\ \cline{2-6} 
 & Cervical esophagus (n=27) &  $73.22 \pm 27.32$ [0.00 97.29] & $12.49 \pm 24.45$ [0.00 75.99] & $60.73 \pm 35.53$ [0.00 97.29] & $1.95E-05$ \\ \cline{2-6} 
 & Anterior eyeball segment - Left (n=60) &  $91.41 \pm 5.92$ [65.73 96.86] & $74.91 \pm 16.19$ [0.00 92.53] & $16.50 \pm 16.25$ [-1.99 84.72] & $4.30E-11$ \\ \cline{2-6} 
 & Anterior eyeball segment - Right (n=60) &  $91.80 \pm 5.91$ [60.73 96.93] & $75.23 \pm 16.56$ [0.00 92.39] & $16.57 \pm 16.34$ [0.46 85.57] & $4.20E-11$ \\ \cline{2-6} 
 & Posterior eyeball segment - Left (n=60) &  $96.39 \pm 1.94$ [87.52 98.51] & $88.55 \pm 13.28$ [0.00 95.37] & $7.84 \pm 12.45$ [0.58 87.52] & $4.20E-11$ \\ \cline{2-6} 
 & Posterior eyeball segment - Right (n=60) &  $96.50 \pm 1.78$ [90.05 98.46] & $88.77 \pm 13.14$ [0.00 95.98] & $7.73 \pm 12.66$ [-0.02 91.40] & $4.20E-11$ \\ \cline{2-6} 
 & Lacrimal gland - Left (n=60) &  $87.25 \pm 9.24$ [42.70 96.43] & $72.75 \pm 17.23$ [0.00 91.97] & $14.50 \pm 15.00$ [-10.20 78.04] & $4.25E-10$ \\ \cline{2-6} 
 & Lacrimal gland - Right (n=60) &  $86.39 \pm 10.50$ [40.00 95.51] & $74.02 \pm 17.12$ [0.00 91.56] & $12.38 \pm 15.43$ [-9.20 93.53] & $4.27E-10$ \\ \cline{2-6} 
 & Submandibular gland - Left (n=59) &  $94.54 \pm 4.58$ [72.29 97.99] & $36.17 \pm 32.89$ [0.00 86.56] & $58.37 \pm 35.18$ [-0.46 97.93] & $4.40E-11$ \\ \cline{2-6} 
 & Submandibular gland - Right (n=60) &  $94.57 \pm 5.43$ [61.36 97.95] & $34.37 \pm 33.06$ [0.00 87.59] & $60.20 \pm 35.36$ [1.30 97.89] & $4.20E-11$ \\ \cline{2-6} 
 & Thyroid gland (n=48) &  $87.37 \pm 17.85$ [0.00 98.05] & $9.85 \pm 23.94$ [0.00 82.29] & $77.52 \pm 28.18$ [0.00 98.05] & $2.66E-09$ \\ \cline{2-6} 
 & Larynx - glottis (n=50) &  $89.67 \pm 6.12$ [73.39 95.92] & $9.75 \pm 20.14$ [0.00 69.32] & $79.92 \pm 24.38$ [6.44 95.92] & $1.78E-14$ \\ \cline{2-6} 
 & Larynx - supraglottic (n=59) &  $93.07 \pm 4.51$ [75.89 98.07] & $21.79 \pm 27.20$ [0.00 76.57] & $71.28 \pm 30.03$ [9.57 98.07] & $4.30E-11$ \\ \cline{2-6} 
 & Lips (n=60) &  $91.71 \pm 8.30$ [39.58 97.14] & $68.98 \pm 14.53$ [24.01 89.54] & $22.74 \pm 16.80$ [-16.70 71.61] & $8.96E-11$ \\ \cline{2-6} 
 & Optic chiasm (n=59) &  $79.67 \pm 9.50$ [50.27 95.40] & $70.57 \pm 16.85$ [0.00 89.74] & $9.10 \pm 18.04$ [-12.18 90.68] & $9.86E-05$ \\ \cline{2-6} 
 & Optic nerve - Left (n=60) &  $85.67 \pm 13.23$ [0.00 96.93] & $66.31 \pm 18.18$ [0.00 90.05] & $19.36 \pm 17.19$ [-1.59 91.17] & $4.40E-11$ \\ \cline{2-6} 
 & Optic nerve - Right (n=60) &  $86.80 \pm 13.12$ [0.00 97.65] & $69.16 \pm 16.91$ [0.00 88.82] & $17.64 \pm 14.90$ [-2.74 74.92] & $4.40E-11$ \\ \cline{2-6} 
 & Parotid gland - Left (n=60) &  $94.88 \pm 5.51$ [64.19 98.55] & $77.46 \pm 11.84$ [23.13 91.64] & $17.42 \pm 11.62$ [-1.06 44.42] & $4.20E-11$ \\ \cline{2-6} 
 & Parotid gland - Right (n=60) &  $95.09 \pm 4.63$ [74.56 98.58] & $78.75 \pm 10.29$ [30.86 90.94] & $16.34 \pm 10.36$ [-1.88 43.70] & $4.20E-11$ \\ \cline{2-6} 
 & Pituitary gland (n=60) &  $88.15 \pm 9.99$ [36.78 97.91] & $76.43 \pm 17.27$ [0.00 95.98] & $11.72 \pm 16.88$ [-8.03 85.52] & $1.45E-09$ \\ \cline{2-6} 
 & Spinal cord (n=60) &  $93.60 \pm 5.68$ [74.25 97.90] & $47.73 \pm 17.65$ [13.66 85.77] & $45.87 \pm 20.18$ [-5.89 83.59] & $4.20E-11$ \\ 
\hline \hline
\multirow{21}{*}{\makecell[c]{Hausdorff Distance\\ (mm) $\downarrow$}} & Carotid artery - Left (n=60) &  $1.65 \pm 2.19$ [1.00 13.91] & $30.96 \pm 16.14$ [1.41 79.95] & $-29.30 \pm 16.43$ [-78.53 1.64] & $6.46E-11$ \\ \cline{2-6} 
 & Carotid artery - Right (n=60) &  $1.33 \pm 0.84$ [1.00 6.48] & $28.15 \pm 15.05$ [1.41 68.78] & $-26.82 \pm 15.12$ [-65.78 -0.41] & $6.46E-11$ \\ \cline{2-6} 
 & Mandible bone (n=60) &  $2.98 \pm 15.11$ [1.00 119.01] & $27.84 \pm 33.37$ [1.41 103.98] & $-24.85 \pm 37.44$ [-102.98 106.20] & $5.24E-10$ \\ \cline{2-6} 
 & Buccal mucosa (n=60) &  $1.73 \pm 1.48$ [1.00 8.06] & $5.00 \pm 2.72$ [1.41 13.38] & $-3.27 \pm 2.66$ [-11.96 0.86] & $6.46E-11$ \\ \cline{2-6} 
 & Oral cavity (n=60) &  $1.50 \pm 1.05$ [1.00 5.39] & $8.85 \pm 6.84$ [1.41 31.00] & $-7.35 \pm 7.06$ [-29.59 0.00] & $6.46E-11$ \\ \cline{2-6} 
 & Lips (n=60) &  $1.51 \pm 1.34$ [1.00 9.91] & $6.54 \pm 5.23$ [1.00 23.09] & $-5.04 \pm 5.44$ [-21.67 4.43] & $5.27E-10$ \\ \cline{2-6} 
 & Optic chiasm (n=59) &  $1.24 \pm 0.46$ [0.00 3.00] & $1.64 \pm 0.91$ [1.00 6.08] & $-0.40 \pm 0.85$ [-4.67 1.00] & $3.41E-04$ \\ \cline{2-6} 
 & Optic nerve - Right (n=59) &  $1.10 \pm 0.38$ [0.00 2.56] & $1.79 \pm 1.39$ [1.00 10.74] & $-0.69 \pm 1.39$ [-9.74 0.82] & $1.25E-06$ \\ \cline{2-6} 
 & Parotid gland - Left (n=60) &  $1.97 \pm 2.64$ [1.00 19.80] & $7.65 \pm 4.29$ [1.73 24.39] & $-5.67 \pm 4.02$ [-13.80 2.52] & $1.14E-10$ \\ \cline{2-6} 
 & Parotid gland - Right (n=60) &  $3.99 \pm 16.17$ [1.00 126.24] & $7.79 \pm 4.49$ [2.24 24.02] & $-3.80 \pm 16.28$ [-23.02 113.55] & $9.18E-09$ \\ \cline{2-6} 
 & Spinal cord (n=60) &  $2.20 \pm 6.38$ [1.00 50.45] & $36.11 \pm 26.70$ [2.24 176.77] & $-33.90 \pm 22.67$ [-126.32 0.21] & $6.46E-11$ \\ 
 \hline
    \end{tabular}
    \caption{Average performance for significantly different OAR performances at 95\% confidence level according to the corrected Wilcoxon signed-rank test between CT only and MRI only predictions of model trained with MD. Results are presented as mean $\pm$ standard deviation [minimum maximum]. *Results are computed only for non NaN values. }
    \label{table:sig_ctonly_vs_mrionly_md}
\end{table}
Table~\ref{table:sig_ctonly_vs_mrionly_md} shows OAR performances that were significantly different at a 95\% confidence level according to the corrected Wilcoxon signed-rank test. Brainstem performance is not shown because performance difference was not significant enough at the 95\% confidence level but it was at the 90\% confidence level. For OARs outside the MRI FOV, MRI predictions HD resulted in NaN and thus, HD metrics for these specific OARs are not shown in this table, even thought they are significantly different than the corresponding ones for CT only predictions.

\subsubsection{Predictions using CT and MRI scans}

Table~\ref{table:mri_ct_pred_md_vs_nomd} shows performance of models trained with and without MD and inferred using CT and MRI scans, for OARs whose performance were significantly different at a 95\% confidence level according to the corrected Wilcoxon signed-rank test. 
\begin{table}[h!]
    \scriptsize
    \centering
    \begin{tabular}{|P{6em}|P{11em}|P{8em}|P{8em}|P{7em}|P{6em}|}
    \hline
    \multirow{4}{*}{\textbf{Metric}} & \multirow{4}{*}{\textbf{OAR}} & \multicolumn{2}{c|}{\textbf{Model performance}} & \multirow{4}{*}{\begin{tabular}[c]{@{}c@{}}\textbf{Paired}\\ \textbf{differences}\end{tabular}} & \multirow{4}{*}{\begin{tabular}[c]{@{}c@{}}\textbf{Corrected}\\ \textbf{p-value}\end{tabular}} \tabularnewline 
    \cline{3-4}
    && \textbf{With Modality Dropout} & \textbf{Without Modality Dropout} & &  \tabularnewline 
    \hline\hline
\multirow{17}{*}{\makecell[c]{Dice Score\\ (\%) $\uparrow$}} & Carotid artery - Left (n=60) &  $91.27 \pm 4.49$ [77.82 96.77] & $91.77 \pm 4.44$ [76.18 97.34] & $-0.51 \pm 1.34$ [-3.45 4.39] & $1.88E-03$ \\ \cline{2-6} 
 & Mandible bone (n=60) &  $97.59 \pm 1.53$ [90.87 99.09] & $97.69 \pm 1.46$ [91.98 99.18] & $-0.11 \pm 0.25$ [-1.11 0.76] & $1.88E-03$ \\ \cline{2-6} 
 & Oral cavity (n=60) &  $96.78 \pm 2.78$ [86.02 98.82] & $96.95 \pm 2.69$ [86.83 99.12] & $-0.18 \pm 0.37$ [-1.28 0.63] & $2.26E-03$ \\ \cline{2-6} 
 & Lacrimal gland - Right (n=60) &  $86.58 \pm 10.19$ [40.00 96.11] & $87.25 \pm 9.95$ [40.78 96.20] & $-0.68 \pm 1.74$ [-5.57 3.66] & $2.25E-02$ \\ \cline{2-6} 
 & Submandibular gland - Left (n=59) &  $94.63 \pm 4.34$ [73.42 97.96] & $94.90 \pm 4.31$ [74.31 98.19] & $-0.27 \pm 0.61$ [-2.27 1.40] & $2.97E-03$ \\ \cline{2-6} 
 & Optic chiasm (n=59) &  $80.20 \pm 8.80$ [46.68 92.79] & $81.61 \pm 9.48$ [48.08 92.59] & $-1.41 \pm 3.49$ [-8.95 9.80] & $2.97E-03$ \\ \cline{2-6} 
 & Parotid gland - Left (n=60) &  $94.97 \pm 5.33$ [64.22 98.51] & $95.19 \pm 5.22$ [64.89 98.49] & $-0.23 \pm 0.62$ [-1.92 2.26] & $1.13E-02$ \\ \cline{2-6} 
 & Parotid gland - Right (n=60) &  $95.14 \pm 4.46$ [74.85 98.52] & $95.45 \pm 4.05$ [80.78 98.50] & $-0.30 \pm 1.10$ [-7.27 2.04] & $7.72E-03$ \\ \cline{2-6} 
 & Spinal cord (n=60) &  $93.97 \pm 5.18$ [76.09 98.23] & $94.40 \pm 4.76$ [76.81 98.00] & $-0.43 \pm 0.88$ [-3.69 1.49] & $2.67E-03$ \\ 

 \hline
    \end{tabular}
    \caption{Average performance for significantly different OAR performances at 95\% confidence level according to the corrected Wilcoxon signed-rank test between predictions using CT and MRI scans of models trained with and without MD. Results are presented as mean $\pm$ standard deviation [minimum maximum].}
    \label{table:mri_ct_pred_md_vs_nomd}
\end{table}

\newpage

\section{Left-Right OAR combination}\label{sec:leftright}

Table~\ref{table:left_right} shows the performance of models trained with and without left and right OARs combined for OARs whose performance was significantly different at a 95\% confidence level according to the corrected Wilcoxon signed-rank test. It can be seen that the model trained with the OAR combination performs better for 11 left/right structures. 
\begin{table}[h!]
    \scriptsize
    \centering
    \begin{tabular}{|P{10em}|P{11em}|P{7em}|P{7em}|P{7em}|P{5em}|}
    \hline
    \multirow{4}{*}{\textbf{Metric}} & \multirow{4}{*}{\textbf{OAR}} & \multicolumn{2}{c|}{\textbf{Model performance}} & \multirow{4}{*}{\begin{tabular}[c]{@{}c@{}}\textbf{Paired}\\ \textbf{differences}\end{tabular}} & \multirow{4}{*}{\begin{tabular}[c]{@{}c@{}}\textbf{Corrected}\\ \textbf{p-value}\end{tabular}} \tabularnewline 
    \cline{3-4}
    && \textbf{With OAR combination} & \textbf{Without OAR combination} & &  \tabularnewline 
    \hline\hline
\multirow{37}{*}{Dice Score (\%) $\uparrow$} & Carotid artery - Left (n=60) &  $91.27 \pm 4.49$ [77.82 96.77] & $90.58 \pm 4.53$ [76.69 96.04] & $0.69 \pm 1.43$ [-2.96 3.68] & $2.45E-03$ \\ \cline{2-6} 
 & Carotid artery - Right (n=60) &  $91.79 \pm 3.82$ [77.40 96.87] & $90.91 \pm 4.12$ [74.52 96.59] & $0.87 \pm 1.17$ [-2.09 4.20] & $7.73E-06$ \\ \cline{2-6} 
 & Arytenoids (n=50) &  $83.66 \pm 8.35$ [59.23 94.44] & $82.46 \pm 8.52$ [57.76 94.07] & $1.20 \pm 3.06$ [-5.00 11.43] & $2.35E-02$ \\ \cline{2-6} 
 & Mandible bone (n=60) &  $97.59 \pm 1.53$ [90.87 99.09] & $97.49 \pm 1.50$ [90.41 99.01] & $0.10 \pm 0.18$ [-0.26 0.49] & $1.23E-03$ \\ \cline{2-6} 
 & Buccal mucosa (n=60) &  $90.24 \pm 6.63$ [68.44 96.46] & $89.25 \pm 6.76$ [67.55 96.17] & $0.99 \pm 1.16$ [-4.20 4.24] & $3.06E-07$ \\ \cline{2-6} 
 & Oral cavity (n=60) &  $96.78 \pm 2.78$ [86.02 98.82] & $96.38 \pm 4.27$ [68.90 98.85] & $0.40 \pm 2.22$ [-1.28 17.12] & $5.19E-03$ \\ \cline{2-6} 
 & Anterior eyeball segment - Left (n=60) &  $91.55 \pm 5.61$ [67.50 96.50] & $91.02 \pm 5.63$ [69.49 97.56] & $0.54 \pm 1.40$ [-2.72 3.81] & $8.61E-03$ \\ \cline{2-6} 
 & Anterior eyeball segment - Right (n=60) &  $91.96 \pm 5.67$ [61.22 96.70] & $91.41 \pm 5.51$ [63.55 96.48] & $0.55 \pm 1.53$ [-3.52 3.98] & $1.22E-02$ \\ \cline{2-6} 
 & Posterior eyeball segment - Left (n=60) &  $96.44 \pm 1.94$ [87.48 98.40] & $96.33 \pm 1.85$ [88.18 98.28] & $0.10 \pm 0.37$ [-0.75 1.01] & $2.81E-02$ \\ \cline{2-6} 
 & Lacrimal gland - Left (n=60) &  $87.56 \pm 9.14$ [40.91 96.80] & $85.90 \pm 9.31$ [39.08 95.38] & $1.66 \pm 2.24$ [-3.71 6.50] & $3.52E-05$ \\ \cline{2-6} 
 & Submandibular gland - Left (n=59) &  $94.63 \pm 4.34$ [73.42 97.96] & $94.29 \pm 4.33$ [72.38 97.81] & $0.34 \pm 0.67$ [-0.92 2.41] & $1.82E-03$ \\ \cline{2-6} 
 & Submandibular gland - Right (n=60) &  $94.50 \pm 5.39$ [62.30 97.91] & $92.73 \pm 13.19$ [0.00 97.54] & $1.77 \pm 10.71$ [-1.71 83.87] & $1.04E-03$ \\ \cline{2-6} 
 & Larynx - glottis (n=50) &  $89.75 \pm 6.08$ [72.95 95.93] & $88.86 \pm 6.37$ [70.64 95.69] & $0.89 \pm 2.21$ [-3.35 9.98] & $8.36E-03$ \\ \cline{2-6} 
 & Larynx - supraglottic (n=59) &  $93.12 \pm 4.39$ [77.49 98.04] & $92.82 \pm 4.31$ [76.84 97.24] & $0.30 \pm 0.66$ [-2.46 1.76] & $2.83E-04$ \\ \cline{2-6} 
 & Lips (n=60) &  $91.74 \pm 8.24$ [40.11 97.12] & $91.24 \pm 8.02$ [40.98 96.51] & $0.51 \pm 0.86$ [-1.50 2.79] & $3.09E-04$ \\ \cline{2-6} 
 & Optic nerve - Left (n=60) &  $85.58 \pm 13.26$ [0.00 96.33] & $84.68 \pm 12.96$ [0.00 94.59] & $0.89 \pm 2.01$ [-4.51 5.67] & $3.43E-03$ \\ \cline{2-6} 
 & Parotid gland - Left (n=60) &  $94.97 \pm 5.33$ [64.22 98.51] & $94.64 \pm 5.54$ [64.14 98.31] & $0.32 \pm 0.98$ [-1.22 7.03] & $1.23E-03$ \\ \cline{2-6} 
 & Parotid gland - Right (n=60) &  $95.14 \pm 4.46$ [74.85 98.52] & $94.98 \pm 4.29$ [77.85 98.13] & $0.17 \pm 0.98$ [-5.11 4.97] & $2.24E-03$ \\ \cline{2-6} 
 & Pituitary gland (n=60) &  $88.19 \pm 9.60$ [37.79 98.02] & $86.98 \pm 9.52$ [38.83 96.62] & $1.21 \pm 3.39$ [-6.41 11.06] & $7.98E-03$ \\
\hline \hline
Hausdorff Distance (mm) $\downarrow$ & Buccal mucosa (n=60) &  $1.73 \pm 1.47$ [1.00 8.06] & $1.91 \pm 1.57$ [1.00 9.17] & $-0.17 \pm 0.35$ [-1.63 0.76] & $9.07E-03$ \\ 
 \hline
    \end{tabular}
    \caption{Average performance for significantly different OAR performances at 95\% confidence level according to the corrected Wilcoxon signed-rank test between predictions of models trained with left right OAR combined or not. Results are presented as mean $\pm$ standard deviation [minimum maximum].}
    \label{table:left_right}
\end{table}